\documentclass[11pt,a4paper]{article}
\usepackage[utf8]{inputenc}
\usepackage[T1]{fontenc}
\usepackage{amsmath}
\usepackage{amsfonts}
\usepackage{amssymb}
\usepackage{graphicx}

\providecommand{\keywords}[1]{\small\textbf{Keywords: } #1}
\usepackage[margin=2cm]{geometry}

\usepackage{authblk}
\usepackage{lipsum}
\usepackage{dsfont}
\usepackage{mathtools}
\usepackage{amsthm}
\usepackage{thmtools}
\usepackage{tcolorbox}
\usepackage{physics}
\usepackage{algorithm}
\usepackage{algpseudocode}

\usepackage{subcaption}
\usepackage{wrapfig}

\def\sizetwobyone{0.48}
\def\sizeonebyone{0.24}

\usepackage{siunitx}
\usepackage{thmtools}
\newtheoremstyle{break}{\topsep}{\topsep}{\itshape}{}{\bfseries}{}{\newline}{}
\theoremstyle{break}

\mathtoolsset{showonlyrefs}

\DeclareMathOperator*{\argmin}{arg\,min}

\usepackage[backend=biber, style=numeric,giveninits, doi=true, url=false, eprint=true, isbn=false]{biblatex}
\DeclareSourcemap{
  \maps{
    \map{
      \step[fieldset=month, null]
      \step[fieldset=address, null]
      \step[fieldset=location, null]
      \step[fieldset=url, null]
      \step[fieldset=isbn, null]
      \step[fieldset=publisher, null]
       \step[fieldset=_publisher, null]
    }
  }
}
\usepackage[colorlinks=true, allcolors=blue, hypertexnames=false]{hyperref}

\title{A deep BSDE approach for the simultaneous pricing and delta-gamma hedging of large portfolios consisting of high-dimensional multi-asset Bermudan options}

\author[1, 2]{Balint Negyesi\thanks{Corresponding author. Email: \href{mailto:B.Negyesi@tudelft.nl}{B.Negyesi@tudelft.nl}. Postal address: P.O. Box 5031, 2600 GA Delft, The Netherlands.}}
\affil[1]{\small Delft Institute of Applied Mathematics (DIAM), Delft University of Technology}

\author[2]{Cornelis W. Oosterlee\thanks{Email: \href{mailto:C.W.Oosterlee@uu.nl}{C.W.Oosterlee@uu.nl}. Postal address: Postbus 80010, 3508 TA Utrecht, The Netherlands.}}
\affil[2]{Mathematical Institute, Utrecht University}

\begin{document}
	\maketitle
	\begin{abstract}
    A deep BSDE approach is presented for the pricing and delta-gamma hedging of high-dimensional Bermudan options, with applications in portfolio risk management. Large portfolios of a mixture of multi-asset European and Bermudan derivatives are cast into the framework of discretely reflected BSDEs. This system is discretized by the One Step Malliavin scheme (Negyesi et al. [2024, 2025]) of discretely reflected Markovian BSDEs, which involves a $\Gamma$ process, corresponding to second-order sensitivities of the associated option prices. The discretized system is solved by a neural network regression Monte Carlo method, efficiently for a large number of underlyings. The resulting option Deltas and Gammas are used to discretely rebalance the corresponding replicating strategies.
    Numerical experiments are presented on both high-dimensional basket options and large portfolios consisting of multiple options with varying early exercise rights, moneyness and volatility. These examples demonstrate the robustness and accuracy of the method up to $100$ risk factors. The resulting hedging strategies significantly outperform benchmark methods both in the case of standard delta- and delta-gamma hedging.
\end{abstract}


\section{Introduction}\label{sec:introduction}

 In this paper we are concerned with the challenging hedging problem of large portfolios of high-dimensional options with early-exercise features. Consider an investor who is in possession of a portfolio consisting of $J$ many options whose values depend on a set of common risk factors $\{X_t\}_{0\leq t\leq T}$ which can be decomposed into a set of tradeable underlying assets $\{S_t=(S_t^{1}, \dots, S_t^{m})\}_{0\leq t\leq T}$ and some other non-tradeable component $\{\nu_t=(\nu_t^1, \dots, \nu_t^{d-m})\}_{0\leq t\leq T}$, and they together form an It\^{o} process $X_t\coloneqq(S_t, \nu_t)$ solving the following stochastic differential equation (SDE)
 \begin{align}\label{eq:introduction:sde}
     X_t = X_0 + \int_0^t \mu(s, X_s)\mathrm{d}s + \int_0^t \sigma(s, X_s)\mathrm{d}W_s,
 \end{align}
 where $\mu: [0, T]\times \mathbb{R}^{d}\to\mathbb{R}^{d}$, $\sigma: [0, T]\times \mathbb{R}^{d}\to\mathbb{R}^{d\times d}$, and $\{W_t\}_{0\leq t\leq T}$ a $d$-dimensional Brownian motion in an appropriate probability space.
 
 Fixing some time horizon $0<T<\infty$, let $\mathcal{R}^j\subseteq [0, T]$ denote the set of early exercise opportunities for each $j=1, \dots, J$. We are concerned with the Markovian framework, i.e. options whose prices are deterministic mappings of the underlying risk factors at each point in time, and we put $v^j: [0, T]\times \mathbb{R}^{d}\to \mathbb{R}$, $j=1, \dots, J$ for each of these functions.
 The investor's objective is to insure her positions against random movements in the underlyings and she achieves this through a hedging strategy. In particular, she constructs a delta hedging replicating portfolio consisting of a short position in all options; long position in all underlyings and a deposit in a bank account
\begin{align}\label{eq:delta_hedging:portfolio_value}
     \mathrm{d}P_t^\Delta = -\sum_{j=1}^J \mathrm{d}v^j(t, X_t)+ \sum_{i=1}^{m}\alpha_t^{i}(\mathrm{d}S_t^{i} + q_t^{i}S_t^{i}\mathrm{d}t) + \mathrm{d}B_t,\quad P_0^\Delta=0,
 \end{align}
 where we allow each underlying $S_t^i$ to pay off dividends continuously with a rate $q^{i}_t$ at time $t$.
 It is well-known that the variance-minimizing first-order conditions $\partial P_t^\Delta / \partial S_t^{i}=0$ result in the optimal hedging weights
 \begin{align}\label{eq:delta_hedging:foc}
     \alpha_t^{i} = \sum_{j=1}^J \frac{\partial v^j}{\partial S_t^{i}}(t, X_t),\quad i=1, \dots, m.
 \end{align}
 In particular, \eqref{eq:delta_hedging:portfolio_value} and \eqref{eq:delta_hedging:foc} together with It\^{o}'s lemma imply that given a machinery which yields simultaneous option prices and deltas, the investor can perfectly offset her exposure in the underlyings, at least in the continuous, complete market setting.

The discussion above motivates to cast our problem into the framework of (decoupled) forward-backward stochastic differential equations (FBSDE). In fact, it is classically known, see e.g. \cite{pardoux_backward_1992, zhang_backward_2017}, that in absence of early exercise rights $\mathcal{R}^j=\{0, T\}$, the $j$th option is intimately related to the following standard (Markovian) BSDE
\begin{align}\label{eq:introduction:bsde}
    Y_t^j = g^j(X_T) + \int_t^T f^j(s, X_s, Y_s^j, Z_s^j)\mathrm{d}s - \int_t^T Z_s^j\mathrm{d}W_s,
\end{align}
where $g^j: \mathbb{R}^{d}\to \mathbb{R}$ denotes the payoff and $f^j:[0, T]\times \mathbb{R}^{d}\times \mathbb{R} \times \mathbb{R}^{1\times d}\to \mathbb{R}$ the driver. Namely, non-linear extensions to the Feynman-Kac relations establish the relations
\begin{align}\label{eq:feynman_kac}
    Y_t^j = v^j(t, X_t),\quad Z_t^j = \nabla_x v^j(t, X_t)\sigma(t, X_t),
\end{align}
in an almost sure sense.
Comparing \eqref{eq:delta_hedging:foc} with \eqref{eq:feynman_kac}, one can conclude that solving the BSDE associated to the option is, in a sense, equivalent with the task of (delta-)hedging. Similar relations hold in case of early-exercise rights $\mathcal{R}^j\setminus \{0, T\}\neq\emptyset$, see section \ref{sec:discretely_reflected_fbsdes} below.
Over the last three decades a vast literature has been developed dealing with the numerical resolution of different types of BSDEs, see e.g. \cite{bouchard_discrete-time_2004, bender_forward_2007, briand_simulation_2014, ma_solving_1994, gobet_regression-based_2005} and the references therein.

However, whenever the aforementioned portfolio \eqref{eq:delta_hedging:portfolio_value} is high-dimensional, i.e. $d$ or $J$ is large, one either has to deal with a high-dimensional BSDE \eqref{eq:introduction:bsde}, or a large number of equations simultaneously, potentially both. This makes classical numerical methods intractable in the context of this work, as they all suffer from the curse of dimensionality. In recent years, initiated by the pioneering paper \cite{han_solving_2018}, a rapidly growing research line has been developed by the numerical analysis community, where BSDEs of the type \eqref{eq:introduction:bsde} are approximated in regression Monte Carlo frameworks using deep neural networks to parameterize the (Markovian) solution pair of \eqref{eq:introduction:bsde}. Without the sake of completeness we mention \cite{hure_deep_2020, germain_deep_2020, becker_pricing_2020, chassagneux_deep_2022, chen_deep_2021}. These methods have shown remarkable empirical results tackling the numerical solution of \eqref{eq:introduction:bsde}, and by now some results are also known about their convergence properties up to universal approximation type, see \cite{germain_approximation_2022, han_convergence_2020, negyesi_one_2024}. Moreover, they have successfully been applied in the context of hedging single options, see \cite{chen_deep_2021, becker_pricing_2020, gnoatto_deep_2024}.

However, all these aforementioned methods solely focus on solving the hedging problem explained by \eqref{eq:delta_hedging:portfolio_value}. Nonetheless, whenever rebalancing is only done over a finite set of dates in time, delta hedging does not achieve a perfect replication and due to the discrete time approximations, the corresponding portfolio entails risk.
In particular, in a volatile economic climate, corresponding to large volatilities in the diffusion component of \eqref{eq:introduction:sde}, the deltas on the right hand side of \eqref{eq:delta_hedging:foc} change rapidly and the corresponding discrete replication error of \eqref{eq:delta_hedging:portfolio_value} also grows accordingly.
In order to mitigate the effect of fluctuating deltas, one can impose additional second-order constraints on top of \eqref{eq:delta_hedging:foc}, which effectively set the second order sensitivities, Gammas, of the accordingly constructed replicating portfolio to zero, in terms of the underlying risk factors. Doing so, one encounters two additional challenges. First, as assets themselves have vanishing gammas, in order to be able to formulate the corresponding second order conditions, one needs to augment the replicating portfolio with \emph{gamma hedging instruments}, whose prices and Greeks are available at all points in time. Second, the resulting second order conditions involve appropriate second order sensitivities of the underlying options that are meant to be hedged. This implies additional modelling error, as in order to effectively carry out the gamma hedging strategy, the investor does not merely have to efficiently model the underlying options' prices and deltas, but also their \emph{gammas}, even in the high-dimensional setting, consisting of many risk factors. For details, we refer to section \ref{sec:gamma_hedging_and_osm} below. 

The main objective of the present paper is to develop a deep BSDE methodology which efficiently tackles the aforementioned challenges in the high-dimensional portfolio framework. In fact, motivated by the ideas in \cite{negyesi_one_2024, negyesi_reflected_2024}, one can derive an additional vector-valued, linear BSDE related to \eqref{eq:introduction:bsde}, whose solution pair involves a \emph{matrix-valued} process corresponding to second-order sensitivities of the underlying option. Consequently, solving this additional BSDE together with \eqref{eq:introduction:bsde} naturally extends the Feynman-Kac relations in \eqref{eq:feynman_kac}, and results in a \emph{triple} of stochastic processes, which coincide with option prices, deltas and gammas, respectively. Given robust and efficient numerical approximations of this stochastic triple, one can mitigate the additional modelling errors arising in delta-gamma hedging, and assess the accumulating discrete replication errors in the sole delta-hedging framework.

Our main contributions are as follows. We propose a deep BSDE methodology for the portfolio hedging problem outlined above. Therefore, as a side result, we first extend the application of deep BSDE methods to the context of delta-hedging large portfolios, instead of only single options as in \cite{chen_deep_2021, becker_pricing_2020, gnoatto_deep_2024}, in the complete Bermudan setting. This is done by casting the method of \cite{hure_deep_2020} into the vector-valued BSDE framework, and thus simultaneously solving all $J$ options' pricing and delta-hedging problems. Thereafter, and most importantly, in order to reduce the discrete replication error of the delta-hedging portfolio, we propose a Gamma hedging strategy, which on top of the first-order conditions in \eqref{eq:delta_hedging:foc}, also aims to offset second-order terms in the portfolio's value by imposing second-order conditions, depending on second-order sensitivities, i.e. \emph{Gammas} of the option's value. Such hedging strategies result in less frequent rebalancing and more accurate replication.
However, by doing so, one exposes herself to additional model risk, as the Gammas must accurately be approximated in the numerical setting.  In order to address this, we use recent results on the One Step Malliavin scheme, first proposed in \cite{negyesi_one_2024} for standard BSDEs and later extended to discretely reflected equations in \cite{negyesi_reflected_2024}, and provide a robust and accurate, fully-implementable deep BSDE method for the simultaneous delta-gamma-hedging of large portfolios. We demonstrate that this novel approach may provide a significant improvement to standard delta-hedging strategies in the discrete time framework, whenever the underlyings' deltas are highly volatile, resulting in less frequent rebalancing, and sharper Profit-and-Loss $(\text{PnL})$ distributions.

The paper is organized as follows. Section \ref{sec:discretely_reflected_fbsdes} gives a short summary of the necessary theoretical basis by establishing the connections between Bermudan options and discretely reflected FBSDEs. In section \ref{sec:gamma_hedging_and_osm}, we present the delta-gamma hedging strategies and their corresponding first- and second-order conditions. Thereafter, we apply the discretizations in \cite{negyesi_one_2024, negyesi_reflected_2024} in the vector-valued portfolio framework, and explain how the resulting approximations are applicable in a portfolio hedging context. In section \ref{sec:deep_bsde} we explain how the previous works from \cite{hure_deep_2020, negyesi_one_2024, negyesi_reflected_2024} extend to vector-valued equations and can be used to approximate the collection of BSDEs corresponding to problem \eqref{eq:delta_hedging:portfolio_value} in a Deep BSDE approach. Finally, we demonstrate the accuracy and robustness of these strategies by numerical experiments performed on high-dimensional portfolios in section \ref{sec:numerical_experiments}.

\section{Bermudan options as discretely reflected FBSDEs}\label{sec:discretely_reflected_fbsdes}
We fix $0\leq T< \infty$ and let $J, j, d, m, k\in \mathbb{N}_+$. Throughout the whole paper we are working on a filtered probability space $(\Omega, \mathcal{F}, \mathbb{P}, \{\mathcal{F}_t\}_{t\in[0, T]})$ with $\mathcal{F}=\mathcal{F}_T$, where $\mathbb{F}$ is the natural filtration generated by a $d$-dimensional Brownian motion $\{W_t\}_{t\in[0, T]}$, augmented by the usual $\mathbb{P}$-null sets. In what follows, all equalities concerning $\mathcal{F}_t$ measureable random variables are meant in the $\mathbb{P}$ almost sure sense, and all expectations are taken under $\mathbb{P}$, unless otherwise stated. As usual, we put $\abs{x}\coloneqq \tr{x^\top x}$, for any $x\in \mathbb{R}^{j\times d}$, and remark that this coincides with the Euclidean norm in case of scalars and vectors. We define $\mathbb{H}^p(\mathbb{R}^{j\times d})$ to be the space of $\mathbb{R}^{j\times d}$ valued, progressively measurable stochastic processes such $Z\in \mathbb{H}^p(\mathbb{R}^{j\times d})$: $\mathbb{E}[(\int_0^T \abs{Z_t}^2\mathrm{d}t)^{p/2}]<\infty$. Similarly, $\mathbb{S}^p(\mathbb{R}^{j\times d})\subset \mathbb{H}^p(\mathbb{R}^{j\times d})$, for which in addition $Y\in\mathbb{S}^p(\mathbb{R}^{j\times d})$ is also continuous and admits to $\mathbb{E}[\sup_{t\in [0, T]} \abs{Y_t}^p]<\infty$. In what follows, for any multivariate $f: [0, T]\times \mathbb{R}^d\to \mathbb{R}^j$ function we set $\nabla_x f$ to be the Jacobian matrix taking values in $\mathbb{R}^{j\times d}$. In particular, for a scalar valued function $v: [0, T]\times \mathbb{R}^d\to \mathbb{R}$, we use $\partial_i v$ to denote the $i$'th partial derivative in space, and $\partial_{ji}^2 v$ for the corresponding element in the Hessian matrix. Given a time partition $\mathcal{N}\coloneqq\{0=t_0<t_1<\dots<t_N=T\}$ we set $\mathbb{E}_n[\cdot]\coloneqq \mathbb{E}[\cdot\vert\mathcal{F}_{t_{n}}]$.

With the above notation at hand, we can now formulate discretely reflected BSDEs. 
Heuristically speaking, a reflected BSDE is a generalization of \eqref{eq:introduction:bsde} such that the solution is also forced to stay above a (Markovian) lower boundary process. The forcing is referred to as \emph{reflection}.
Discretely reflected BSDEs are special cases of reflected BSDEs, see e.g. \cite{karoui_reflected_1997, zhang_backward_2017}, where reflection can only occur over a finite set of points in time $\mathcal{R}^j=\{r_i^j, i=0, \dots, R^j\vert r_0^j=0, r_{R^{j}}^j=T\}$.
The solution to a discretely reflected BSDE indexed by $j$ is a pair of stochastic processes $(Y^j, Z^j)\in \mathbb{S}^2(\mathbb{R})\times \mathbb{H}^2(\mathbb{R}^{1\times d_j})$ such that
\begin{align}\label{eq:bsde:discretely_reflected}
    \begin{aligned}
    Y_T^j &= \widetilde{Y}_T^j \coloneqq g^j(X_T),\\
    \widetilde{Y}_t^j &= Y_{\overline{r}^j_t}^j + \int_t^{\overline{r}^j_t} f^j(s, X_s, \widetilde{Y}_s^j, Z_s^j)\mathrm{d}s - \int_t^{\overline{r}^j_t}Z_s^j\mathrm{d}W_s,\quad t\in [0, T],\\
    Y_t^j&\coloneqq \widetilde{Y}_t^j + \mathbf{1}_{t\in\mathcal{R}^j\setminus \{0, T\}}\mathbf{1}_{l^j(X_t)>\widetilde{Y}^j_t}[l^j(X_t)-\widetilde{Y}_t^j]\eqqcolon \mathfrak{R}^j_y(t, X_t, \widetilde{Y}_t^j),
    \end{aligned}
\end{align}
where $\underline{r}_t^j\coloneqq \sup\{r\in \mathcal{R}^j: r\leq t\}$, $\overline{r}_t^j\coloneqq \inf\{r\in \mathcal{R}^j: r>t\}$ and $g^j, l^j: \mathbb{R}^{d_{j}}\to\mathbb{R}$ are deterministic mappings corresponding to the terminal condition and (Markovian) lower boundary process, respectively. Equation \eqref{eq:bsde:discretely_reflected} together with \eqref{eq:introduction:sde} forms a discretely reflected FBSDE system which, under appropriate assumptions, admits a unique solution triple, see e.g. \cite{karoui_reflected_1997}. Reflected BSDEs are inherently related to second-order, semi-linear parabolic, free-boundary PDEs. In fact, under suitable assumptions, they retain a Feynman-Kac type relation similar to \eqref{eq:feynman_kac} in the standard case, see \cite{karoui_reflected_1997}.
From the financial mathematics perspective, the main relevance of such equations is that they are a natural model for optimal stopping problems, such as Bermudan options, where $\mathcal{R}^j$ corresponds to the set of early-exercise dates, $g^j\equiv l^j$ to the instantaneous payoff, $\widetilde{Y}_t^j, Y_t^j, Z_t^j$ to the continuation value, option price and option delta at time $t$, respectively. Notice that \eqref{eq:bsde:discretely_reflected} includes the standard, Markovian BSDE framework associated with European options by letting $\mathcal{R}^j=\{0, T\}$. Furthermore, from the numerical point of view, a suitable discretization of \eqref{eq:bsde:discretely_reflected} is an approximation of American options in the asymptotic $R^j\to\infty$. Hence, in what follows we refer to all of the options in \eqref{eq:delta_hedging:portfolio_value} as solutions to a discretely reflected FBSDE.

Similar to \eqref{eq:introduction:sde}, we define $g\coloneq (g^1;\dots;g^J): \mathbb{R}^d\to\mathbb{R}^J$, $g\coloneqq (f^1; \dots; f^J):[0, T]\times \mathbb{R}^d\times \mathbb{R}^J\times \mathbb{R}^{J\times d}\to \mathbb{R}^J$, $\widetilde{Y}_t\coloneqq (\widetilde{Y}_t^1; \dots; \widetilde{Y}_t^J)$, $Y_t\coloneqq (Y_t^1; \dots; Y_t^J)$ and $Z_t\coloneqq (Z_t^1; \dots; Z_t^J)$ by the row-wise concatenated solutions of \eqref{eq:bsde:discretely_reflected} for each $j=1, \dots, J$. 
Therefore, $(Y, Z)\in \mathbb{S}^2(\mathbb{R}^J)\times \mathbb{H}^2(\mathbb{R}^{J\times d})$ satisfy a \emph{collection} of discretely reflected BSDEs.
It is important to notice that even though in this formulation we cast the problem of $J$ options into the framework of vector-valued, discretely reflected BSDEs, this indeed is of mere formal convenience, and the corresponding system is only a collection not a system, i.e. the solution pair $(Y^i, Z^i)$ does not enter the dynamics of $(Y^j, Z^j)$ for $i\neq j$. This is of fundamental importance, since the well-posedness of multi-dimensional reflected BSDEs remains to be a challenging open problem due to the lack of comparison principles in the vector-valued setting, see \cite{zhang_backward_2017} and the references therein. Nevertheless, in the context of our work, there is no cross-dependence between equations of the type \eqref{eq:bsde:discretely_reflected}, and the resulting collections can safely be treated without the aforementioned theoretical obstacles. Henceforth the system of discretely reflected BSDEs simultaneously representing all options in \eqref{eq:delta_hedging:portfolio_value} reads as follows
\begin{align}\label{eq:bsde:discretely_reflected:collection}
    \begin{aligned}
    Y_T &= \widetilde{Y}_T \coloneqq (g^1(X_T), \dots, g^J(X_T)),\\
    \widetilde{Y}_t &= \begin{aligned}[t]
        (Y_{\overline{r}^1_t}^1; \dots; Y_{\overline{r}^J_t}^J) &+ \Big(\int_t^{\overline{r}^1_t} f^1(s, X_s, \widetilde{Y}_s^1, Z_s^1)\mathrm{d}s; \dots; \int_t^{\overline{r}^J_t} f^J(s, X_s, \widetilde{Y}_s^J, Z_s^J)\mathrm{d}s\Big)\\
        &- \Big(\int_t^{\overline{r}^1_t}Z_s^1\mathrm{d}W_s; \dots; \int_t^{\overline{r}^J_t}Z_s^J\mathrm{d}W_s\Big),\quad t\in [\underline{r}^j_t, \overline{r}^j_t),
    \end{aligned}\\
    Y_t&\coloneqq (\mathfrak{R}^1_y(t, X_t, \widetilde{Y}_t^1; \dots; \mathfrak{R}_y^J(t, X_t, \widetilde{Y}_t^J)\eqqcolon \mathfrak{R}(t, X_t, \widetilde{Y}_t).
    \end{aligned}
\end{align}

The standard Euler discretization of \eqref{eq:bsde:discretely_reflected} is done in two steps. First, in case no analytical solution is available, one needs to approximate the forward diffusion in \eqref{eq:introduction:sde} via suitable discrete time approximations, e.g. an Euler-Maruyama scheme such as below
\begin{align}\label{eq:euler-maruyama:sde}
    X_0^\pi = x_0,\quad X_{n+1}^\pi=X_n^\pi + \mu(t_n, X_n^\pi)\Delta t_n + \sigma(t_n, X_n^\pi)\Delta W_n,\quad \text{for}\quad n=0, \dots, N-1.
\end{align}Thereafter, one gathers discrete time approximations to \eqref{eq:bsde:discretely_reflected} through a backward recursion of conditional expectations, over a discrete time partition $0=t_0<t_1<\dots<t_N=T$, starting from $n=N-1, \dots, 0$
\begin{align}\label{eq:bsde:euler}
    \begin{aligned}
        Y_N^{j, \pi} &= \widetilde{Y}_N^{j, \pi}= g^j(X_N^{\pi}),\quad
        Z_n^{j, \pi} = 1/\Delta t_n \mathbb{E}_n[Y_{n+1}^{j, \pi}\Delta W_n^\top],\\
        \widetilde{Y}_n^{j, \pi} &= \Delta t_n f^j(t_n, X_n^{j, \pi}, \widetilde{Y}_n^{j, \pi}, Z_n^{j, \pi}) + \mathbb{E}_n[Y_{n+1}^{j, \pi}],\quad
        Y_n^{j, \pi} = \mathfrak{R}^j_y(t_n, X_n^{j, \pi}, \widetilde{Y}_n^{j, \pi}),
    \end{aligned}
\end{align}
where $\Delta t_n=t_{n+1}-t_n$, $\Delta W_n = W_{t_{n+1}} - W_{t_{n}}$. Given an appropriate machinery which approximates the conditional expectations above, one subsequently gathers numerical approximations of the solution pair of \eqref{eq:bsde:discretely_reflected}. As for the corresponding discrete time approximations errors, and the convergence of the Euler scheme \eqref{eq:bsde:euler} we refer to \cite{ma_representations_2005, bouchard_discrete-time_2004} and the references therein.

Given the discrete time approximations in \eqref{eq:bsde:euler} for each option in the portfolio $j=1, \dots, J$, one can gather discrete time approximations for the delta-hedging portfolio described by \eqref{eq:delta_hedging:portfolio_value}. In fact, combining the Feynman-Kac relations \eqref{eq:feynman_kac} with the first-order conditions in \eqref{eq:delta_hedging:foc}, yields the following discrete time approximations for the vector of delta-hedging weights
\begin{align}\label{eq:bsde:foc:formula}
    \alpha_n^{\pi} = \big(\sum_{j=1}^J Z_n^{j, \pi}\big)\sigma^{-1}(t_n, X_n^\pi).
\end{align}
Plugging \eqref{eq:bsde:foc:formula} into \eqref{eq:delta_hedging:portfolio_value} gives a discretely rebalanced approximation of the self-financing replicating portfolio. In the above, and for the rest of the paper, we assume a constant risk-free rate of $r$.

\section{Delta-gamma-hedging through One Step Malliavin schemes}\label{sec:gamma_hedging_and_osm}
In order to improve the replication accuracy for a fixed number of rebalancing dates, one needs to offset higher order sensitivities of the associated portfolio.
In the following section we extend the delta-hedging strategy of \eqref{eq:delta_hedging:portfolio_value} to the case where the second-order terms are also offset, involving second-order Greeks, Gammas, of each option. First, we formulate the general delta-gamma-hedging strategies, thereafter we present a method to deal with the additional model error induced by the presence of Gammas. The latter is done by building on the discrete time approximation schemes presented in \cite{negyesi_one_2024, negyesi_reflected_2024}, where on top of \eqref{eq:bsde:discretely_reflected} an additional, linear BSDE is solved at each point in time, corresponding to the Malliavin derivatives of the solution pair in \eqref{eq:bsde:discretely_reflected}, involving a stochastic version of the Gamma process, similarly to \eqref{eq:feynman_kac}.

\subsection{Delta-gamma hedging}
In a discrete time setting, one cannot perfectly hedge her exposure in the underlyings solely by offsetting the first order terms in \eqref{eq:delta_hedging:portfolio_value}. However, assets themselves have vanishing Gammas making them unsuitable for the purpose of Gamma hedging. This motivates to expand the hedging portfolio with a set of Gamma-hedging instruments issued on the same underlyings, whose Gammas are not equal to zero. Henceforth, the augmented delta-gamma-hedging portfolio consists of the additional long positions in $k=1, \dots, K$ gamma-hedging instruments with weights $\beta_t^{k}$: $+\beta_t^k u^k(t, X_t)$, where $K$ is a constant to be fixed, and $u^k(t, X_t)$ denotes the price of the $k$th gamma-hedging instrument at time $t$. Note that each gamma-hedging instrument is allowed (but not required) to depend on all risk factors in the portfolio. The value of the portfolio described above evolves according to the SDE below
\begin{align}\label{eq:gamma_hedging:portfolio_value}
		\mathrm{d}P_t^{\Gamma} =-\sum_{j=1}^J \mathrm{d}v^j(t, X_t)+ \sum_{i=1}^{m} \alpha_t^{i}(\mathrm{d}S_t^{i} + q_t^{i}S_t^{i}\mathrm{d}t) + \sum_{k=1}^K \beta_t^k \mathrm{d}u^k(t, X_t)  + \mathrm{d}B_t,\quad P_0^{\Gamma}=0.
\end{align}	
The first- and second-order conditions require $\partial P_t^{\Gamma}/\partial S_t^{i}(t, X_t)=0$ and $\partial^2 P_t^{\Gamma}/(\partial S_t^{l}\partial S_t^{i})=0$ for a pair of $1\leq i, l\leq m$, implying that the optimal hedging weights solve the following linear system \eqref{eq:delta_hedging:foc}
\begin{subequations}\label{eq:gamma_hedging:foc-soc}\noeqref{eq:gamma_hedging:soc, eq:gamma_hedging:foc}
    \begin{align}
    \sum_{k=1}^K\beta_t^k \partial_{li}^2 u^k(t, X_t)&= \sum_{j=1}^J\partial_{li}^2 v^j(t, X_t),\label{eq:gamma_hedging:soc} && il\in\mathcal{I},\\ \alpha^{i}_t &= \sum_{j=1}^J \partial_i v^j(t, X_t) - \sum_{k=1}^K \beta_t^{k}\partial_i u^k(t, X_t), &&1\leq i\leq d.\label{eq:gamma_hedging:foc}
\end{align}
\end{subequations}
Note that \eqref{eq:gamma_hedging:soc} is a $\abs{\mathcal{I}}\times K$ sized linear system whose solution, at each point in time, is a vector in $\mathbb{R}^K$. While optimizing her hedging weights according to the contraints established by \eqref{eq:gamma_hedging:foc-soc}, the investor has two degrees of freedom: 
\begin{itemize}
    \item the index set $\mathcal{I}$ in \eqref{eq:gamma_hedging:soc}, with which she can decide which elements in the corresponding $\Gamma$ matrices she would like to offset;
    \item the number (and type) of gamma-hedging instruments $K$. 
\end{itemize}
Given these choices the resulting linear system in \eqref{eq:gamma_hedging:soc} may be under- or over-determined. The choice of $\mathcal{I}$ allows for the freedom to offset gammas and cross-gammas of particularly volatile assets only that may have more severe effects on the options values $v^j$ in the portfolio. 

\paragraph{Risk measures.} Due to the finite number of rebalancing dates the resulting portfolio is not riskless in neither \eqref{eq:delta_hedging:portfolio_value} nor \eqref{eq:gamma_hedging:portfolio_value}. In order to assess the quality of a hedging strategy, one can assess the distribution of the relative Profit-and-Loss ($\text{PnL}$) which is an $\mathcal{F}_t$ measurable random variable defined by
\begin{align}\label{eq:def:pnl}
	\text{PnL}^\Delta_t \coloneqq \frac{e^{-rt}P_t^\Delta}{\sum_{j}^J v^j(0, X_0)},\qquad \text{PnL}^{\Gamma}_t \coloneqq \frac{e^{-rt}P_t^{\Gamma}}{\sum_{j=1}^J v^j(0, X_0)},\qquad \text{ for all } 0\leq t\leq T.
\end{align}
In particular, some of the most common risk measures include Value-at-Risk (VaR), Expected Shortfall (ES) and semivariance, which are defined below
\begin{align}\label{eq:def:risk_measures}
    \begin{aligned}
        \text{VaR}_{\alpha}(\text{PnL}) &\coloneqq \inf\{x: \mathbb{P}[\text{PnL} < x]> \alpha\},\quad\text{ES}_\alpha(\text{PnL})\coloneqq \mathbb{E}[\text{PnL}\vert \text{PnL}<\text{VaR}_\alpha],\\
        \text{SVar}_{-}(\text{PnL})&\coloneqq \mathbb{E}[(\text{PnL} - \mathbb{E}[\text{PnL}])^2\vert \text{PnL}<\mathbb{E}[\text{PnL}]].
    \end{aligned}
\end{align}
\begin{wrapfigure}{r}{0.5\textwidth}
    \centering
    \begin{subfigure}[t]{\sizeonebyone\textwidth}
        \includegraphics[width=\textwidth]{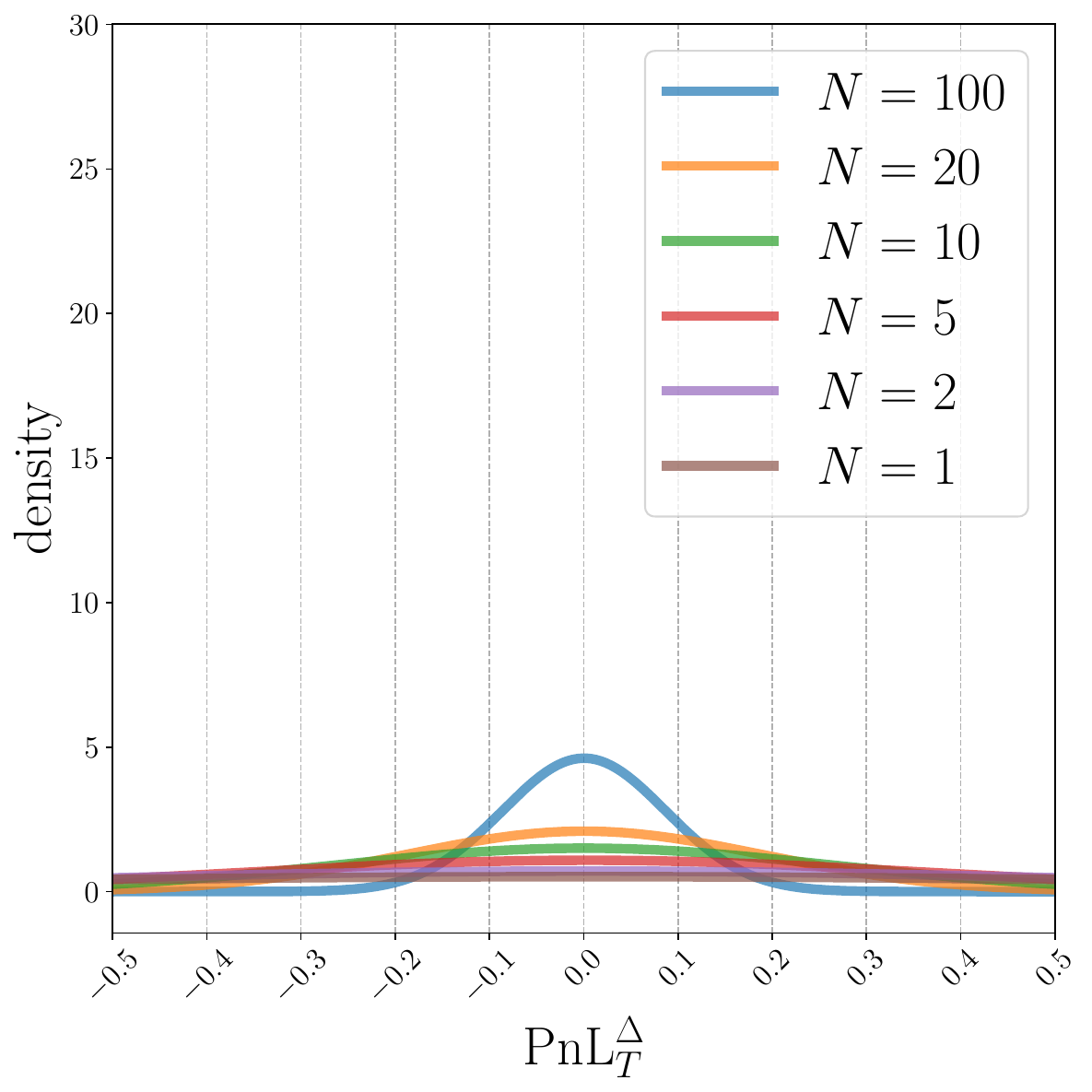}
        \caption{delta hedging}
    \end{subfigure}
    \begin{subfigure}[t]{\sizeonebyone\textwidth}
        \includegraphics[width=\textwidth]{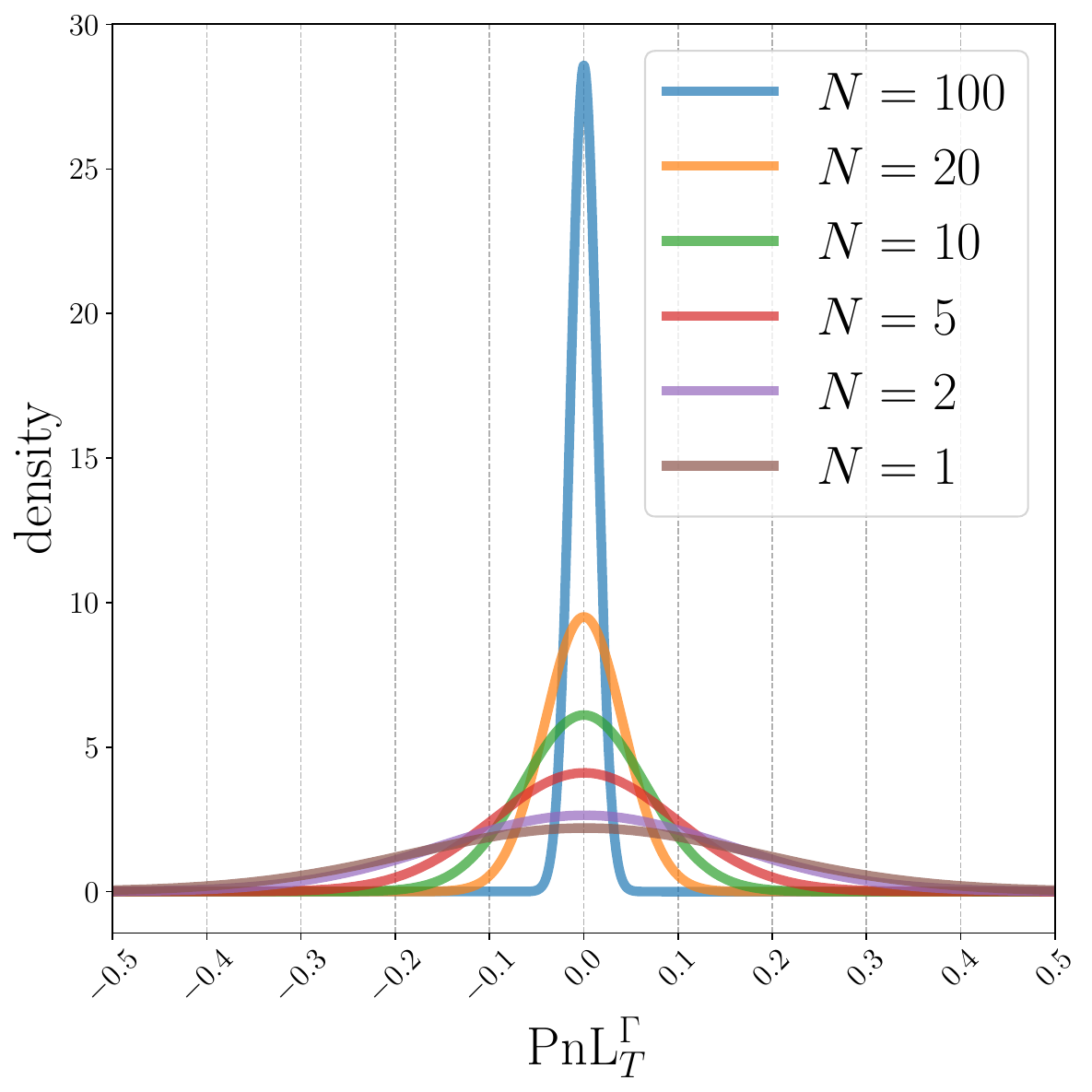}
        \caption{delta-gamma hedging}
    \end{subfigure}
    \caption{Comparison between delta- and delta-gamma hedging, given the number of discrete rebalancing dates $N$. Portfolio consisting of a single European vanilla call with maturity $T=1$, strike $K=100$, and $r=0$, $\sigma=0.25$. European vanilla put used as Gamma hedging instrument with maturity $2T$, and the same strike.}
    \label{fig:black_scholes:delta_vs_gamma}
\end{wrapfigure}

As an illustration between the accuracy of delta-gamma compared to delta hedging in the discrete time rebalancing framework, figure \ref{fig:black_scholes:delta_vs_gamma} depicts the distribution of the PnL at maturity corresponding to strategies \eqref{eq:delta_hedging:portfolio_value} and \eqref{eq:gamma_hedging:portfolio_value}, when the corresponding portfolio consists of a single, European vanilla call. In fig. \ref{fig:black_scholes:delta_vs_gamma} the distribution of the PnL is collected for several equidistant rebalancing dates. As can be seen, by including second-order sensitivities, one gains roughly an order of magnitude accuracy in the variance of the PnL distribution.


\subsection{One Step Malliavin schemes}
As illustrated by figure \ref{fig:black_scholes:delta_vs_gamma}, 
equation \eqref{eq:gamma_hedging:portfolio_value} brings an improvement to standard delta hedging in the discrete time framework by offsetting second-order terms in the series expansion of the corresponding portfolio value. However, unlike in the case of a vanilla Black-Scholes option, this comes with an additional modelling challenge. In exchange for the additional discrete replication accuracy, \eqref{eq:gamma_hedging:foc-soc} exposes an investor to additional model risk, namely, accurate approximations of the options' Gammas, which is necessary in order to be able to compute the gamma-hedging weights $\beta_t^k$ according to the second-order constraints -- see right-hand side of \eqref{eq:gamma_hedging:soc}. Discretely reflected BSDEs and the Feynman-Kac relations \eqref{eq:feynman_kac} only provide access to the Deltas and not to Gamma. 
In order to address this gap in the FBSDE context, we exploit ideas presented in \cite{negyesi_one_2024, negyesi_reflected_2024} and use the One Step Malliavin scheme to collect the necessary approximations of second-order Greeks.

In particular, it turns out that under sufficient conditions, see \cite[Proposition 5.1]{bouchard_discrete-time_2008} that the $Z$ part of the solution pair of the (discretely) reflected BSDE \eqref{eq:bsde:discretely_reflected} can be represented as the solution to a linear, vector-valued BSDE, corresponding to the Malliavin derivatives $\{(D_t\widetilde{Y}_u, D_tZ_u)\}_{0\leq t\leq r\leq T}$
\begin{align}\label{eq:bsde:discretely_reflected:malliavin}
    \begin{aligned}
        \widetilde{Z}_t\coloneqq D_t\widetilde{Y}_u^j &= \begin{aligned}[t]
            &D_t\widetilde{Y}_u^j + \mathbf{1}_{u\in \mathcal{R}^j\setminus \{0, T\}} \mathbf{1}_{l^j(X_u)>\widetilde{Y}_u^j}\\
            &+\int_{\underline{r}_{t}^{j}}^{\overline{r}_{t}^{j}} \begin{aligned}[t]
                \big[&\nabla_x f(s, X_s, \widetilde{Y}_s^j, Z_s^j)D_tX_s + \nabla_y f(s, X_s, \widetilde{Y}_s^j, Z_s^j)D_t\widetilde{Y}_s^j\\
                &+\nabla_z f(s, X_s, \widetilde{Y}_s^j, Z_s^j)D_tZ_s^j\big]\mathrm{d}s-\int_{\underline{r}_{t}^{j}}^{\overline{r}_{t}^{j}} \big(D_tZ_s^j\big)^T \mathrm{d}W_s,\quad t\in [0, T], u\in [\underline{r}_t^j, \overline{r}_t^j),
            \end{aligned}
        \end{aligned}\\
        Z_t^j &= D_t\widetilde{Y}_t^j+ \mathbf{1}_{t\in\mathcal{R}^{j}}\mathbf{1}_{l^j(X_t)>\widetilde{Y}_t^{j}}\big[D_t l^j(X_t) - D_t\widetilde{Y}_t^j\big]\eqqcolon \mathfrak{R}^{j}_z(t, X_t,\widetilde{Y}_t^j, D_t\widetilde{Y}_t^j).
    \end{aligned}
\end{align}
Most importantly, the representation in \eqref{eq:bsde:discretely_reflected:malliavin} does not only provide a way to compute the Deltas in form of a linear BSDE, but through the $DZ$ process and the Malliavin chain rule, the solution of \eqref{eq:bsde:discretely_reflected:malliavin} also includes second-order derivatives of the $j$'th option price, i.e. Gammas. In light of the Malliavin chain rule -- see e.g. \cite{nualart_malliavin_2006} -- the $DZ$ process coincides with the formal derivative of the $Z$ process, and one has
\begin{align}\label{eq:osm:gamma_derivative_z}
    D_tZ_t\sigma^{-1}(t, X_t)=\gamma(t, X_t)=\nabla_x z(t, X_t),
\end{align}
where $Z_t=z(t, X_t)$ is the Markovian mapping of the $Z$ process given by the Feynman-Kac formula \eqref{eq:feynman_kac}.
Subsequently, this leads to the following expression recovering the Hessian matrix of the value function
\begin{align}\label{eq:hessian_from_gamma}
    \text{Hess } v^j(t, X_t) = (\sigma^{-1}(t, X_t))^T (\gamma(t, X_t) - \nabla v^j(t, X_t) \nabla \sigma(t, X_t)).
\end{align}
The One Step Malliavin (OSM) scheme is a discretization which simultaneously solves the \emph{pair of BSDEs} \eqref{eq:bsde:discretely_reflected} and \eqref{eq:bsde:discretely_reflected:malliavin}. Henceforth, it involves a sequence of discrete time approximations, which -- in case of appropriate approximations of the resulting nested, backward recursion of conditional expectations -- gives approximations for the stochastic counterparts of options' prices, Deltas and also Gammas, i.e. all components needed in the discrete time rebalancing of the delta-gamma hedging portfolio in \eqref{eq:gamma_hedging:portfolio_value}. In particular, given a simultaneous discrete time solution of \eqref{eq:bsde:discretely_reflected} and \eqref{eq:bsde:discretely_reflected:malliavin} one can immediately solve the corresponding gamma hedging constraints in \eqref{eq:gamma_hedging:foc-soc}.

In order to be able to formulate the corresponding deep BSDE approximations, we briefly outline the time discretization of the OSM scheme.
The forward diffusion is approximated according to an Euler-Maruyama scheme as in \eqref{eq:euler-maruyama:sde}. Thereafter, the Malliavin derivative of the forward diffusion is again approximated by an Euler-Maruyama scheme leading to the following discrete time approximations
\begin{align}\label{eq:euler-maruyama:sde:malliavin}
\begin{split}
    D_nX_n^\pi&=\sigma(t_n, X_n^\pi),\\
    D_nX_{n+1}^\pi &= D_nX_{n}^\pi +\Delta t_n \nabla_x \mu(t_n, X_n^\pi)D_nX_n^\pi + \sum_{k=1}^d\nabla_x \sigma^k(t_n, X_n^\pi)D_nX_n^\pi\Delta W_n^k,
\end{split}
\end{align}
where $\sigma^k$ is the $k$th column of the diffusion matrix, and $\Delta W_n^k$ is the $k$th element of the Brownian increment vector -- see e.g. \cite{turkedjiev_two_2015}.
Then a backward recursive sequence of conditional expectations can be formulated for the discrete approximations of the backward equations \eqref{eq:bsde:discretely_reflected}, \eqref{eq:bsde:discretely_reflected:malliavin} as follows
\begin{align}\label{eq:scheme:osm:time}
\begin{split}
    Y_N^{j, \pi} &= \widetilde{Y}_N^{j, \pi} = g^j(X_N^\pi),\quad Z_N^{j, \pi} = \widetilde{Z}_N^{j, \pi} =\nabla g(X_N^\pi)\sigma(T, X_N^\pi),\\
    \Gamma_n^{j, \pi}\sigma(t_n, X_n^\pi)&\coloneqq D_nZ_n^{j, \pi} =\begin{aligned}[t]
        \frac{1}{\Delta t_n}\mathbb{E}\big[\Delta W_n \big(D_nY_{n+1}^{j, \pi} + \Delta t_n f^{D, j}(t_{n+1}, \mathbf{X}_{n+1}^{j, \pi}, \mathbf{D}_n\mathbf{X}_{n+1, n}^{j, \pi}\big)\big]
    \end{aligned}\\
    \widetilde{Z}_n^{j, \pi}=D_n\widetilde{Y}_n^{j, \pi} &= \begin{aligned}[t]
        \mathbb{E}\big[D_nY_{n+1}^{j, \pi} + \Delta t_n f^{D, j}(t_{n+1}, \mathbf{X}_{n+1}^{j, \pi}, \mathbf{D}_n\mathbf{X}_{n+1, n}^{j, \pi}\big],
    \end{aligned}\\
    \widetilde{Y}_n^{j, \pi} &=\begin{aligned}[t]
        &\vartheta_y \Delta t_n f^j(t_n, X_n^\pi, \widetilde{Y}_n^{j, \pi}, Z_n^{j, \pi}) + \mathbb{E}_n[Y_{n+1}^{j, \pi} + (1-\vartheta_y)\Delta t_n f^j(t_{n+1}, X_{n+1}^\pi, \widetilde{Y}_{n+1}^{j, \pi}, Z_{n+1}^{j, \pi})],
    \end{aligned}\\
    Z_n^{j, \pi}&=\mathfrak{R}^j_z(t_n, X_n^\pi,\widetilde{Y}_n^{j, \pi}, D_n\widetilde{Y}_n^{j, \pi}), \quad Y_n^{j, \pi} = \mathfrak{R}^j_y(t_n, X_n^\pi, \widetilde{Y}_n^{j, \pi}),
\end{split}
\end{align}
with the following approximation motivated by the Malliavin chain rule and the Feynman-Kac formula
\begin{align}
    D_nY_{n+1}^{j, \pi} = Z_{n+1}^{j, \pi}\sigma^{-1}(t_{n+1}, X_{n+1}^\pi) D_nX_{n+1}^\pi.
\end{align}
\sloppy Similarly to \cite{negyesi_one_2024, negyesi_reflected_2024}, in the above we use the short hand notations $\mathbf{X}_n^{j, \pi}\coloneqq (X_n^\pi, \widetilde{Y}_n^{j, \pi}, Z_n^{j, \pi})$, $\mathbf{D}_n\mathbf{X}_{n+1, n}^{j, \pi}=(D_nX_{n+1}^\pi, D_n\widetilde{Y}_{n+1}^{j, \pi}, D_nZ_n^{j, \pi})$ and
\begin{align}
    f^{D, j}(t, \mathbf{X}_n^{j, \pi}, \mathbf{D}_n\mathbf{X}_{n+1, n}^{j, \pi})=\nabla_x f(t, \mathbf{X}_n^{j, \pi})D_nX_{n+1}^\pi +\nabla_y f(t, \mathbf{X}_n^{j, \pi})D_n\widetilde{Y}_{n+1}^{j, \pi} + \nabla_z f(t, \mathbf{X}_n^{j, \pi})D_nZ_{n}^\pi.
\end{align}
In particular, combining the Feynman-Kac relations in \eqref{eq:feynman_kac} with the Malliavin chain rule, one gathers the following one on one correspondence between the stochastic processes in \eqref{eq:scheme:osm:time} and the prices and Greeks of the options in the gamma-hedging portfolio in \eqref{eq:gamma_hedging:foc-soc}. Indeed, for each option $j$, $\widetilde{Y}^{j}, Y^j$ describe the continuation value and price; $\widetilde{Z}^j$ -- derivative of the continuation value; $Z^j$ the Delta an $\Gamma$ the second-order Greeks.
In order to recover option Deltas and Gammas, from these processes, one combines \eqref{eq:feynman_kac} with \eqref{eq:hessian_from_gamma}, to obtain
\begin{align}
    \text{Delta}^{j, \pi}(t_n, X_n^\pi) = Z_n^{j, \pi} \sigma^{-1}(t_n, X_n^\pi),\quad
    \text{Gamma}^{j, \pi}(t_n, X_n^\pi) &= (\sigma^{-1}(t, X_t))^T (\gamma(t, X_t) - \nabla v^j(t, X_t) \nabla \sigma(t, X_t)).
\end{align}
As implied by these expressions above, the One Step Malliavin scheme for discretely reflected BSDEs in \eqref{eq:scheme:osm:time} provides discrete time approximations for the simultaneous pricing and delta-gamma hedging for each option $j$ in the portfolio \eqref{eq:gamma_hedging:portfolio_value}. In fact, given appropriate approximations of the conditional expectations in \eqref{eq:scheme:osm:time}, one can quantify the right-hand sides of \eqref{eq:gamma_hedging:foc-soc} corresponding to first- and second-order Greeks of all associated options. Then the solution of the linear system in \eqref{eq:gamma_hedging:foc-soc} provides the evolution of the corresponding discretely rebalanced replicating portfolio as follows
\begin{align}
    \begin{split}
    P_0^{\Gamma, \pi}=0, 
    \qquad P_{n}^{\Gamma, \pi} = -\sum_{j=0}^J Y^{j, \pi}_n + \sum_{i=1}^m \alpha_n^{i, \pi}\big( S_n^{i, \pi} + q_{t_{n}}^iS_n^{i, \pi}\Delta t_n\big) +\sum_{k=1}^K \beta^{k, \pi}_n u^k(t_n, X_n^\pi) + B_{n}^\pi,
    \end{split}
\end{align}
for $n=0,\dots,N$.
In the above, $B_n^\pi = e^{r\Delta t_n}B_{n-1}^\pi -\sum_{i=1}^m S_n^{i, \pi} (\alpha_n^{i,\pi} - \alpha_{n-1}^{i, \pi}) - \sum_{k=1}^K (\beta^{k, \pi}_n -\beta^{k, \pi}_{n-1})u^k(t_n, X_n^\pi),$ when $n\geq 1$,
as the portfolio is self-financing. However, in order to make the scheme implementable in a high-dimensional framework, i.e. whenever the number of underlying risk factors collected in $X$, or the number of options in the portfolio $J$ is large, one needs to have a method which accurately approximates the conditional expectations in \eqref{eq:scheme:osm:time}. This is discussed in the following section.

\section{Deep BSDE approximations on the portfolio level}\label{sec:deep_bsde}
In order to compute the conditional expectations in \eqref{eq:bsde:euler} and \eqref{eq:scheme:osm:time}, accurately and robustly in a high-dimensional framework when $d$ in \eqref{eq:introduction:sde} or $J$ in \eqref{eq:delta_hedging:portfolio_value}-\eqref{eq:gamma_hedging:portfolio_value} are large, we present a methodology based on deep neural network Monte Carlo regressions which is capable of dealing with such high-dimensional problems. Our formulation is a variant of the backward deep BSDE method. 
In what follows we extend this backward deep BSDE methodology in \cite{negyesi_one_2024, negyesi_reflected_2024} to the delta \eqref{eq:delta_hedging:foc} and delta-gamma hedging \eqref{eq:gamma_hedging:foc-soc} framework of the portfolio problem in \eqref{eq:delta_hedging:portfolio_value} and \eqref{eq:gamma_hedging:portfolio_value}, i.e. the simultaneous approximation of all option prices, Deltas and Greeks $j=1, \dots, J$. 
In the rest of the paper, $\pi\coloneqq \{0=t_0<t_1<\dots<t_N=T\}$ denotes a partition of the finite time interval $[0, T]$. Without loss of generality, we assume $\cup_{j=1}^J \mathcal{R}^j\subseteq \pi$, i.e. that the potential early exercise dates are included in the discrete time partition. We put $X_{t_{n}}^{\pi}=X_n^\pi$ for all discrete time time approximations over the time partition, and similarly for other processes.  We define the notations $\Delta t_n \coloneqq t_{n+1}-t_{n}$, $\Delta W_n\coloneqq W_{t_{n+1}} - W_{t_{n}}$.

\subsection{Deep BSDE approximations for the OSM scheme}\label{section:deep_bsde}
\begin{algorithm}[t]
\caption{Deep BSDE approximations with OSM schemes}\label{algorithm:osm}
\begin{algorithmic}[t]
\Require $\pi(N), \varrho(R)\subset [0, T], I\in \mathbb{N}_+, \eta: \mathbb{N}\to\mathbb{R}$
\Require $\varphi(\cdot\vert\theta^y): \mathbb{R}^d\to\mathbb{R}^J$, $\psi(\cdot\vert\theta^z):\mathbb{R}^d\to\mathbb{R}^{J\times d}$, $\chi(\cdot\vert\theta^\gamma):\mathbb{R}^d\to \mathbb{R}^{J\times d\times d}$ \Comment{neural networks}
\Ensure $\varrho(R)\subseteq \pi(N)$ \Comment{all discrete reflection dates are contained in the discretization}
\State $\widehat{Y}_N^\pi \gets g(X_N^\pi)$, $\widehat{Z}_N^\pi\gets \nabla_x g(X_N^\pi)\sigma(T, X_N^\pi)$ \Comment{collect terminal conditions of the BSDEs \eqref{eq:bsde:discretely_reflected}, \eqref{eq:bsde:discretely_reflected:malliavin}}
\For{$n=N-1, \dots, 0$}
    \If{$n=N-1$}
        \State $(\theta^{y, (0)}_n, \theta^{z, (0)}_n, \theta_n^{\gamma, (0)}) \gets \text{random initialization}$
    \Else{}
        \State $(\theta^{y, (0)}_n, \theta^{z, (0)}_n, \theta^{\gamma, (0)}_n) \gets (\widehat{\theta}_{n+1}^y, \widehat{\theta}^{z}_{n+1}, \widehat{\theta}_{n+1}^{\gamma})$ \Comment{transfer learning in \eqref{eq:deep_bsde:transfer_learning}}
    \EndIf
    \For{$i=0, \dots, I-1$}
        \State $\{X_n^\pi\}_{0\leq n\leq N}$ \Comment{Euler-Maruyama approximation of \eqref{eq:introduction:sde}}
        \State $\{D_nX_n^\pi, D_nX_{n+1}^\pi\}$ \Comment{Malliavin derivative approximations in \eqref{eq:euler-maruyama:sde:malliavin}}
        \State $\widehat{\mathcal{L}}^z_n(\theta^{z, (i)}_n, \theta^{\gamma, (i)}_n)$ \Comment{empirical version of \eqref{eq:osm:loss:z}}
        
        \State $(\theta_n^{z, (i+1)}, \theta_n^{\gamma, (i+1)})\gets (\theta_n^{z, (i)}, \theta_n^{\gamma, (i)}) - \eta(i)\nabla \widehat{\mathcal{L}}_n^z(\theta_n^{z, (i)}, \theta_n^{\gamma, (i)})$ \Comment{SGD step}
    \EndFor
    \State $(\widehat{\theta}_n^z, \widehat{\theta}_n^{\gamma})\gets (\theta_n^{z, (i+1)}, \theta_n^{\gamma, (i+1)})$
    \State $(\hat{\tilde{Z}}_n^{\pi}, \widehat{\Gamma}_n^\pi\sigma(t_n, X_n^\pi))\gets (\psi(X_n^\pi\vert\widehat{\theta}_n^z),  \chi(X_n^\pi\vert \widehat{\theta}_n^\gamma)\sigma(t_n, X_n^\pi))$ \Comment{approximations $\widetilde{Z}, \Gamma$}
    
    \For{$i=0, \dots, I-1$}
        \State $\widehat{\mathcal{L}}_n^y(\theta^y)$ \Comment{empirical version of \eqref{eq:osm:loss:y}}
        \State $\theta_n^{y, (i+1)}\gets \theta_n^{y, (i)} - \eta(i)\nabla \widehat{\mathcal{L}}_n^y(\theta_n^{y, (i)})$ \Comment{SGD step}
    \EndFor
    \State $\widehat{\theta}_n^y\gets \theta_n^{y, (i+1)}$
    \State $\hat{\tilde{Y}}_n^\pi\gets \varphi(X_n^\pi\vert\widehat{\theta}_n^y)$ \Comment{approximation continuation value}
    \State $\widehat{Z}_n^\pi\gets \big(\mathfrak{R}^1_z(t_n, X_n^\pi, \hat{\tilde{Y}}_n^{1, \pi} \hat{\tilde{Z}}_n^{1, \pi}), \dots,\mathfrak{R}^J_z(t_n, X_n^\pi, \hat{\tilde{Y}}_n^{J, \pi} \hat{\tilde{Z}}_n^{J, \pi}) \big)$ \Comment{reflection Delta \eqref{eq:deep_bsde:reflection:z}} 
    
    \State $\widehat{Y}_n^\pi \gets \big(\mathfrak{R}^1_y(t_n, X_n^\pi, \hat{\tilde{Y}}^{1, \pi}_n), \dots, \mathfrak{R}^J_y(t_n, X_n^\pi, \hat{\tilde{Y}}^{J, \pi}_n)\big)$ 
    \Comment{reflection price process \eqref{eq:deep_bsde:reflection:y}}
\EndFor
\end{algorithmic}
\end{algorithm}
In order to address the second-order terms appearing in \eqref{eq:gamma_hedging:foc-soc}, we present the deep BSDE methodology built on the OSM scheme \eqref{eq:scheme:osm:time}, providing sufficient gamma estimates that can be used in the context of delta-gamma hedging. To this end, let us put $\varphi(\cdot\vert\theta^y): \mathbb{R}^d\to \mathbb{R}^J$, $\psi(\cdot\vert \theta^z):\mathbb{R}^d\to\mathbb{R}^{J\times d}$ and $\chi(\cdot\vert \theta^\gamma):\mathbb{R}^d\to \mathbb{R}^{J\times d\times d}$ for neural networks depending on some parameter sets $\theta^y,\theta^z, \theta^\gamma$. These parametrizations are supposed to approximate the conditional expectations corresponding to the left hand sides of \eqref{eq:scheme:osm:time}, for each option $j=1, \dots, J$ in the portfolios \eqref{eq:delta_hedging:portfolio_value}, \eqref{eq:gamma_hedging:portfolio_value}. In particular, we emphasize that the output of the \emph{pricing} network $\varphi$ is a vector containing the continuation value of each option in the portfolio; the output \emph{delta} network $\psi$ contains each option's delta, concatenated row by row; whereas the output of the \emph{gamma} network $\chi$ is a tensor including each option's Gamma matrix.

In order to find appropriate parameter sets, such that $\varphi, \psi, \chi$ accurately approximate the conditional expectations for $\widetilde{Y}, \widetilde{Z}, DZ$ in \eqref{eq:bsde:discretely_reflected:malliavin}, respectively, we define the following $L^2$ loss functions motivated by the martingale representation theorem -- see \cite{negyesi_one_2024} --
\begin{align}\label{eq:osm:loss:z}
    \mathcal{L}_n^z (\theta^z, \theta^\gamma)\coloneqq \begin{aligned}[t]
        \frac{1}{J}\sum_{j=1}^{J}\mathbb{E}\Big[\vert &D_nY_{n+1}^{j, \pi}+ \Delta t_n \nabla_x f(t_{n+1}, \widehat{\mathbf{X}}_{n+1}^{j, \pi})D_nX_{n+1}^\pi+\nabla_y f(t_{n+1}, \widehat{X}_{n+1}^{j, \pi}) D_n\widetilde{Y}_{n+1}^{j, \pi}\\
        & + \nabla_z f(t_{n+1}, \widehat{\mathbf{X}}_{n+1}^{j, \pi}) \chi^j(X_n^\pi\vert \theta^\gamma)\sigma(t_n, X_n^\pi)\\
        &- \psi^j(X_n^\pi\vert\theta^z)-\big(\chi^j(X_n^\pi\vert\theta^\gamma)\sigma(t_n, X_n^\pi) \big)^T\Delta W_n\vert^2\Big],
    \end{aligned}
\end{align}
where $\psi^j$ and $\chi^j$ are the $j$'th row, $j$th element of the first axis of the output of the corresponding neural network. One can think of the loss function in \eqref{eq:osm:loss:z} as a mean-squared error in the Frobenius matrix norm for the row-wise concatenated system consisting of \eqref{eq:bsde:discretely_reflected} for each $j=1, \dots, J$ -- corresponding to the collection in \eqref{eq:bsde:discretely_reflected:collection}. This way, by minimizing \eqref{eq:osm:loss:z}, one gathers simultaneous deltas and gammas for each option, without having to run separate optimization problems for each $j=1, \dots, J$. A suitable minimizer of $\mathcal{L}_n^z$ denoted by $(\widehat{\theta}^z_n, \widehat{\theta}_n^\gamma)$ can be obtained by a Stochastic Gradient Descent (SGD) type optimization. Subsequently, the resulting parametrizations approximate the conditional expectation in \eqref{eq:scheme:osm:time} for each $j=1, \dots, J$ in the following manner
\begin{align}
    \widetilde{Z}_n^{j, \pi} \approx \psi^j(X_n^\pi\vert \widehat{\theta}^z),\quad \Gamma_n^{j, \pi}\approx \chi^j(X_n^\pi\vert \widehat{\theta}^\gamma).
\end{align}
Thereafter, one can use the reflection operators defined in \eqref{eq:bsde:discretely_reflected}, \eqref{eq:bsde:discretely_reflected:malliavin}, to get an approximation for the option Delta, and not just the derivative of the continuation value. In fact, given an approximation of the continuation value $\widetilde{Y}_n^{j, \pi}$, the reflection in the $Z$ process reads as follows
\begin{align}\label{eq:deep_bsde:reflection:z}
    Z_n^{j, \pi}\begin{aligned}[t]
        &=\mathfrak{R}^j_z(t_n, X_n^\pi, \varphi^j(X_n^\pi\vert \theta^y), \psi^j(X_n^\pi\vert \theta^z))\\
        &= \psi^j(X_n^\pi\vert \theta^z) +\mathbf{1}_{t_n\in \mathcal{R}^{j}\setminus\{0, T\}}\mathbf{1}_{\varphi^j(X_n^\pi\vert \theta^y)>g^j(X_n^\pi)}\left[\nabla_x g^j(X_n^\pi)\sigma(t_n, X_n^\pi) - \psi^j(X_n^\pi\vert \theta^z)\right].
    \end{aligned}
\end{align}
Similarly, given the same approximation for the continuation value, the corresponding option price can be approximated by combining the continuation value with the discrete early exercise strategy implied by the reflection taking place in \eqref{eq:bsde:discretely_reflected}. The resulting approximations read as follows
\begin{align}\label{eq:deep_bsde:reflection:y}
    Y_n^{j, \pi} \begin{aligned}[t]
        &\approx \mathfrak{R}^j_y(t_n, X_n^\pi, \varphi^j(X_n^\pi\vert \theta^y))= \varphi^j(X_n\vert \theta^y) + \mathbf{1}_{t_n\in \mathcal{R}^{j}\setminus\{0, T\}}\mathbf{1}_{\varphi^j(X_n^\pi\vert \theta^y)>g^j(X_n^\pi)}\left[g^j(X_n^\pi) - \varphi^j(X_n^\pi\vert \theta^y)\right].
    \end{aligned}
\end{align}
The approximations in \eqref{eq:deep_bsde:reflection:z} and \eqref{eq:deep_bsde:reflection:y} can easily be vectorized over $j=1, \dots, J$, so that they act simultaneously on the whole row-wise concatenated system of discretely reflected BSDEs in \eqref{eq:bsde:discretely_reflected:collection}. Then,
combining \eqref{eq:deep_bsde:reflection:z} with the reflection in the continuation value defined in \eqref{eq:deep_bsde:reflection:y}, one can subsequently formulate the following loss function approximating the last conditional expectation in the discrete time recursion in \eqref{eq:scheme:osm:time}.
In particular, we define the following loss function for the $Y$ part of \eqref{eq:bsde:discretely_reflected}
\begin{align}\label{eq:osm:loss:y}
    \mathcal{L}_n^y(\theta^y)\coloneqq \begin{aligned}[t]
    \mathbb{E}\Big[\vert&\widehat{Y}_{n+1}^\pi +(1-\vartheta_y)\Delta t_n f(t_{n+1}, \mathbf{X}_{n+1}^{\pi})+ \vartheta_y\Delta t_n f(t_n, X_n^\pi, \varphi(X_n^\pi\vert\theta^y), \widehat{Z}_n^\pi)\\
    &- \varphi(X_n^\pi\vert\theta^y)  - \widehat{Z}_n^\pi\Delta W_n\vert^2\Big].
    \end{aligned}
\end{align}
We emphasize that the approximation above, and its corresponding conditional expectation in \eqref{eq:scheme:osm:time}, is implicit in the continuation value, not just through the $\vartheta_y>0$ parameter, but also through the reflection occuring in \eqref{eq:deep_bsde:reflection:z}. In fact, the approach in \eqref{eq:osm:loss:z}, \eqref{eq:osm:loss:y} approximates the BSDEs in \eqref{eq:bsde:discretely_reflected:malliavin} and \eqref{eq:bsde:discretely_reflected}, respectively, meaning that the resulting approximations $\varphi, \psi$ correspond to the continuation value and its gradient, respectively. In order to then approximate the option prices, and their derivatives, one needs to approximate the reflection associated with early exercising in the discrete time framework. This is achieved by \eqref{eq:deep_bsde:reflection:z} and \eqref{eq:deep_bsde:reflection:y} accordingly.

The deep BSDE method outlined above explains all necessary steps one needs to take at a given time step $t_n$ locally. Thereafter, the method given by the loss functions \eqref{eq:osm:loss:z} and \eqref{eq:osm:loss:y}, is made fully implementable by a backward recursion starting at terminal time, executed as follows.
First, given a suitable discretization of the forward SDE and its Malliavin derivative, e.g. an Euler-Maruyama scheme such as \eqref{eq:euler-maruyama:sde} and \eqref{eq:euler-maruyama:sde:malliavin}, one needs to collect the terminal condition $\widehat{Y}_N^\pi=\hat{\tilde{Y}}_N^{\pi}=(g^1(X_N^\pi), \dots, g^J(X_N^\pi))$ and $Z_N^\pi=\tilde{Z}_N^\pi=(\nabla_x g^1(X_N^\pi), \dots, \nabla_x g^J(X_N^\pi))\sigma(T, X_N^\pi)$ as in \eqref{eq:scheme:osm:time}.
Then, in a backward recursion going from $n=N-1$ to $0$, one parametrizes the solution pair of the Malliavin BSDE in \eqref{eq:bsde:discretely_reflected:malliavin} at time step $n$ according to $\psi(\cdot\vert\theta^z), \chi(\cdot\vert\theta^\gamma)$. These parametrizations are optimized according to the loss function $\mathcal{L}_n^z(\theta^z, \theta^\gamma)$ defined in \eqref{eq:osm:loss:z}. Through a suitable minimization procedure such as stochastic gradient descent, one then subsequently gathers appropriate approximations of the optimal parameter set $(\widehat{\theta}_n^z, \widehat{\theta}_n^\gamma)\in \argmin_{\theta^z, \theta^\gamma} \mathcal{L}_n^z(\theta^z, \theta^\gamma)$. Setting
\begin{align}\label{eq:deep_bsde:z_tilde:approx}
    \widehat{\Gamma}_n^\pi = \chi(X_n^\pi\vert \widehat{\theta}_n^\gamma),\quad \hat{\tilde{Z}}_n^\pi=\psi(X_n^\pi\vert\widehat{\theta}_n^z)
\end{align}
provides approximations for the first and second conditional expectations in \eqref{eq:scheme:osm:time}. Combining these approximations with the estimations of the discrete reflections given by \eqref{eq:deep_bsde:reflection:z} and \eqref{eq:deep_bsde:reflection:y} then gives loss function $\mathcal{L}_n^y(\theta^y)$ at time step $t_n$, that is meant to measure the approximation quality in the continuation values in the discretely reflected BSDE of \eqref{eq:bsde:discretely_reflected}. A second stochastic gradient descent optimization then estimates the optimal parameter set $\widehat{\theta}_n^y\in\argmin_{\theta^y} \mathcal{L}_n^y(\theta^y)$, giving approximations for the continuation value at time $t_n$, defined by the last conditional expectation in \eqref{eq:scheme:osm:time}
\begin{align}\label{eq:deep_bsde:y_tilde:approx}
    \hat{\tilde{Y}}_n^\pi = \varphi(X_n^\pi\vert\widehat{\theta}_n^y).
\end{align}
Finally, combining the approximations of the continuation value $\hat{\tilde{Y}}_n^\pi$ and its gradient $\hat{\tilde{Z}}_n^\pi$ with the early exercise decision given by the vectorized expressions in \eqref{eq:deep_bsde:reflection:y} and \eqref{eq:deep_bsde:reflection:z} for all $j=1, \dots, J$, gives the final discrete time approximations for all processes in the pair of backward SDEs in \eqref{eq:bsde:discretely_reflected} and \eqref{eq:bsde:discretely_reflected:malliavin}
\begin{align}\label{eq:deep_bsde:z_y:approx}
    \widehat{Z}_n^\pi = \mathfrak{R}_z(t_n, X_n^\pi, \hat{\tilde{Y}}_n^\pi, \hat{\tilde{Z}}_n^\pi),\quad \widehat{Y}_n^\pi = \mathfrak{R}_y(t_n, X_n^\pi, \hat{\tilde{Y}}_n^\pi).
\end{align}
The approximations given by \eqref{eq:deep_bsde:z_tilde:approx}, \eqref{eq:deep_bsde:y_tilde:approx}, \eqref{eq:deep_bsde:z_y:approx} complete the processing of time step $t_n$. The initialization of a stochastic gradient descent optimization has a substantial impact on the speed of convergence and also the final accuracy of the resulting approximations. Argued by continuity in time of the stochastic processes $\widetilde{Y}, \widetilde{Z}$ in \eqref{eq:bsde:discretely_reflected} and \eqref{eq:bsde:discretely_reflected} we therefore initialize the parameter sets of the loss functions $\mathcal{L}_{n-1}^z$ and $\mathcal{L}_{n-1}^y$ according to the transfer learning trick
\begin{align}\label{eq:deep_bsde:transfer_learning}
    (\theta^z, \theta^\gamma)\gets (\widehat{\theta}_n^z, \widehat{\theta}_n^\gamma),\quad \theta^y\gets \widehat{\theta}_n^y,\qquad \text{ for each } n=N-1, \dots, 1.
\end{align}
With these initial parameter guesses, all points $n=N-2, \dots, 0$ in time are more efficiently optimized. One then completes the algorithm by carrying out the same procedure in a backward iteration terminating at $n=0$.
The complete fully-implementable backward deep BSDE algorithm is collected in algorithm \ref{algorithm:osm}.

\paragraph{The backward deep BSDE method of Huré et al. \cite{hure_deep_2020}.}
When only the Deltas are of interest, alternative deep BSDE schemes are applicable in the high-dimensional context, see \cite{chen_deep_2021, hure_deep_2020}.
The RDBDP scheme \cite[sec. 3.3]{hure_deep_2020} is concerned with the numerical approximation of variational inequalities, which can be considered as a continuous asymptotics for \eqref{eq:bsde:discretely_reflected} in the case $R\to\infty$. Even though, their scheme is only given for a single ($J=1$) continuously reflected BSDE, in what follows, we naturally extend this to the collection of discretely reflected equations given in \eqref{eq:bsde:discretely_reflected:collection}. Let us put $\varphi(\cdot\vert\theta^y): \mathbb{R}^d\to \mathbb{R}^J$ and $\psi(\cdot\vert \theta^z): \mathbb{R}^d\to \mathbb{R}^{J\times d}$ for two feedforward, fully-connected neural networks, depending on some potentially non-disjoint parameter sets. In fact, in \cite{chen_deep_2021}, $\psi(\cdot\vert\theta^z)=\nabla_x \varphi(\cdot\vert\theta^y)\sigma(t_n, \cdot)$, with $\theta^z\equiv\theta^y$. These neural networks are parametrizations of the Markovian conditional expectations of the collection of conditional expectations in \eqref{eq:bsde:euler} for every $j=1, \dots, J$ in \eqref{eq:bsde:discretely_reflected:collection}. Using these parametrizations, we define the $L^2$ loss function 
\begin{align}\label{eq:loss:hure}
    \mathcal{L}_n^\text{Euler}(\theta^y, \theta^z) \coloneqq 
    \begin{aligned}[t]
    \mathbb{E}\big[\vert\widehat{Y}_{n+1}^{\pi} - \varphi(X_n^\pi\vert \theta^y) - \Delta t_n f(t_n, X_n^\pi, \varphi(X_n^\pi\vert\theta^y), \psi(X_n^\pi\vert \theta^z))+ \psi(X_n^\pi\vert\theta^z)\Delta W_n\vert^2\big],
    \end{aligned}
\end{align}
which is a function of the total parameter set $\Theta\coloneqq (\theta^y, \theta^z)$. Herein, we denote the row-wise concatenated solutions of each discretely reflected BSDE as in \eqref{eq:bsde:discretely_reflected:collection}. As shown in \cite[thm. 4.1, 4.4]{hure_deep_2020}, a minimizer $(\widehat{\theta}^y, \widehat{\theta}^z)$ of the loss function defined by \eqref{eq:loss:hure} provides a simultaneous approximation of the conditional expectations $\widetilde{Y}_n^{j, \pi}$ and $Z_n^{j, \pi}$ in \eqref{eq:bsde:euler} given by the following expressions
$\hat{\tilde{Y}}_n^{j, \pi}\approx \varphi^j(X_n^\pi\vert \theta^y),\quad \widehat{Z}_n^{j, \pi}\approx \psi^j(X_n^\pi\vert \widehat{\theta}^z)$,
where $\varphi^j, \psi^j$ denote the $j$'th row of the output layer of each neural network. In fact, by collecting each discretely reflected BSDE into the system \eqref{eq:bsde:discretely_reflected:collection}, the loss function \eqref{eq:loss:hure} takes a mean-squared loss of all equations in \eqref{eq:bsde:euler}, for every $j=1, \dots, J$, at the same time.
The scheme is made fully-implementable by a similar backward recursion as for the OSM scheme.
The main differences between the OSM approximation in algorithm \ref{algorithm:osm} and those of \cite{hure_deep_2020, chen_deep_2021} is that the One Step Malliavin scheme does not only provide option prices and Deltas, like \cite{hure_deep_2020}, but also second-order Greeks, Gammas, throughout the entire spacetime by the process $DZ$. This makes the deep BSDE approximations for the OSM scheme suitable for delta-gamma hedging in \eqref{eq:gamma_hedging:foc-soc}. Additionally, as shown in \cite{negyesi_one_2024, negyesi_reflected_2024}, the OSM scheme provides more accurate approximations for the $Z$ process in regression Monte Carlo frameworks when the time step size $\Delta t_n$ is small or the volatility is high. As we show below -- see fig. \ref{fig:bs:osm_vs_hure}, \ref{fig:heston:pnl:osm_vs_hure} in particular --, this in fact results in more accurate Deltas, leading to better delta replication with the use of the OSM scheme.

\subsection{Delta hedging with OSM}
Given the deep BSDE approximations in algorithm \ref{algorithm:osm}, one can subsequently solve the delta hedging problem of our investor, which comes down to the discrete time approximation of the hedging weights $\alpha$ in \eqref{eq:delta_hedging:portfolio_value} where $t_n$ is in a set of finite rebalancing dates $\mathcal{S}$. Without loss of generality, we assume that the deep BSDE approximation of the corresponding collection of discretely reflected BSDEs in \eqref{eq:bsde:discretely_reflected:collection} is given over a time partition that includes all rebalancing dates, i.e. $\mathcal{S}\subseteq \pi(N)$. Therefore, combining the first-order condition of the delta hedging weights given by \eqref{eq:delta_hedging:foc} with the Feynman-Kac formula in \eqref{eq:feynman_kac}, one immediately gathers the following approximations for all option Deltas in the portfolio
\begin{align}\label{eq:deep_bsde:delta:approx}
    \widehat{\text{Delta}}^\pi_n = (\widehat{\text{Delta}}^{1, \pi}_n,\dots,\widehat{\text{Delta}}_n^{J, \pi}) =(\widehat{Z}_n^{1, \pi}\sigma^{-1}(t_n, X_n^\pi);\dots; \widehat{Z}_n^{J, \pi}\sigma^{-1}(t_n, X_n^\pi)) =\widehat{Z}_n^\pi\sigma^{-1}(t_n, X_n^\pi),
\end{align}
for all $t_n\in \pi(N)$, and in particular in $\mathcal{S}$. We remark that $\widehat{\text{Delta}}_n^\pi\in\mathbb{R}^{J\times d}$ as it is the row-wise collection of each contract's Delta in the portfolio.
Plugging this into the first-order condition given by \eqref{eq:delta_hedging:foc}, one subsequently gathers the discrete time approximations of the optimal hedging weights at each rebalancing date $t_n\in\mathcal{S}$
\begin{align}\label{eq:deep_bsde:delta_hedging:alpha}
    \widehat{\alpha}_n^{\pi} = \sum_{j=1}^J \widehat{\text{Delta}}_n^{j, \pi}.
\end{align}
Given \eqref{eq:deep_bsde:delta_hedging:alpha}, on top of the deep BSDE approximations for the option prices in the portfolio $\widehat{Y}_n^{j, \pi}$, the corresponding discretely rebalanced delta hedged replicating portfolio's value evolves according to the following recursion
\begin{align}\label{eq:deep_bsde:delta:portfolio_update}
    \widehat{P}_0^{\Delta, \pi}=0,\quad \widehat{P}_{n}^{\Delta, \pi} = -\sum_{j=1}^J \widehat{Y}^{j, \pi}_n + \sum_{i=1}^m \widehat{\alpha}_n^{i, \pi}\big( S_n^{i, \pi} + q_{t_{n}}^iS_n^{i, \pi}\Delta t_n\big) + B_{n}^\pi,\quad&&\text{for } n=0,\dots,N,\\
    B_0^{\Delta, \pi} = 0,\qquad B_n^\pi = e^{r\Delta t_n}B_{n-1}^\pi -\sum_{i=1}^m S_n^{i, \pi} (\widehat{\alpha}_n^{i,\pi} - \widehat{\alpha}_{n-1}^{i, \pi}),&&\text{for } n=1,\dots, N.
\end{align}
Notice that all necessary ingredients for the discrete time rebalancing of the hedging portfolio are included in the OSM approximations of the discretely reflected BSDE \eqref{eq:bsde:discretely_reflected}, i.e. option prices and Deltas. 


\subsection{Delta-gamma hedging with OSM}
The main motivation behind the One Step Malliavin scheme simultaneously solving the BSDEs \eqref{eq:bsde:discretely_reflected} and \eqref{eq:bsde:discretely_reflected:malliavin}, is that through the numerical resolution of the latter, it includes a $\Gamma$ process which corresponds to second-order Greeks of the associated option prices. In particular, by modeling all options in the portfolio by the One Step Malliavin scheme performed on the row-wise concatenated collection of reflected BSDEs \eqref{eq:bsde:discretely_reflected:collection}, one simultaneously has approximations of all options' prices, Deltas and Gammas through $\widehat{Y}_n^\pi, \widehat{Z}_n^\pi$ and $\widehat{\Gamma}_n^\pi$, respectively, with the latter being recovered from the Markovian approximations $\Gamma_n^{j, \pi}=\gamma^{j, \pi}_n(X_n^\pi)$. Through a discrete time expression analogous to \eqref{eq:hessian_from_gamma}, we find
\begin{align}\label{eq:deep_bsde:gamma:approx}
    \widehat{\text{Gamma}}_n^{j, \pi} = (\sigma^T(t_n, X_n^\pi))^{-1} \left(\widehat{\Gamma}_n^{j, \pi} - \widehat{\text{Delta}}^{j, \pi}_n\nabla \sigma(t_n, X_n^\pi)\right),
\end{align}
where the approximations $\widehat{\text{Delta}}^{j, \pi}_n$ are recovered identically to \eqref{eq:deep_bsde:delta:approx}. Equation \eqref{eq:deep_bsde:gamma:approx} provides discrete time approximations for the right-hand side of the linear system established by the second-order conditions of delta-gamma hedging in \eqref{eq:gamma_hedging:soc}. Given a set of Gamma hedging instruments whose Gammas are also available in (semi-)closed form, this makes the discrete time approximation of all terms in the delta-gamma hedging portfolio \eqref{eq:gamma_hedging:portfolio_value} possible, after solving the following $\abs{\mathcal{I}}\times K$ sized linear system
\begin{subequations}\label{eq:gamma_hedging:deep_bsde:foc-soc}\noeqref{eq:gamma_hedging:deep_bsde:soc, eq:gamma_hedging:deep_bsde:foc}
    \begin{align}
    \sum_{k=1}^K\widehat{\beta}_n^{k, \pi} \partial_{li}^2 u^k(t, X_t)&= \sum_{j=1}^J(\widehat{\text{Gamma}}_n^{j, \pi})^{li},\label{eq:gamma_hedging:deep_bsde:soc} && il\in\mathcal{I},\\ \widetilde{\alpha}^{\pi}_n &= \sum_{j=1}^J \widehat{\text{Delta}}^{j, \pi}_n - \sum_{k=1}^K \widetilde{\beta}_t^{k}\nabla u^k(t, X_t), &&1\leq i\leq d.\label{eq:gamma_hedging:deep_bsde:foc}
\end{align}
\end{subequations}
Given the deep BSDE approximations of the whole Gamma matrix, the linear system imposed by \eqref{eq:gamma_hedging:deep_bsde:foc-soc} has a solution pair $\widehat{\beta}_n^{\pi}, \widehat{\alpha}_n^\pi\in\mathbb{R}^{K}\times \mathbb{R}^m$, which solve the discrete time version of the first- and second-order conditions of delta-gamma hedging in \eqref{eq:gamma_hedging:foc-soc}. Consequently, after having solved the associated collection of discretely reflected FBSDEs by the OSM scheme, the investor's task at a rebalancing date is to compute the right-hand sides of \eqref{eq:gamma_hedging:deep_bsde:foc-soc}, which can easily be done by evaluating the corresponding neural networks $\varphi, \psi, \chi$ at a given realization of the Brownian motion. Subsequently, the investor needs to buy and sell the tradeable risk factors, and the corresponding gamma-hedging securities according to the differences in the hedging weights between two rebalancing dates. The discretely rebalanced, deep BSDE approximated delta-gamma hedging portfolio evolves according to the following discrete time recursion
\begin{align}
    \widehat{P}_{n}^{\Gamma, \pi} &= -\sum_{j=0}^J \widehat{Y}^{j, \pi}_n + \sum_{i=1}^m \widehat{\alpha}_n^{i, \pi}\big( S_n^{i, \pi} + q_{t_{n}}^i S_n^{i, \pi}\Delta t_n\big) +\sum_{k=1}^K \widehat{\beta}^{k, \pi}_n u^k(t_n, X_n^\pi) + B_{n}^\pi,&&\text{for } n=0,\dots,N,\label{eq:deep_bsde:gamma:portfolio_update}\\
    B_n^\pi &= e^{r\Delta t_n}B_{n-1}^\pi -\sum_{i=1}^m S_n^{i, \pi} (\widehat{\alpha}_n^{i,\pi} - \widehat{\alpha}_{n-1}^{i, \pi}) - \sum_{k=1}^K (\widehat{\beta}^{k, \pi}_n - \widehat{\beta}_{n-1}^{k, \pi})u^k(t_n, X_n^\pi),&&\text{for } n=1,\dots, N,\label{eq:deep_bsde:gamma:bank_account_update}
\end{align}
with $B_n^\pi=0$.
Unlike in the delta case, the replicating portfolio is rebalanced not just by buying and selling the tradeable risk factors $S^i$ according to the approximated hedging weights $\widehat{\alpha}_n^{i, \pi}$ but also by trading each Gamma hedging instruments according to the difference $\widehat{\beta}_n^{k, \pi}-\widehat{\beta}_{n-1}^{k, \pi}$.
The complete algorithm is collected in algorithm \ref{algorithm:delta_gamma_hedging}.

\paragraph{Alternative approaches to Gamma hedging.}
   Given a differentiable function approximation of the associated option prices, one could in principle use \emph{automatic differentiation} to approximate the corresponding Deltas and Gammas in a discrete time framework. For instance, using the deep backward dynamic programming approach of Huré et al. in \cite{hure_deep_2020}, one has a differentiable approximation of $Z$ in \eqref{eq:bsde:discretely_reflected} in the form of the vector-valued function $\psi(\cdot\vert\widehat{\theta}^z)$. Computing the Jacobian matrix by means of automatic differentiation yields an approximation of the \textit{"derivative $Z$ process"}, which is analogous to $\Gamma$ in \eqref{eq:bsde:discretely_reflected:malliavin} -- see also \eqref{eq:osm:gamma_derivative_z}. Thereafter, plugging $\nabla_x \psi(\cdot\vert \widehat{\theta}^z)$ in the place of $\Gamma$ in the formulae \eqref{eq:hessian_from_gamma}, one could obtain a comparable numerical representation of the right-hand side of \eqref{eq:gamma_hedging:soc}. We remark that a similar approach is taken in \cite{chen_deep_2021}, where motivated by the Feynman-Kac formula the authors approximate the $Z$ by automatic differentiation on the parametrization of the approximation of the option prices, i.e. $\psi=\nabla_x \varphi$.
    The problem regarding the automatic differentiation approximations outlined above is two-fold. First, as the Jacobian matrix of $\psi$ does not form part of the loss function in \eqref{eq:loss:hure}, there is no guarantee that $\nabla \psi$ is an accurate approximation of $\nabla_x Z$ -- even when $\psi$ approximates $Z$ arbitrarily well. In order to ensure for this to be the case, one would have to augment the loss function in order to account for this, which would result in a similar representation formula as \eqref{eq:bsde:discretely_reflected:malliavin} in the OSM scheme. Regardless of the lack of theoretical guarantees, one can carry out the corresponding automatic differentations and check if the corresponding results yield meaningful Gammas. We found that this in fact is not the case, rendering $\nabla_x \psi(\cdot\vert\widehat{\theta}^z)$ inapplicable in the context of delta-gamma hedging. According to our findings, this is already case for low-dimensional problems, and the accuracy of the automatic differentiated Gammas further decreases as $d$ grows. In particular, we refer to figure 3 in \cite{negyesi_one_2024}, and figure \ref{fig:heston:pnl:osm_vs_hure} in our upcoming numerical experiments, which both demonstrate this phenomenon.

\begin{algorithm}[t]
\caption{Delta-gamma hedging on the portfolio level}\label{algorithm:delta_gamma_hedging}
\begin{algorithmic}[1]
\Require Deep BSDE approximations of \eqref{eq:bsde:discretely_reflected:collection} and \eqref{eq:bsde:discretely_reflected:malliavin} for each $j=1, \dots, J$ \Comment{OSM in algorithm \ref{algorithm:osm}}
\Require $\{W_{t_{n}}\}_{0\leq n\leq N}$
\State $P_0^{\Gamma, \pi} \gets 0$
\For{$n=0, \dots, N-1$}
    \State $(X_n^\pi, \hat{\tilde{Y}}{}_n^\pi, \hat{\tilde{Z}}{}_n^\pi)$ \Comment{deep BSDE approximations}
    \State compute $\widehat{\beta}_n^{\pi, k}, k=1, \dots, K$
    \Comment{solution to linear system in \eqref{eq:gamma_hedging:deep_bsde:soc}}
    
    \State compute $\widehat{\alpha}_n^{\pi}$ \Comment{via \eqref{eq:gamma_hedging:deep_bsde:foc}}
    \If{$n=0$}
        \State $B_n^\pi \gets \widehat{Y}_n^{j, \pi} - \sum_{i=1}^m \widehat{\alpha}_n^{\pi, i} S_n^{\pi, i} - \sum_{k=1}^K \widehat{\beta}_n^{k, \pi}u^k(t_n, X_n^\pi)$
    \Else{}
        \State update bank account $B_n^\pi$ \Comment{rebalancing with \eqref{eq:deep_bsde:gamma:bank_account_update}}
    \EndIf
    \State update portfolio value $P_{n+1}^{\Delta,\pi} $ \Comment{according to \eqref{eq:deep_bsde:gamma:portfolio_update}}
\EndFor
\State $\text{PnL}^\Delta \gets e^{-rT}P_N^{\Delta, \pi} / (\sum_{j=1}^J \widehat{Y}_0^{j, \pi})$ \Comment{compute PnL in \eqref{eq:def:pnl}}
\end{algorithmic}
\end{algorithm}

\subsection{About the linear system of second-order constraints}\label{sec:subsec:linear_system}
The optimal weights with which one has to hold the Gamma hedging instruments $u^k, k=1, \dots, K$ in \eqref{eq:gamma_hedging:portfolio_value} is determined by the solution of the linear system in \eqref{eq:gamma_hedging:soc}. The main driver of the computational complexity of gamma hedging stems from the numerical solution of this linear system, which depends not only on the number of gamma hedging instruments $K$ contained in the replicating portfolio, but also on the type of contracts used as hedging instruments.
In fact, the linear system in \eqref{eq:gamma_hedging:soc} is a $\abs{\mathcal{I}}\times K$ sized rectangular system. In the special case when one chooses to offset all upper triangular elements in the Gamma matrix, assigning a single gamma hedging instrument to each gamma and cross-gamma, this results in a system of size $(m(m+1)/2)\times (m(m+1)/2)$, depending on the number of tradable risk factors. The memory requirements of storing the coefficient matrix of one instance of such a linear system would thus scale $\mathcal{O}(m^4)$, and a naive direct solution of such a linear system would require $\mathcal{O}(m^6)$ floating point operations, depending on the number of spatial dimensions. These scaling factors grow substantially in case of large number of risk factors, especially compared to the simple matrix-vector multiplication determining the delta hedging weights in \eqref{eq:deep_bsde:delta_hedging:alpha}. 

Nonetheless, one can choose gamma hedging instruments which preserve a special structure of the coefficient matrix in \eqref{eq:gamma_hedging:soc}. In fact, in the upcoming numerical experiments for each higher dimensional problem we choose the gamma hedging instruments to be European exchange options. Henceforth, given by the Margrabe formula \cite{margrabe_value_1978}, these options admit an analytical closed-form expression not only for the prices, but also for all Greeks up to second order in the Black-Scholes framework -- see appendix \ref{sec:appendix:margrabe}. Consequently, the coefficient matrix in \eqref{eq:gamma_hedging:soc} is computed analytically for each Brownian path.
Moreover, the second order derivatives $\partial_{ij}^2 u^k$ determine the shape of the coefficient matrix multiplying $\beta^k, k=1,\dots, K$. The choice of exchange options implies that each row (or column) of the coefficient matrix in \eqref{eq:gamma_hedging:foc-soc} only includes at most $3$ non-zero elements, leading to a sparse linear system which can efficiently be solved by sparse numerical linear algebra methods. In what follows we use the sparse least squares method developed by \cite{paige_lsqr_1982}.\footnote{In particular, the \texttt{scipy} implementation, see \href{https://docs.scipy.org/doc/scipy/reference/generated/scipy.sparse.linalg.lsqr.html}{documentation}.} As solving such a large linear system, for a cloud of Monte Carlo simulations -- in order to be able to assess the distribution of the corresponding PnL distribution -- is computationally expensive, this provides a substantial computational improvement for the delta-gamma hedging strategy in algorithm \ref{algorithm:delta_gamma_hedging}.

\paragraph{Replication errors and computational complexity.}
The main source of computational complexity of delta-gamma hedging is the deep BSDE optimization of the OSM scheme in algorithm \ref{algorithm:osm}. Nevertheless, we remark that once the replicating portfolio is constructed and the corresponding options are fixed, this step only needs to be done once. It can be done offline, and the resulting discrete time approximations can be used throughout the whole spacetime. With respect to the convergence of the discrete time approximations we refer to \cite{hure_deep_2020, negyesi_one_2024} where it is shown that the deep BSDE approximations converge to the continuous solution triple of the BSDEs with an $L^2$ rate of $\mathcal{O}(\abs{\pi}^{1/2})$. These results can be generalized to the vector-valued setting corresponding to the collection of BSDEs in \eqref{eq:bsde:discretely_reflected:collection} without any substantial difficulty. In case the forward SDE's solution is given in closed-form, the replication error of the corresponding delta hedging strategy coincides with those of the deep BSDE approximations. Whereas, when the forward diffusion is approximated by, e.g., an Euler-Maruyama discretization the product term in \eqref{eq:delta_hedging:portfolio_value} will hamper the convergence rate by an appropriate application of Young's inequality.
For a theoretical assessment of the convergence rates of the tracking errors induced by delta \eqref{eq:delta_hedging:portfolio_value} and delta-gamma hedging \eqref{eq:gamma_hedging:portfolio_value}, we refer to \cite{gobet_tracking_2012}. Therein, the authors show that the convergence rate of the hedging portfolio depends on the fractional regularity of the payoff $g$ -- at least in the European options' context. In particular, with the use of an equidistant time their results imply the convergence of the variance of the PnL distribution with a rate of $\mathcal{O}(h^{0.5})$ in case of delta-, and a rate of $\mathcal{O}(h^{0.75})$ for delta-gamma hedging of European put and call options. Consequently, including the second-order Greeks enabled by the One Step Malliavin scheme in the replicating portfolio \eqref{eq:gamma_hedging:portfolio_value} does not only improve replication accuracy by a constant, but in special cases may also result in a higher order convergence rate.

\section{Numerical experiments}\label{sec:numerical_experiments}
In order to demonstrate the accuracy and robustness of the proposed FBSDE based delta-gamma hedging strategies, numerical experiments are presented. 
In what follows, we formulate a (discretely reflected) FBSDE system corresponding to each portfolio below. Subsequently, each of these systems is discretized using a fine time grid containing $N'=100$ equally sized intervals, and solved by deep BSDE approximations as in algorithm \ref{algorithm:osm}, including approximations for the associated options' prices, deltas and gammas throughout the entire spacetime. This step is the most time consuming part of our approach, however, it needs to be emphasized that the training of these neural networks only needs to be done once, offline. Thereafter, the resulting approximations for prices, deltas and gammas can be simply evaluated for each rebalancing date in the updates of the hedging strategies \eqref{eq:deep_bsde:delta:approx} and \eqref{eq:deep_bsde:gamma:approx}, which is fast and efficient. We choose a very fine time partition for the numerical resolution of the FBSDE system in order to be able to use sub-points of it for each rebalancing frequency we consider. In order to ease the presentation, for the OSM scheme in alg. \ref{algorithm:osm}, we fix $\vartheta_y=1/2$ and remark, that results are very similar in case of other choices of $\vartheta_y\in[0, 1]$.

For each experiment presented below we use fully-connected, feedforward neural networks of $L=4$ hidden layers with $50$ neurons in each layer and hyperbolic tangent activations. For the networks at time step $n=N-1$, batch normalization is applied in between each pair of layers, whose parameters are thereafter frozen for preceding time-steps $n<N-1$. The historical mean and standard deviation are reset at each time step, in order to preserve adaptivity of the solutions of the BSDEs. Adam optimization is deployed with $I_{N-1}=2^{16}$ SGD steps at the training of time step $n=N-1$ and thereafter -- motivated by the transfer learning trick as in \eqref{eq:deep_bsde:transfer_learning} -- this is reduced to $I_n=2^{12}$ for $n<N-1$.
The OSM method in alg. \ref{algorithm:osm} has been implemented in TensorFlow 2.15. The library used in this paper will be made publicly accessible on the first author's personal \href{https://github.com/balintnegyesi/OSM-delta-gamma-hedging.git}{github repository}\footnote{\href{https://github.com/balintnegyesi/OSM-delta-gamma-hedging.git}{https://github.com/balintnegyesi/OSM-delta-gamma-hedging.git}}. All experiments below were run on a Dell Alienware Aurora R10 machine, equipped with an AMD Ryzen 9 3950X CPU (16 cores, 64Mb cache, 4.7GHz) and an Nvidia GeForce RTX 3090 GPU (24Gb).

For the computations of the hedging strategies in alg. \ref{algorithm:delta_gamma_hedging}, we use equally spaced rebalancing dates with $N=1, 2, 5, 10, 20, 100$, which roughly correspond to \emph{yearly}, \emph{quarterly}, \emph{monthly}, \emph{fortnightly}, \emph{weekly} and \emph{daily} rebalancing of each portfolio. This choice ensures that each rebalance date is included in the time partition used to solve the discretely reflected FBSDEs. In each example below, we assume that options can only be exercised on rebalancing dates. In the replication of Bermudan options below, we denote the optimal stopping time at which each option is exercised by $\tau$, which is approximated as follows
\begin{align}\label{def:tau}
    \tau^j \coloneqq \argmin_{0\leq n\leq N} g^j(X_n^\pi) > \hat{\tilde{Y}}_n^{j, \pi}.
\end{align}
for a given path of the Brownian motion. In other words, $\tau^j$ denotes the first time the continuation value is reflected in the discretely reflected FBSDE system in \eqref{eq:bsde:discretely_reflected}. We approximate continuous Profit-and-Loss densities with a Gaussian kernel density estimate on the discrete Monte Carlo sample.\footnote{KDE is implemented by \texttt{seaborn}, using a smoothing parameter of $1.8$ -- see  \href{https://seaborn.pydata.org/generated/seaborn.kdeplot.html}{documentation}.}
 
\subsection{Example 1: two-dimensional stochastic volatility model}\label{sec:numerical_experiments:ex1}
\begin{table}[t]
    \centering
    \begin{tabular}{l|cccc}
         & \#1: Vega & \#2: Gamma & \#3: Vomma & \#4: Vonna \\
         \hline
       $T$  & $0.3$ & $0.4$ & $0.3$ & $0.25$\\
       $K$ & $10$ & $10$ & $9$ & $11$\\
       $R$ & $1$ & $1$ & $1$ & $1$\\
       $N$ & $60$ & $80$ & $60$ & $60$
    \end{tabular}
    \caption{Example 1. Hedging instruments.}
    \label{tab:heston:instruments}
\end{table}
Our first example is a  single Bermudan put option ($J=1$), issued on a single asset ($m=1$) whose price is driven by the well-known Heston model. Since multi-dimensional extensions to the Heston model are known to suffer from the phenomenon of vanishing correlations, we restrict our illustration to the case $m=1, d=2$, and remark that the OSM scheme would be similarly applicable in the context appropriate higher-dimensional stochastic volatility models, such as the Wishart model -- see e.g. \cite{gauthier_efficient_2009, gourieroux_continuous_2006}. Assuming the market price of volatility risk to be $\lambda=0$, the coefficients of the discretely reflected FBSDE system read as follows
\begin{align}\label{eq:heston:sde:coefficients}
    \begin{split}
        \mu(t, x=(s; \nu)) &= \left(\bar{\mu}s; \kappa(\bar{\nu} - \nu)\right),\qquad \sigma(t, x=(s; \nu))=\begin{pmatrix}
			\rho\sqrt{\nu}s & \sqrt{1-\rho^2}\sqrt{\nu}\\
			0 & \eta\sqrt{\nu}
	\end{pmatrix},\\
		f(t, x=(s; \nu), y, z) &= -ry - (\bar{\mu} - (r - q))z\sigma^{-1}(t, x) s,\qquad l(x)\equiv g(x=(s; \nu))=\max\left[K - s, 0\right],
    \end{split}
\end{align}
with $x\in\mathds{R}^2$. The two risk factors in \eqref{eq:introduction:sde} are the asset price $S=x^{1}$, and its stochastic volatility process $\nu=x^{2}$. The parameters are chosen according to set A in \cite[sec.7]{ruijter_two-dimensional_2012}, i.e. $\bar{\mu}=r=0.1$, $q=0$, $\kappa=5, \bar{\nu}=0.16, \rho=0.1$ and $\eta=0.9$. The Feller condition $2\kappa\bar{\nu}\geq \eta^2$ is satisfied and hence the volatility process does not attain zero. As in \cite{ruijter_two-dimensional_2012}, the initial condition of the forward SDE in \eqref{eq:introduction:sde} is $X_0=(S_0, \nu_0)=(10; 0.0625)$, and set the option strike to $K=10$. We consider a maturity of $T=0.25$ and equidistant early exercise dates with $R=10$, i.e. $\mathcal{R}=\{0, 0.025, 0.05, \dots, 0.25\}$.

In order to demonstrate the method's robustness with respect to the approximation of the risk factors, we adopt a modified Euler approximation directly on the asset prices and not on log-prices. The coefficients in \eqref{eq:heston:sde:coefficients} are truncated according to \cite{higham_convergence_2005, lord_comparison_2010}, in order to ensure that the discrete time approximations of the volatility process do not go below zero either
\begin{align}
    X_0=(x_0, \nu_0),\qquad X_{n+1}^\pi = (S_{n+1}^\pi, \nu_{n+1}^\pi)=\abs{(S_{n}^\pi; \nu_n^\pi) + \mu(t_n, X_n^\pi)\Delta t_n + \sigma(t_n, (S_n^\pi; \abs{\nu_n^\pi}))\Delta W_n},
\end{align}
whereas Malliavin derivative is approximated according to \eqref{eq:euler-maruyama:sde:malliavin}.
Given the approximations of the forward diffusion, we solve the discretely reflected BSDE \eqref{eq:bsde:discretely_reflected} and \eqref{eq:bsde:discretely_reflected:malliavin} corresponding to \eqref{eq:heston:sde:coefficients} by the OSM scheme in algorithm \ref{algorithm:osm} once, with $N'=50$ equally sized time intervals, which through the relations \eqref{eq:deep_bsde:delta:approx} and \eqref{eq:deep_bsde:gamma:approx} provide approximations for all first- and second-order Greeks. In particular, due to volatility uncertainty, the OSM scheme provides \emph{Delta}, \emph{Vega}, \emph{Gamma}, \emph{Vomma} and \emph{Vanna} approximations.
So that we can hedge the uncertain volatility, similar to \eqref{eq:gamma_hedging:portfolio_value}, we  augment the replicating portfolio with $4$ additional instruments, that offset Vega, Gamma, Vomma and Vanna exposure, respectively. These instruments are all European put options issued on the same asset, with maturities and strikes as in table \ref{tab:heston:instruments}. In order to sketch the potential of the OSM scheme outside of the Black-Scholes framework, we use algorithm \ref{algorithm:osm} as reference prices and Greeks for the coefficient matrices in the linear system described by \eqref{eq:delta_vega_hedging:foc} and \eqref{eq:delta_vega_hedging:soc}. Algorithm \ref{algorithm:osm} is run on each European option's associated BSDE separately, with an equidistant time grid using $N$ intervals as in table \ref{tab:heston:instruments}. We consider $N=1, 2, 5, 10, 25, 50$ rebalancing dates, for all cases.

\begin{figure}[t]
    \centering
    \begin{subfigure}[t]{\sizeonebyone\textwidth}
        \includegraphics[width=\textwidth]{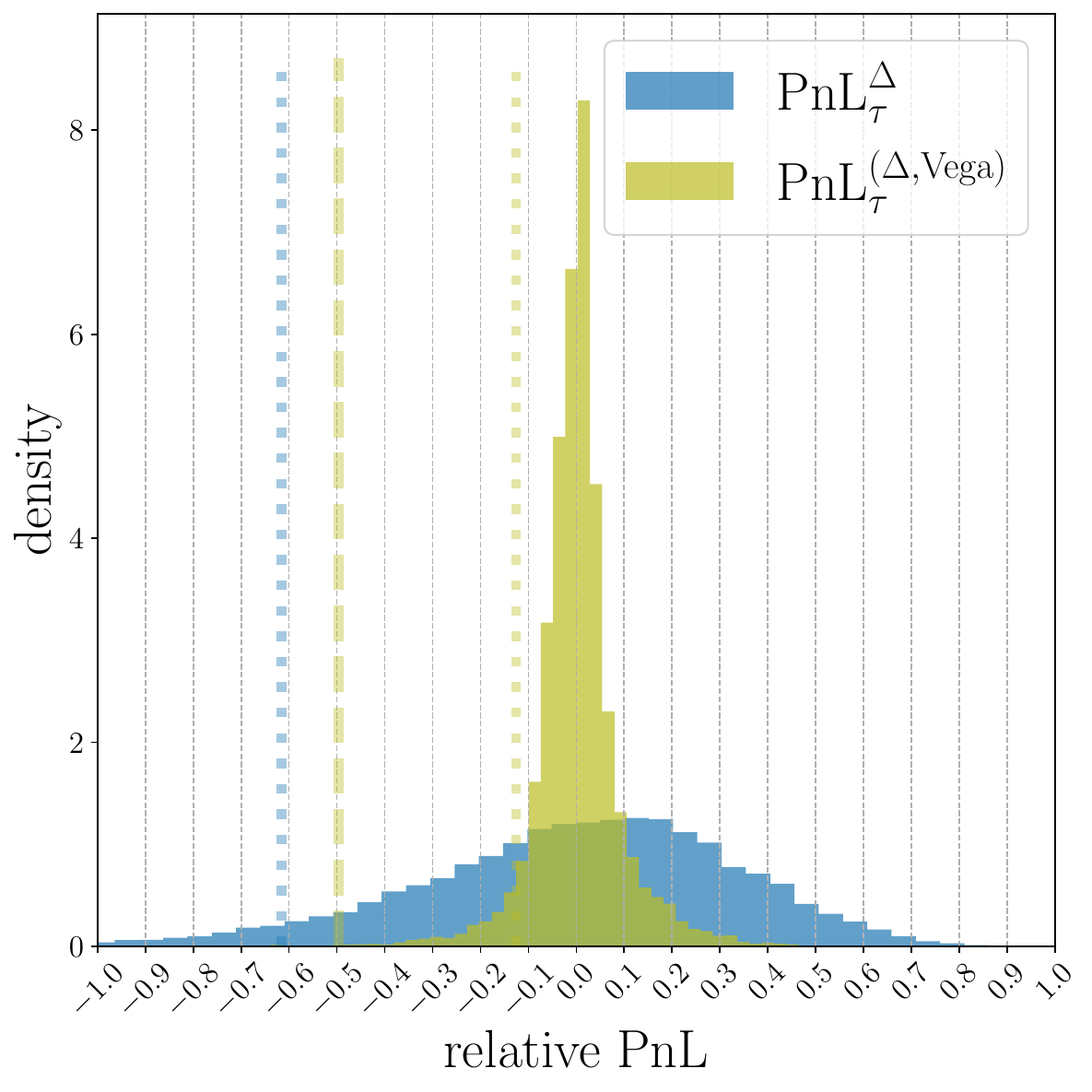}
        \caption{Delta vs. Vega}
        \label{fig:heston:pnl:first_vs_second:delta_vs_vega}
    \end{subfigure}
    \begin{subfigure}[t]{\sizeonebyone\textwidth}
        \includegraphics[width=\textwidth]{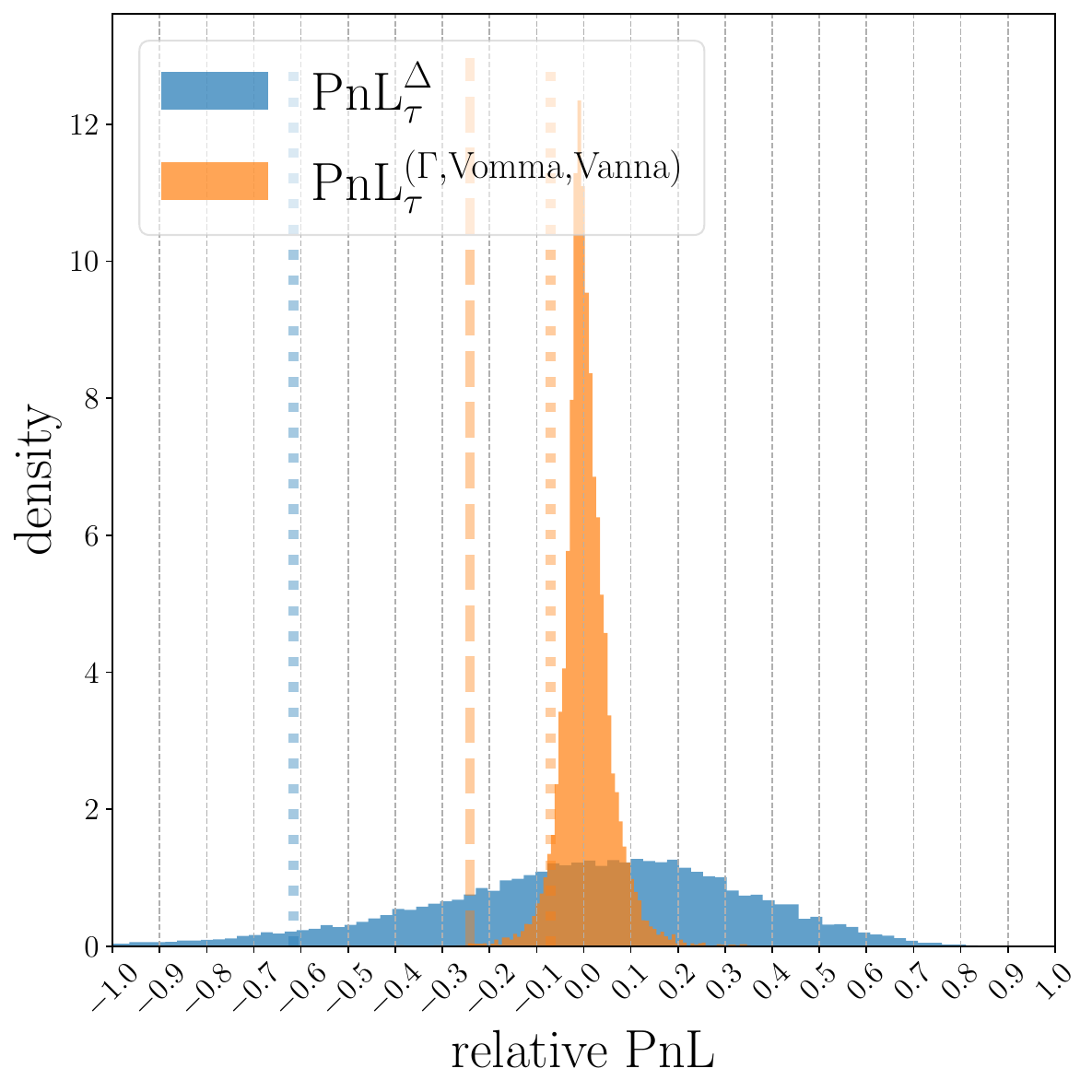}
        \caption{Delta vs. Gamma-Vomma-Vanna}
        \label{fig:heston:pnl:first_vs_second:delta_vs_2ndorder}
    \end{subfigure}
    \begin{subfigure}[t]{\sizeonebyone\textwidth}
        \includegraphics[width=\textwidth]{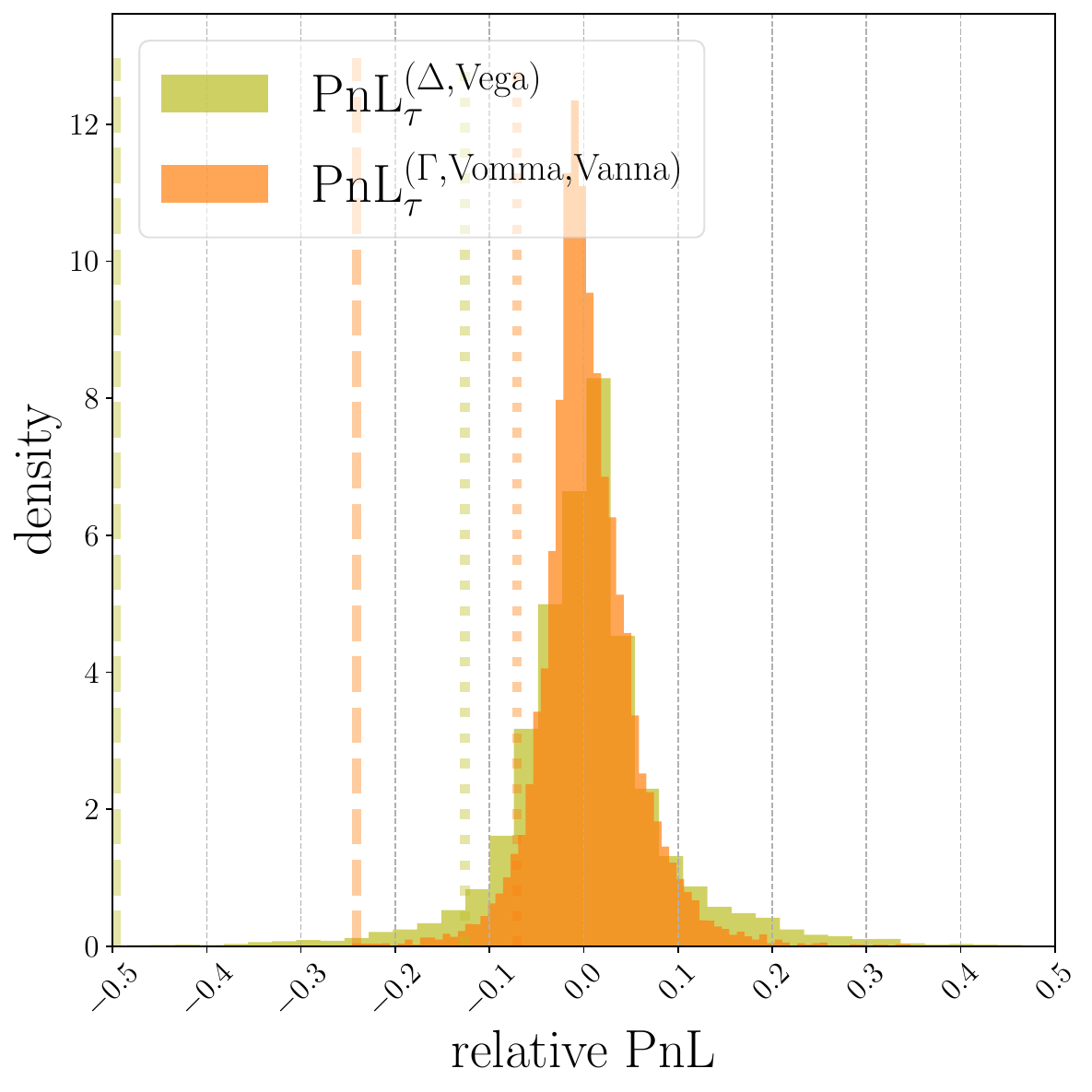}
        \caption{Delta-Vega vs. Gamma-Vomma-Vanna}
        \label{fig:heston:pnl:first_vs_second:delta-vega_vs_2ndorder}
    \end{subfigure}
    \caption{Example 1. Comparison of first- and second-order hedging strategies in \eqref{eq:delta_hedging:foc}, \eqref{eq:delta_vega_hedging:foc} and \eqref{eq:delta_vega_hedging:soc}, respectively. $N=50$ rebalancing dates. Dashed and dotted vertical lines corresponding to $\text{VaR}_{95}$ and $\text{ES}_{99}$, respectively. Hedging instruments as in table \ref{tab:heston:instruments}. PnL approximated through an independent Monte Carlo sample of size $2^{14}$.}
    \label{fig:heston:pnl:first_vs_second}
\end{figure}

\begin{figure}[t]
    \centering
    \begin{subfigure}[t]{\sizeonebyone\textwidth}
        \includegraphics[width=\textwidth]{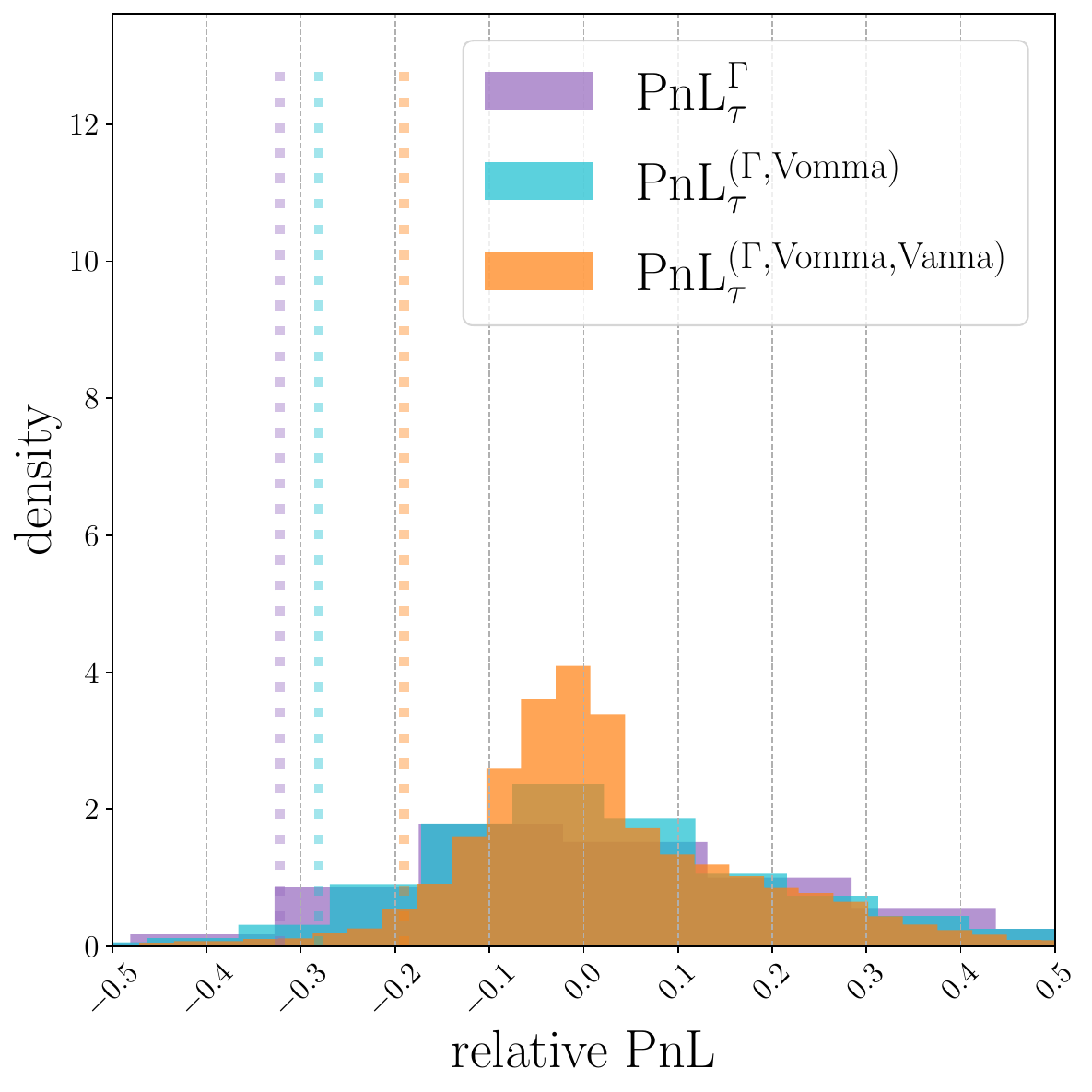}
        \caption{Comparison on the set of second-order Greeks accounted for in \eqref{eq:delta_vega_hedging:soc} with quarterly rebalancing ($N=2$).}
        \label{fig:heston:pnl:2ndorder_comparison}
    \end{subfigure}
    \quad
    \begin{subfigure}[t]{\sizetwobyone\textwidth}
        \includegraphics[width=0.494\textwidth]{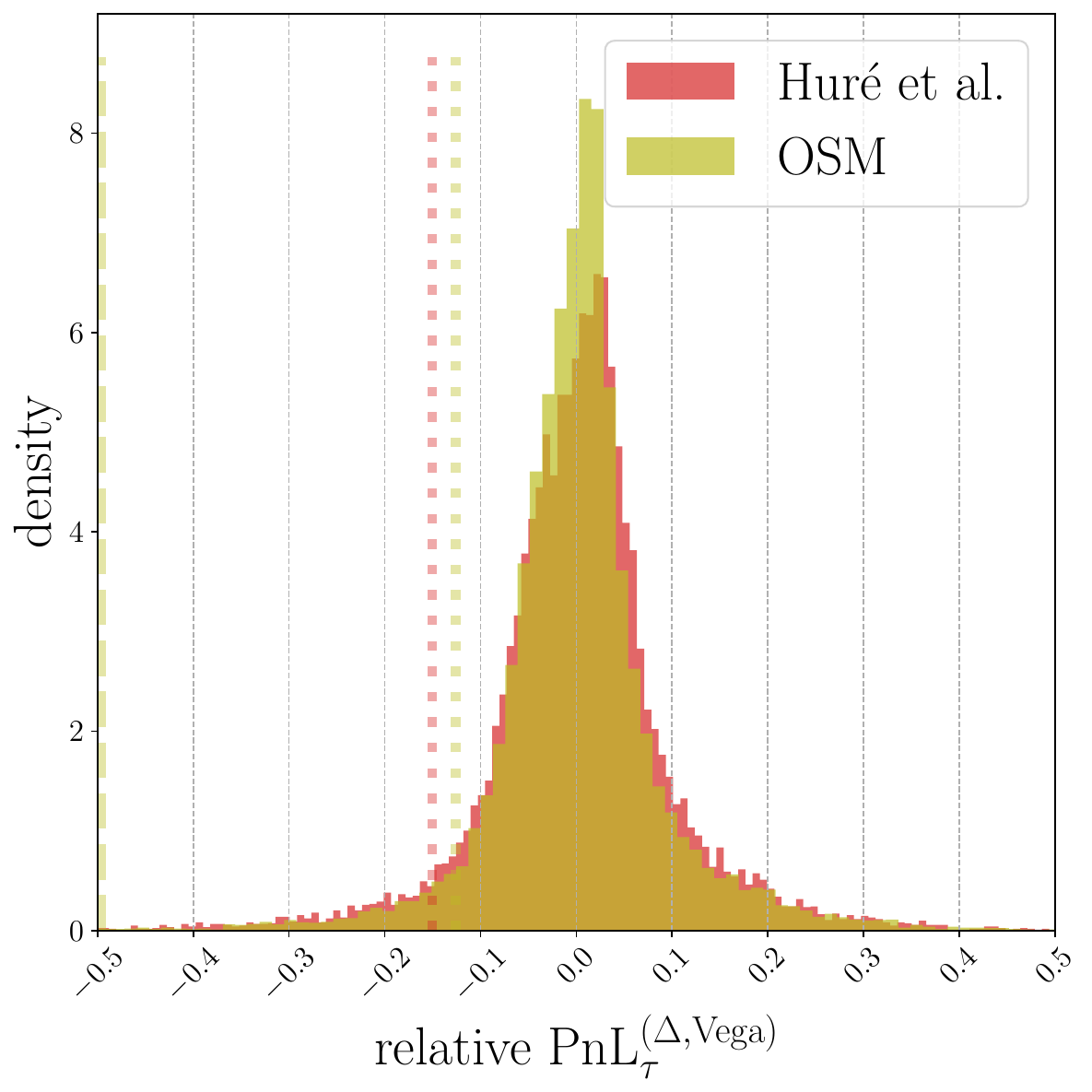}
        \includegraphics[width=0.494\textwidth]{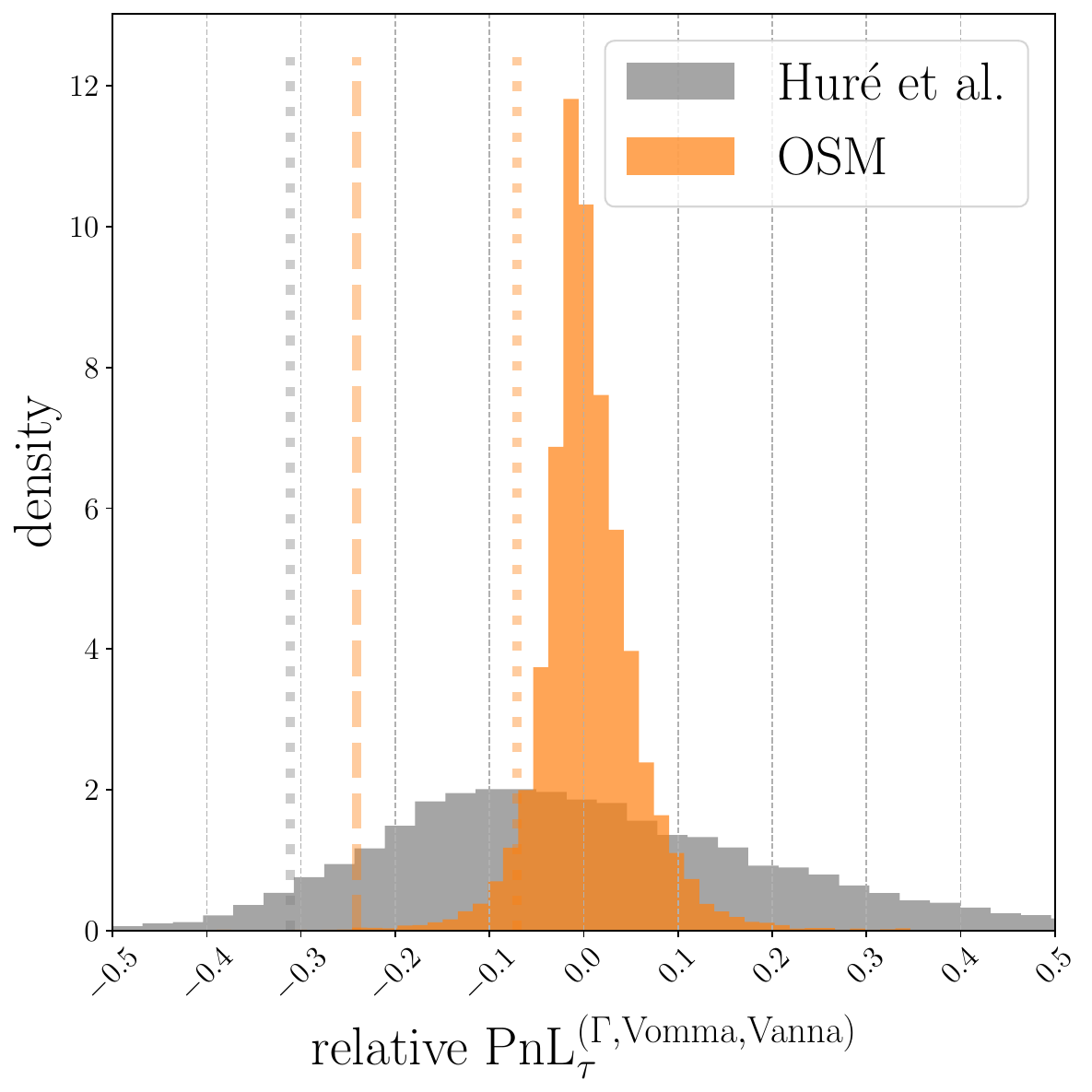}
        \caption{Comparison between the One Step Malliavin scheme (alg.\ref{algorithm:osm}) and the RDBDP approach of \cite{hure_deep_2020}. First- \eqref{eq:delta_vega_hedging:foc} and second-order \eqref{eq:delta_vega_hedging:soc} hedging on the left and right, respectively.}
        \label{fig:heston:pnl:osm_vs_hure}
    \end{subfigure}
    \caption{Example 1. Dashed and dotted vertical lines corresponding to $\text{VaR}_{95}$ and $\text{ES}_{99}$, respectively. Hedging instruments as in table \ref{tab:heston:instruments}. PnL approximated through an independent Monte Carlo sample of size $2^{14}$.}
\end{figure}

The impact of volatility uncertainty on the hedging with only first-order constraints in \eqref{eq:deep_bsde:delta_hedging:alpha} is illustrated by figure \ref{fig:heston:pnl:first_vs_second:delta_vs_vega}. Herein, we find that augmenting the replicating portfolio with an additional Vega hedging instrument, offsetting first-order volatility uncertainty according to \eqref{eq:delta_vega_hedging:foc} results in a significantly sharper PnL distribution. 
Nevertheless, as demonstrated by figures \ref{fig:heston:pnl:first_vs_second:delta_vs_2ndorder} and \ref{fig:heston:pnl:first_vs_second:delta-vega_vs_2ndorder}, the delta-vega replicating portfolio entails substantial risk due to the neglection of second-order sensitivities. Augmenting the replicating portfolio with the additional Gamma, Vomma and Vanna hedging securities, one can further improve the replication of the Bermudan put option at $\tau$, resulting in sharper profit and loss distributions centered around $0$. As seen in fig. \ref{fig:heston:pnl:first_vs_second:delta-vega_vs_2ndorder}, offsetting all second-order sentivities with respect to the underlying risk factors improves $\text{VaR}_{95}$ by $30$ percentage points.
The impact on the choices of second-order Greeks to be offset, determined by the index set $\mathcal{I}$ in \eqref{eq:delta_vega_hedging:soc} is depicted in figure \ref{fig:heston:pnl:2ndorder_comparison}. Three cases are compared depending on the number of second-order Greeks accounting for \textit{(i)} Gamma hedging ($\mathcal{I}=\{11\}$); \textit{(ii)} Gamma-Vomma hedging ($\mathcal{I}=\{11, 22\}$); \textit{(iii)} Gamma-Vomma-Vanna hedging ($\mathcal{I}=\{11, 12, 22, 21\}$).
As demonstrated by fig.\ref{fig:heston:pnl:2ndorder_comparison}, these gradually result in better replication, as more second-order Greeks are taken into account.
Table \ref{tab:heston:risk_measures} collects the mean and variance for the aforementioned replication strategies in case of $N=25$ rebalancing dates. In line with the discussion above, we see that offsetting second-order Greeks not only with respect to the asset price but also for the uncertain volatility considerably improves the replication accuracy across all risk metrics.

Finally, as discussed in section \ref{section:deep_bsde}, let us compare the deep BSDE approximations of all above considered option Greeks provided by the One Step Malliavin scheme in alg. \ref{algorithm:osm} to the RDBDP method of \cite{hure_deep_2020}. The left plot of fig. \ref{fig:heston:pnl:osm_vs_hure} compares the delta-vega hedging strategies where the corresponding right-hand sides of \eqref{eq:delta_vega_hedging:foc} are computed by the OSM and RDBDP schemes, respectively. As we can see, even in this low-dimensional case the OSM scheme brings a marginal improvement in the replication accuracy, indicating that the corresponding first-order Greeks (Deltas and Vegas) are more accurately approximated by the OSM scheme. More importantly, the right side of \ref{fig:heston:pnl:osm_vs_hure} demonstrates the need for the Malliavin representation formula \eqref{eq:bsde:discretely_reflected:malliavin} and its corresponding discrete time approximation by the OSM scheme in order to accurately capture second-order Greeks. Herein, the second-order hedging corresponding to Gamma-Vomma-Vanna hedging \eqref{eq:delta_vega_hedging:soc} are compared across the OSM and RDBDP schemes, where, for the latter, the right-hand side of \eqref{eq:delta_vega_hedging:soc} is computed by automatic differentiation. As we can see, automatic differentiation of \cite{hure_deep_2020} does not provide accurate Gammas, Vommas and Vannas, and is rendered inapplicable in the context of second-order hedging, whereas the OSM approximations bring an improvement in the replication accuracy compared to the first-order conditions.

\begin{table}[t]
    \centering
    \begin{tabular}{l|ccccc}
       risk measure  & $\text{PnL}_{\tau}^{\Delta}$ & $\text{PnL}_{\tau}^{\Delta,\text{Vega}}$ & $\text{PnL}_{\tau}^{\Gamma}$ & $\text{PnL}_{\tau}^{\Gamma,\text{Vomma}}$ & $\text{PnL}_{\tau}^{\Gamma,\text{Vomma},\text{Vanna}}$\\
       \hline
        mean & \num{1.2E-02} & \num{1.8E-02} & \num{2.0E-02} & \num{1.9E-02} & \num{1.9E-02}\\
        variance & \num{1.5E-01} & \num{9.0E-02} & \num{1.0E-02} & \num{8.5E-03} & \num{7.6E-03}
    \end{tabular}
    \caption{Example 1. Risk measures for different types of replicating portfolios and $N=25$ rebalancing dates.}
    \label{tab:heston:risk_measures}
\end{table}

\subsection{Example 2: single high-dimensional option}
The high-dimensional examples are given on $d=m$ assets driven by Black-Scholes dynamics under the physical measure. First, we investigate a single Bermudan geometric call option $(J=1)$, with varying early exercise rights. The coefficients of the corresponding discretely reflected FBSDE system read as follows\footnote{we denote element-wise multiplication by $\odot$}
\begin{align}\label{eq:example1:black_scholes:physical}
    \mu(t, x) &= \bar{\mu} \odot x,\quad \sigma(t, x)=\text{diag}(\bar{\sigma}\odot x)\Sigma,\\
    f(t, x, y, z) &= -ry - \big(\frac{\bar{\mu} - (r-q)}{\bar{\sigma}}\big)^T z\Sigma^{-1},\quad
    l(x)\equiv g(x)=\max\left[\omega\left(\big(\prod_{i=1}^d x^i\big)^{1/d} - K\right), 0\right],
\end{align}
	with $\omega\in\{-1, 1\}$ for put and call options, respectively. In the above, $\bar{\mu}, \bar{\sigma}, q\in\mathbb{R}^m$ are the drift, volatility and continuous dividend yield parameters of each asset, $r$ is the risk free rate and $C=\Sigma^T\Sigma$ is the correlation structure between the assets with $\Sigma$ being its Cholesky decomposition. This example appears often in the literature, see e.g. \cite{chen_deep_2021, hure_deep_2020}, due to its special property in that the general $d$-dimensional problem can be reduced to a scalar one -- for full details see e.g. \cite{negyesi_reflected_2024}. We set the parameters in \eqref{eq:example1:black_scholes:physical} according to \cite{chen_deep_2021}, and consider a fixed $T=2, X_0=100, r=0.0, q=(0.02, \dots, 0.02)$, $\omega=1$, $c_{ij}=0.75, i\neq j$.
    
    We train a deep BSDE solver once, offline for each parameter setting with $N'=100$ equally sized time intervals according to algorithm \ref{algorithm:osm} and use the resulting approximations to recover option prices, Deltas in \eqref{eq:deep_bsde:delta:approx} and Gammas according to \eqref{eq:deep_bsde:gamma:approx}. We emphasize that the OSM scheme provides \emph{all Gammas} in the high-dimensional framework, including all \emph{cross-gammas}, i.e. the Gamma in \eqref{eq:deep_bsde:gamma:approx} takes values in $\mathbb{R}^{d\times d}$. Due to the high-correlation and similarly to example \ref{sec:numerical_experiments:ex1} -- see fig. \ref{fig:heston:pnl:2ndorder_comparison} in particular --, we found that it is not sufficient to remove pure second-order sensitivities with respect to each individual asset $(\mathcal{I}=\{ii: 1\leq i\leq m\})$, as there is a substantial cross-gamma exposure. Therefore, we choose to hedge the whole upper triangular part of the Hessian matrix in \eqref{eq:gamma_hedging:soc} in case of delta-gamma hedging, i.e. $\mathcal{I}=\{ij: 1\leq i\leq j\leq d\}$. In order to be able to compute the coefficient matrix on the left-hand side of the second-order condition \eqref{eq:gamma_hedging:soc}, we augment the delta-gamma hedging portfolio \eqref{eq:gamma_hedging:portfolio_value} with $K=d(d+1)/2$ Gamma hedging instruments, as follows
    \begin{itemize}
        \item for the diagonal elements in the Gamma matrix $(i=j)$ we choose standard European put options with the same strike and maturity $\widetilde{T}=2T$;
        \item for the cross Gamma indices $(i\neq j)$ we choose European exchange calls with strike $K^{ij}=1$ and maturity $\widetilde{T}=2T$. This in particular implies the corresponding hedging instruments' prices, Delta and Gammas are all available in closed form due to the Margrabe formula \cite{margrabe_value_1978} -- see appendix \ref{sec:appendix:margrabe}.
    \end{itemize}
    As discussed in section \ref{sec:subsec:linear_system}, these choices imply that the linear system corresponding to \eqref{eq:gamma_hedging:foc-soc} has a sparse coefficient matrix, and thus the linear system in \eqref{eq:gamma_hedging:deep_bsde:soc} can be stored and solved efficiently for a large number of Monte Carlo simulations.
    We split the discussion of the example into four key aspects: the impact of moneyness (determined by strike $K$), early exercising ($R$), strength of volatility $(\bar{\sigma})$ and dimensionality $(d=m)$.
    
\begin{figure}
        \centering
        \begin{subfigure}[t]{\sizeonebyone\textwidth}
        \centering
        \includegraphics[width=\textwidth]{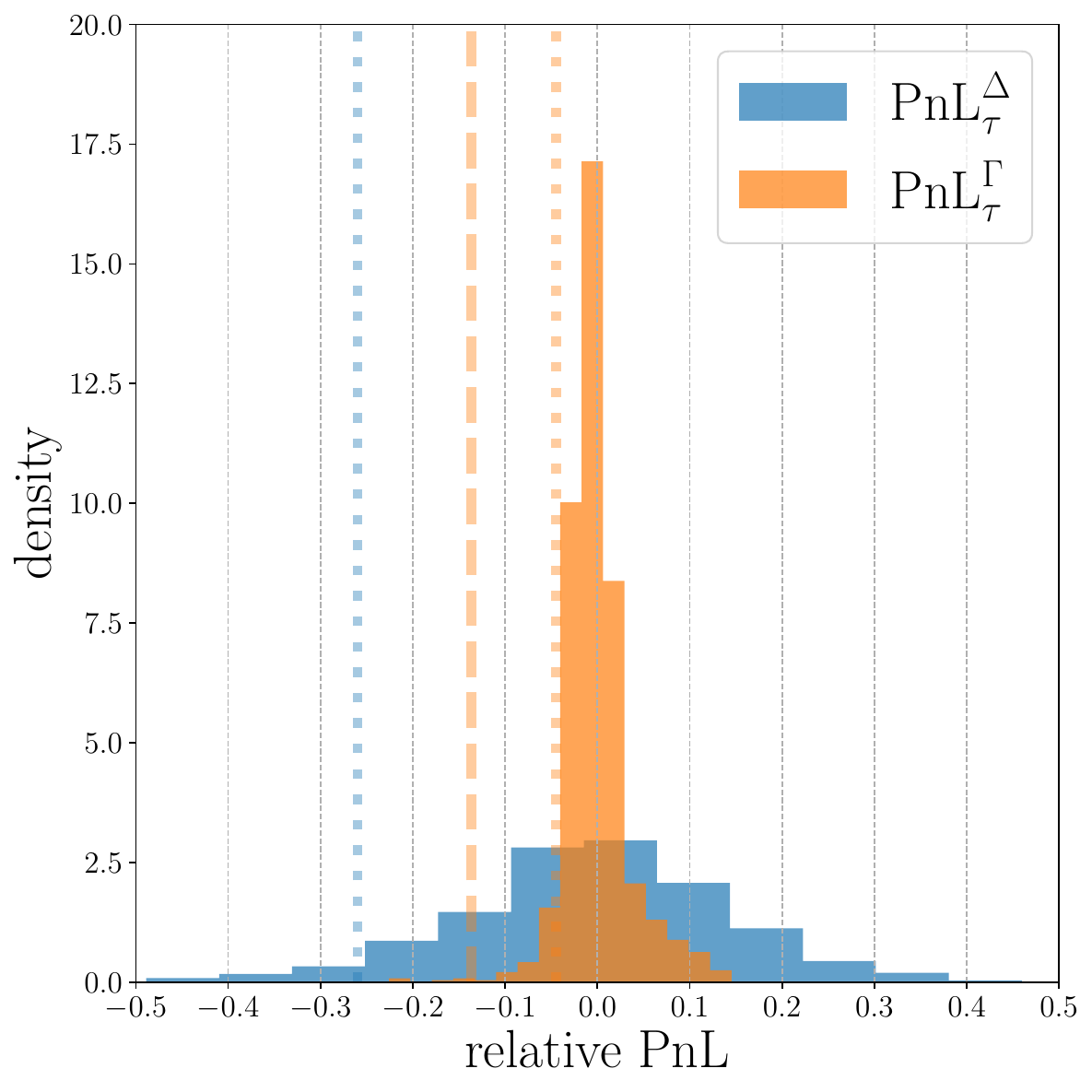}
        \caption{$K=90$ (ITM)}
        \label{fig:bs:strike_comparison_N=20:itm}
        \end{subfigure}
        \begin{subfigure}[t]{\sizeonebyone\textwidth}
        \centering
        \includegraphics[width=\textwidth]{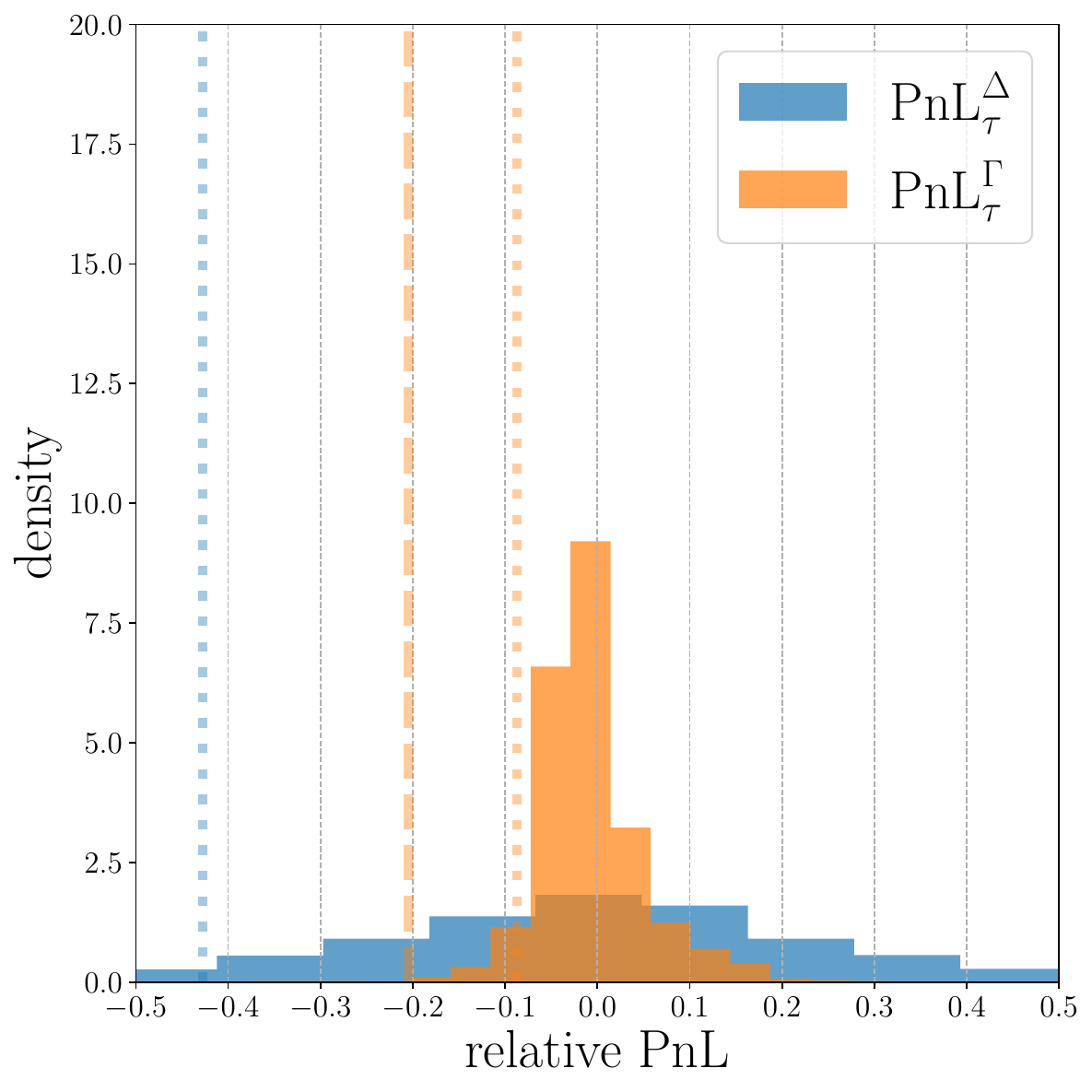}
        \caption{$K=100$ (ATM)}
        \label{fig:bs:strike_comparison_N=20:atm}
        \end{subfigure}
        \begin{subfigure}[t]{\sizeonebyone\textwidth}
        \centering
        \includegraphics[width=\textwidth]{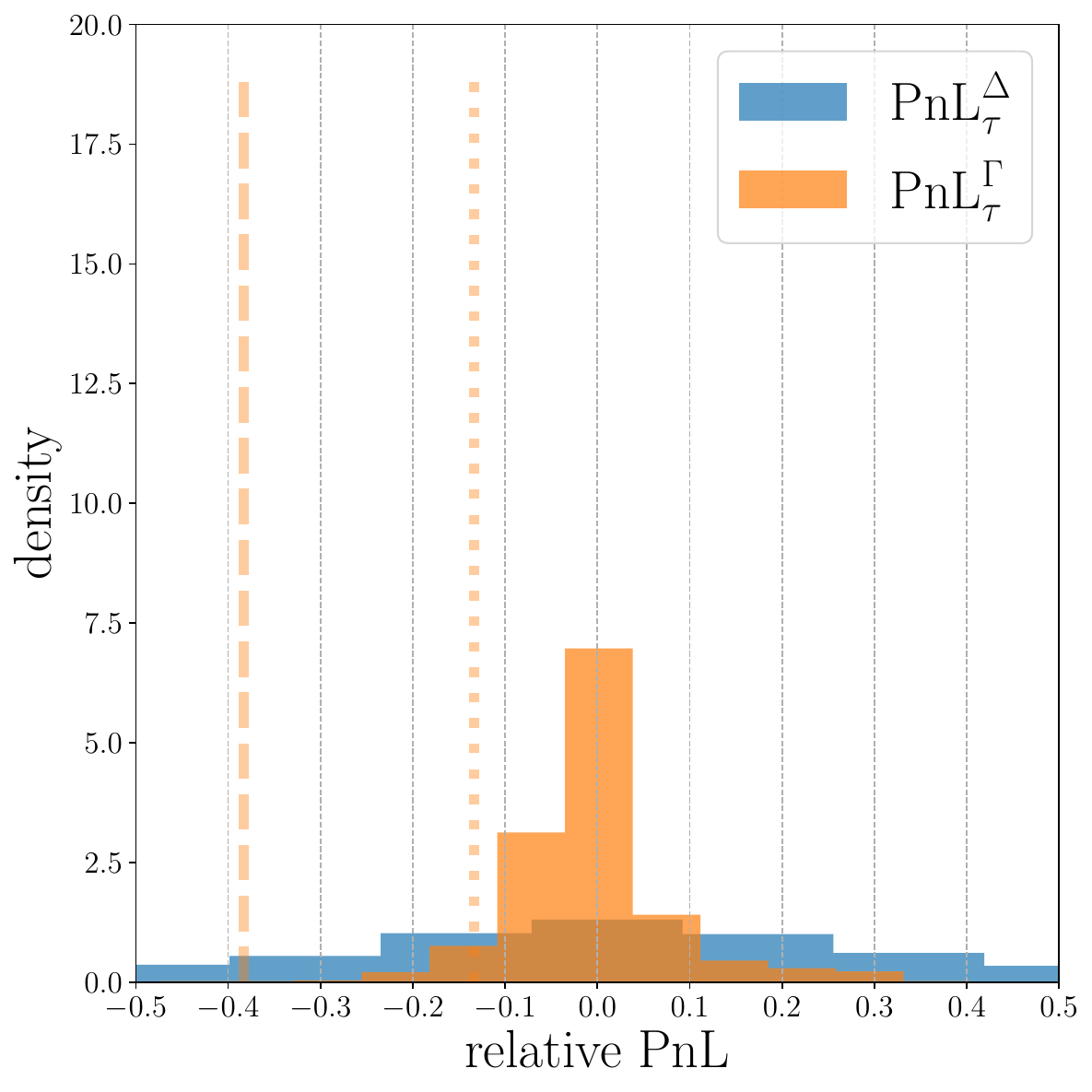}
        \caption{$K=110$ (OTM)}
        \label{fig:bs:strike_comparison_N=20:otm}
        \end{subfigure}
        \caption{Example 2 in \eqref{eq:example1:black_scholes:physical}. Moneyness comparison with weekly rebalancing ($N=20$). Example 2. Histograms for delta \eqref{eq:delta_hedging:foc} and delta-gamma replication \eqref{eq:gamma_hedging:foc-soc}. Dotted and dashed vertical lines corresponding to $\text{VaR}_{95}$ and $\text{ES}_{99}$, respectively.}
        \label{fig:bs:strike_comparison_N=20}
    \end{figure}

    \paragraph{OTM/ATM/ITM.} We fix a European contract $(R=1, \mathcal{R}=\{0, T\})$ on $d=m=50$ assets, each with volatility $\bar{\sigma}=(0.25, \dots, 0.25)$ and an initial condition of $X_0=(100,\dots, 100)$. We consider in-, at- and out-of-the-money strikes with $K=90, 100, 110$, respectively, to investigate the impact of moneyness on the discrete replication accuracy both in case of delta and delta-gamma hedging. Results are collected in figure \ref{fig:bs:strike_comparison_N=20} and table \ref{tab:bs:moneyness}. We see that Gamma hedging significantly outperforms delta hedging in terms of replication accuracy, irrespective of the moneyness.  Across all values of moneyness, we found that a fortnightly rebalanced ($N=10$) delta-gamma hedging strategy achieves the same replication accuracy as a daily rebalanced $(N=100)$ delta strategy. In particular, we find that once rebalanced at least weekly $(N=20)$ the delta-gamma hedging strategy achieves an expected shortfall of less than $-0.4$ for each moneyness, whereas delta rebalancing with the same frequency yields a much wider Profit-and-Loss distribution with more profound tail risk.  
    All risk measures are collected in table \ref{tab:bs:moneyness} for fortnightly $(N=10)$ rebalanced portfolios in \eqref{eq:delta_hedging:portfolio_value} and \eqref{eq:gamma_hedging:portfolio_value}. In line with fig. \ref{fig:bs:strike_comparison_N=20}, we find that the discrete replication becomes more difficult as the option is going out-of-the-money. Nonetheless, even in the OTM case, when the position is delta-gamma hedged with higher than monthly ($N=5$) frequency, the corresponding replication achieves a $\text{VaR}_{95}$ below $-0.5$, which is only attained by daily delta rebalancing. 
    The delta-gamma hedged portfolio using alg. \ref{algorithm:osm} and \ref{algorithm:delta_gamma_hedging} achieves an order of magnitude higher replication accuracy across all risk measures consistently, regardless of the moneyness, in the high-dimensional setting ($d=m=50$).
    
        
        

    \begin{table}[t]
        \centering
        \begin{tabular}{l|ccc|ccc}
             & \multicolumn{3}{c}{Delta} & \multicolumn{3}{c}{Delta-Gamma}\\
             & $K=90$ & $K=100$ & $K=110$ & $K=90$ & $K=100$ & $K=110$\\
             \hline
            mean & \num{-8.8E-03} & \num{-7.4E-03} & \num{-2.7E-02} & \num{-7.1E-03} & \num{-1.4E-02} & \num{-4.1E-03}\\
            variance & \num{4.4E-02} & \num{1.3E-01} & \num{3.3E-01} & \num{3.7E-03} & \num{7.9E-03} & \num{2.7E-02}\\
            $\text{VaR}_{95}$ & \num{-3.7E-01} & \num{-6.3E-01} & \num{-1.1E+00} & \num{-7.7E-02} & \num{-1.2E-01} & \num{-2.0E-01}\\
            $\text{ES}_{95}$ & \num{-5.2E-01} & \num{-8.6E-01} & \num{-1.5E+00} & \num{-1.4E-01} & \num{-1.9E-01} & \num{-3.3E-01}\\
            semivariance & \num{2.4E-02} & \num{6.8E-02} & \num{2.0E-01} & \num{1.7E-03} & \num{2.8E-03} & \num{1.0E-02}
        \end{tabular}
        \caption{Example 2 in \eqref{eq:example1:black_scholes:physical}. Comparison of risk measures across moneyness. Fortnightly rebalancing $(N=10$).}
        \label{tab:bs:moneyness}
    \end{table}

\begin{figure}
        \centering
        \begin{subfigure}[t]{\textwidth}
        \centering
        \includegraphics[width=\sizeonebyone\textwidth]{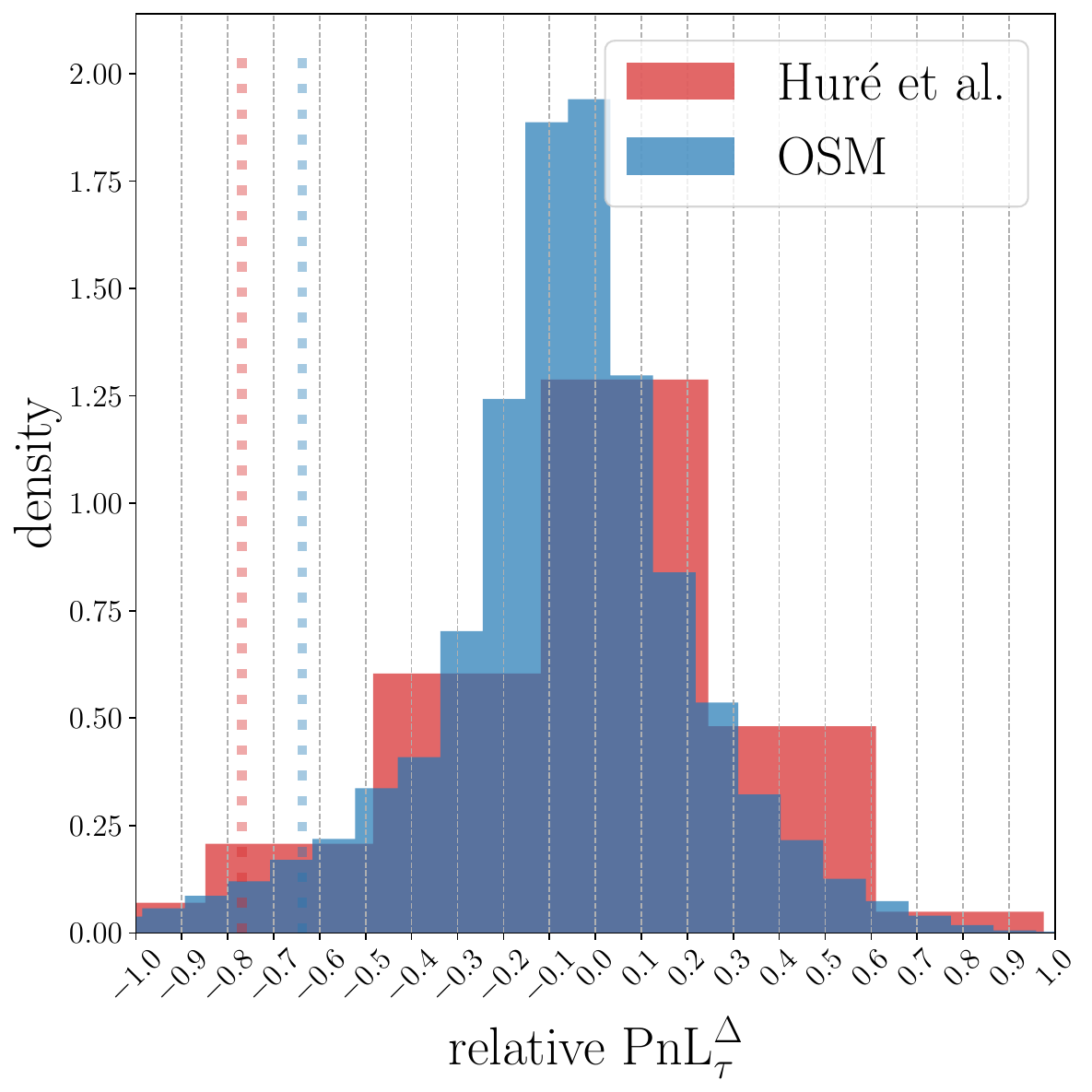}
        \includegraphics[width=\sizeonebyone\textwidth]{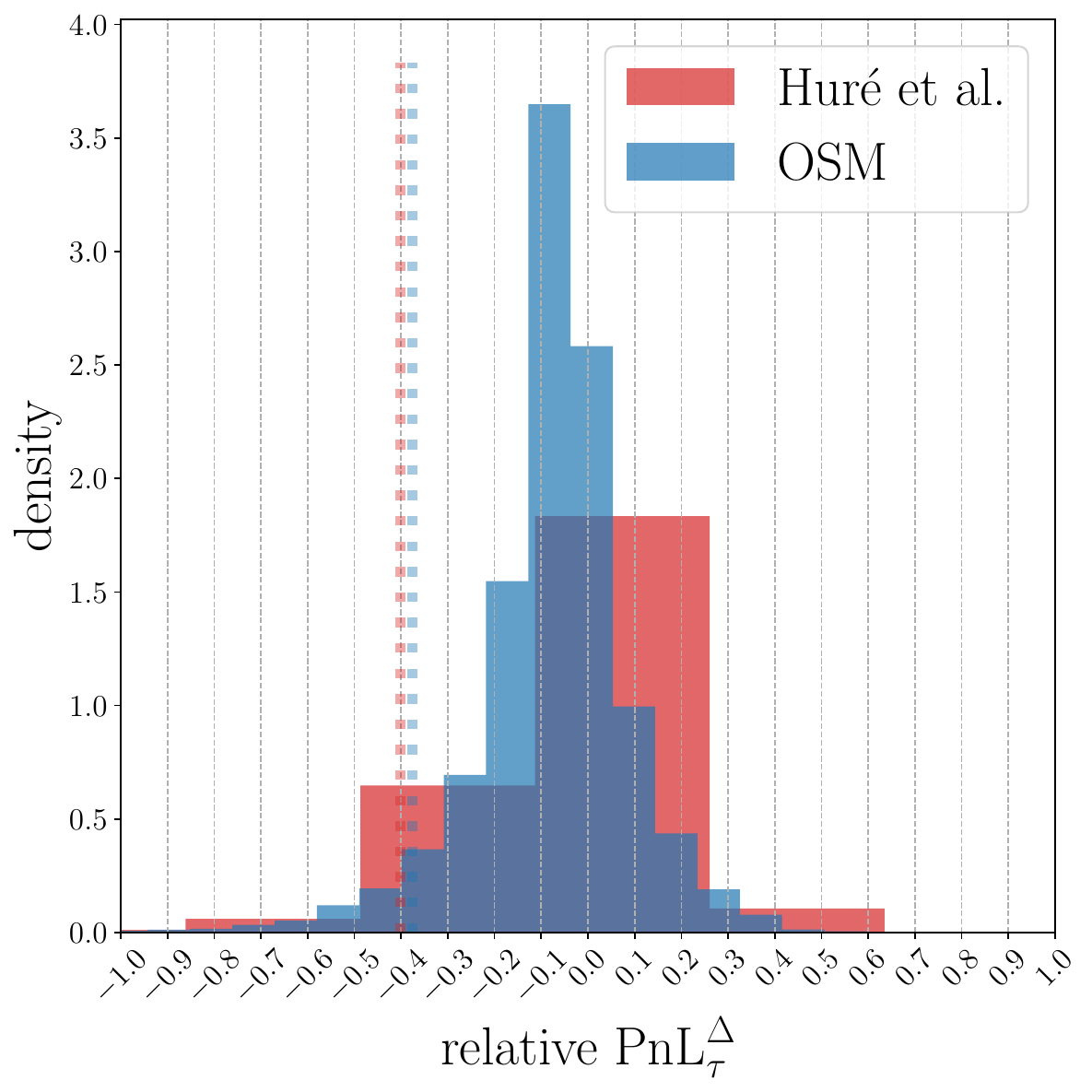}
        \caption{Comparison between Deltas approximated by the OSM scheme (alg.\ref{algorithm:osm}) and Huré et al. \cite{hure_deep_2020}. Left weekly $(N=20)$, right daily ($N=100$) rebalancing.}
        \label{fig:bs:osm_vs_hure:delta}
        \end{subfigure}
        
        \begin{subfigure}[t]{\textwidth}
        \centering
        \includegraphics[width=\sizeonebyone\textwidth]{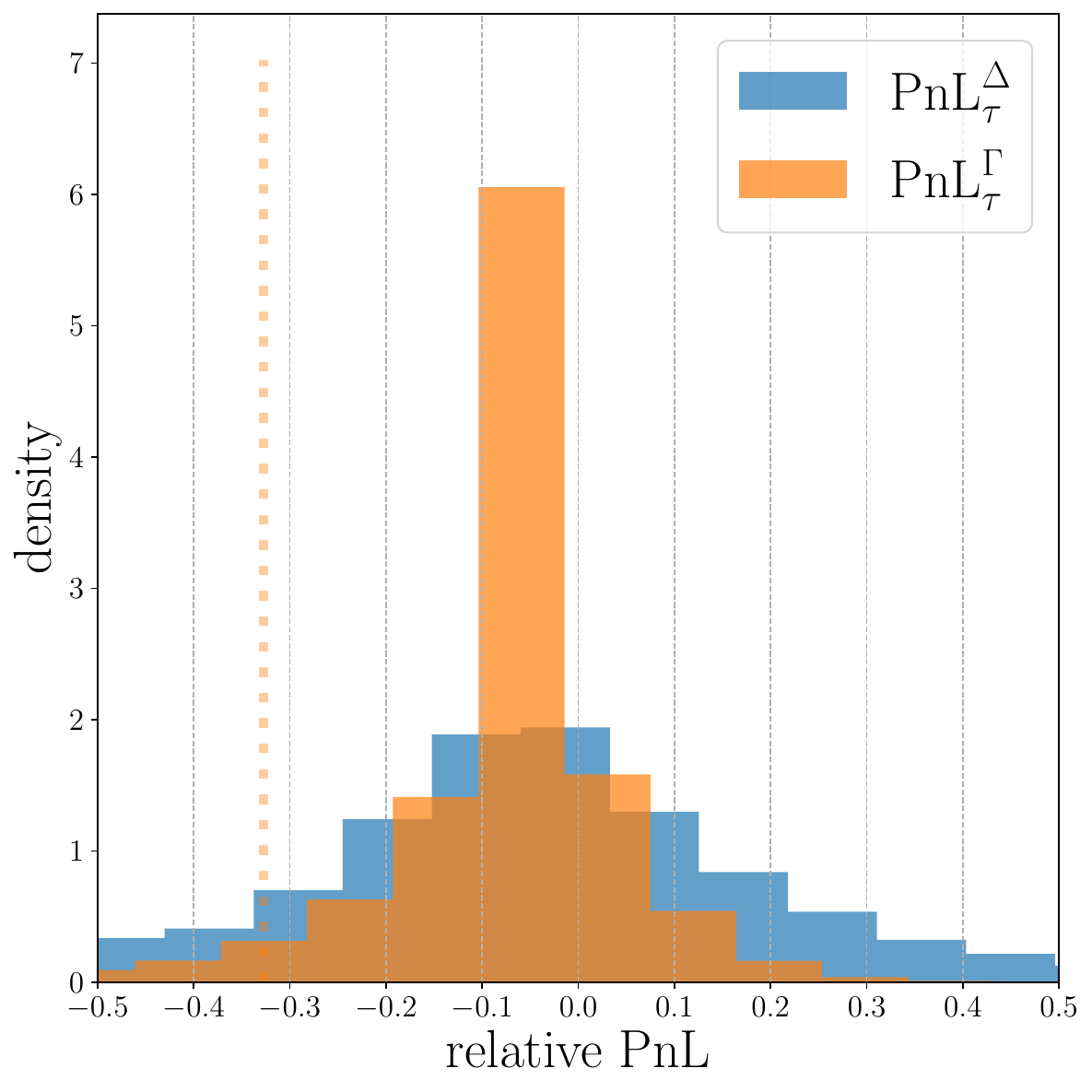}
        \includegraphics[width=\sizeonebyone\textwidth]{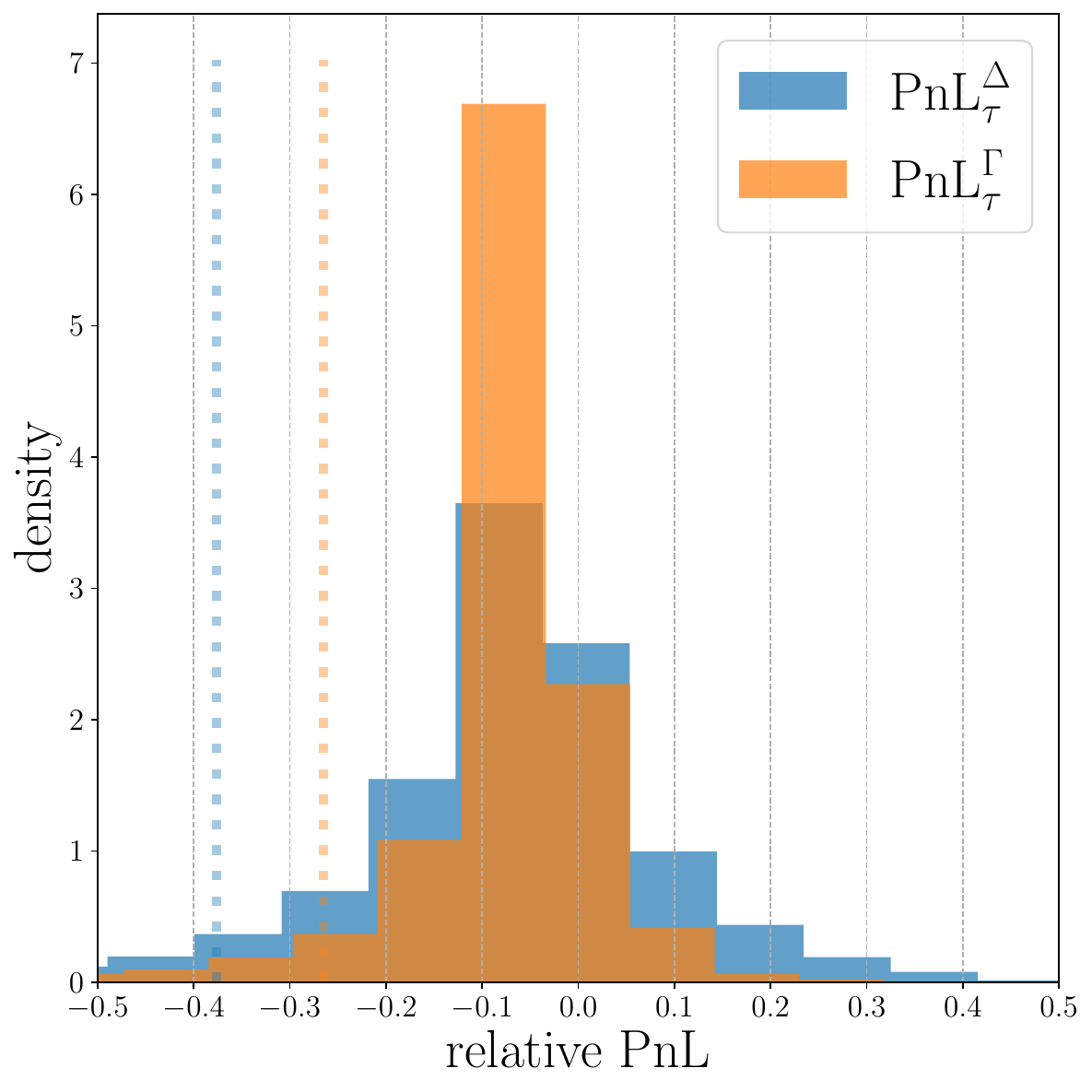}
        \caption{Comparison of delta \eqref{eq:delta_hedging:foc} and delta-gamma replication \eqref{eq:gamma_hedging:foc-soc}. Left weekly $(N=20)$, right daily ($N=100$) rebalancing.}
        \label{fig:bs:osm_vs_hure:delta_gamma}
        \end{subfigure}
        
        \caption{Example 2 in \eqref{eq:example1:black_scholes:physical}. Volatility impact comparison for $\bar{\sigma}=(0.75, \dots, 0.75)$.  Dotted and dashed vertical lines corresponding to $\text{VaR}_{95}$ and $\text{ES}_{99}$, respectively.}
        \label{fig:bs:osm_vs_hure}
    \end{figure}

    \paragraph{High volatility (OSM versus Huré et al. \cite{hure_deep_2020}).} Next, we assess the impact of the strength of the volatility parameter $\bar{\sigma}$ in \eqref{eq:example1:black_scholes:physical}. In order to do so, we fix $R=1$, consider an ATM strike of $K=100$ and vary $\bar{\sigma}$ between $0.25, 0.5$ and $0.75$ across all $d=m=50$ assets, uniformly. The numerical results are given in figure \ref{fig:bs:osm_vs_hure} and table \ref{tab:bs:osm_vs_hure}. As indicated by \cite{negyesi_one_2024, negyesi_reflected_2024}, the One Step Malliavin scheme in alg.\ref{algorithm:osm} outperforms the reference methods \cite{hure_deep_2020, chen_deep_2021} in the approximation accuracy of the $Z$ process in \eqref{eq:bsde:discretely_reflected}, especially in settings of high-volatility and small time steps. This is demonstrated by figure \ref{fig:bs:osm_vs_hure:delta}. We find that as the strength of the volatility increases the OSM scheme results in gradually more accurate Delta approximations compared to the reference method \cite{hure_deep_2020}, which results in sharper $\text{PnL}^\Delta$ distributions. We found similar results in case of other references, such as \cite{chen_deep_2021}. This effect becomes more profound as $\bar{\sigma}$ increases, indicating that the One Step Malliavin scheme is not only useful in the context of second-order sensitivities, but may also result in higher first-order replication accuracy for highly volatile assets.
    Nonetheless, as shown by figure \ref{fig:bs:osm_vs_hure:delta_gamma}, one can further improve the delta replication accuracy even by offsetting the corresponding second-order sensitivities in \eqref{eq:gamma_hedging:portfolio_value} -- even for large $\bar{\sigma}$. We find that offsetting all option Gammas enabled by the OSM scheme, further improves the replication accuracy resulting in an approximately $30$ percentage point improvement in value-at-risk and a sharper PnL distribution around $0$. Unsurprisingly, as the volatility increases, the replication accuracy decreases, and one gains even more by delta-gamma hedging. All risk measures are collected in table \ref{tab:bs:osm_vs_hure}, when the corresponding discretely rebalanced portfolios are updated daily. Comparing the delta hedging strategies between deep BSDE approximations provided by the OSM scheme and \cite{hure_deep_2020}, we find that algorithm \ref{algorithm:osm} provides higher accuracy in the Delta approximations when the volatility parameter is high, which results in higher replication accuracy. In particular, using the OSM scheme, one gains roughly $3$ times lower variance for the PnL distribution around $0$. Additionally, see $\bar{\sigma}=(0.5, \dots, 0.5)$, offsetting the associated Gammas results in an additional order of magnitude accuracy in the variance of the profit and loss distribution.
    
    \begin{table}[t]
        \centering
        \begin{tabular}{l|cc||cc|cc}
             & \multicolumn{2}{c}{Huré et al. \cite{hure_deep_2020}} & \multicolumn{4}{c}{OSM (alg. \ref{algorithm:osm})}\\
             & \multicolumn{2}{c}{Delta} & \multicolumn{2}{c}{Delta} & \multicolumn{2}{c}{Gamma}\\
             & $\bar{\sigma}=0.5$ & $\bar{\sigma}=0.75$ & $\bar{\sigma}=0.5$ & $\bar{\sigma}=0.75$& $\bar{\sigma}=0.5$ & $\bar{\sigma}=0.75$\\
             \hline
             mean &  \num{-4.1E-03} & \num{-3.7E-02} & \num{-6.4E-03} & \num{-8.0E-02} & \num{-5.2E-03} & \num{-8.0E-02}\\
             variance & \num{2.0E-02} & \num{1.3E-01} & \num{1.4E-02} & \num{3.8E-02} & \num{1.4E-03} & \num{2.3E-02}\\
             $\text{VaR}_{95}$ & \num{-2.4E-01} & \num{-4.0E-01} & \num{-2.1E-01} & \num{-3.8E-01} & \num{-6.8E-02} & \num{-2.6E-01}\\
             $\text{ES}_{95}$ & \num{-3.4E-01} & \num{-7.2E-01} & \num{-3.0E-01} & \num{-6.0E-01} & \num{-9.2E-02} & \num{-5.1E-01}\\
             semivariance & \num{1.1E-02} & \num{8.5E-02} & \num{7.9E-03} & \num{3.8E-02} & \num{7.7E-04} & \num{4.4E-02}
        \end{tabular}
        \caption{Example 2 in \eqref{eq:example1:black_scholes:physical}. Comparison of risk measures across different levels of volatility. Delta hedging with OSM in alg. \ref{algorithm:osm} versus Huré et al. in \cite{hure_deep_2020}. Daily rebalancing $(N=100$).}
        \label{tab:bs:osm_vs_hure}
    \end{table}

\begin{figure}[t]
        \centering
        \begin{subfigure}[t]{\textwidth}
        \centering
        \includegraphics[width=\sizeonebyone\textwidth]{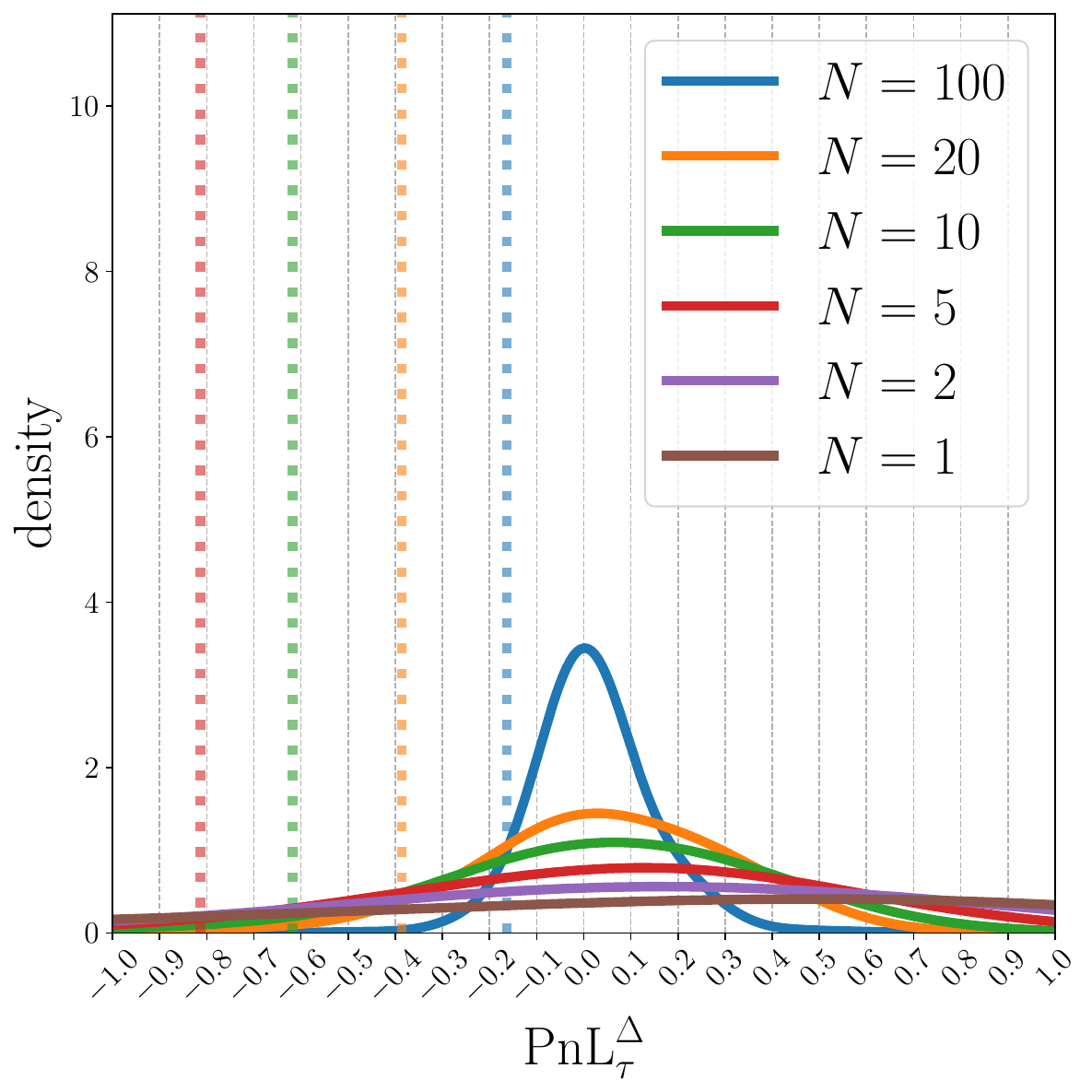}
        \includegraphics[width=\sizeonebyone\textwidth]{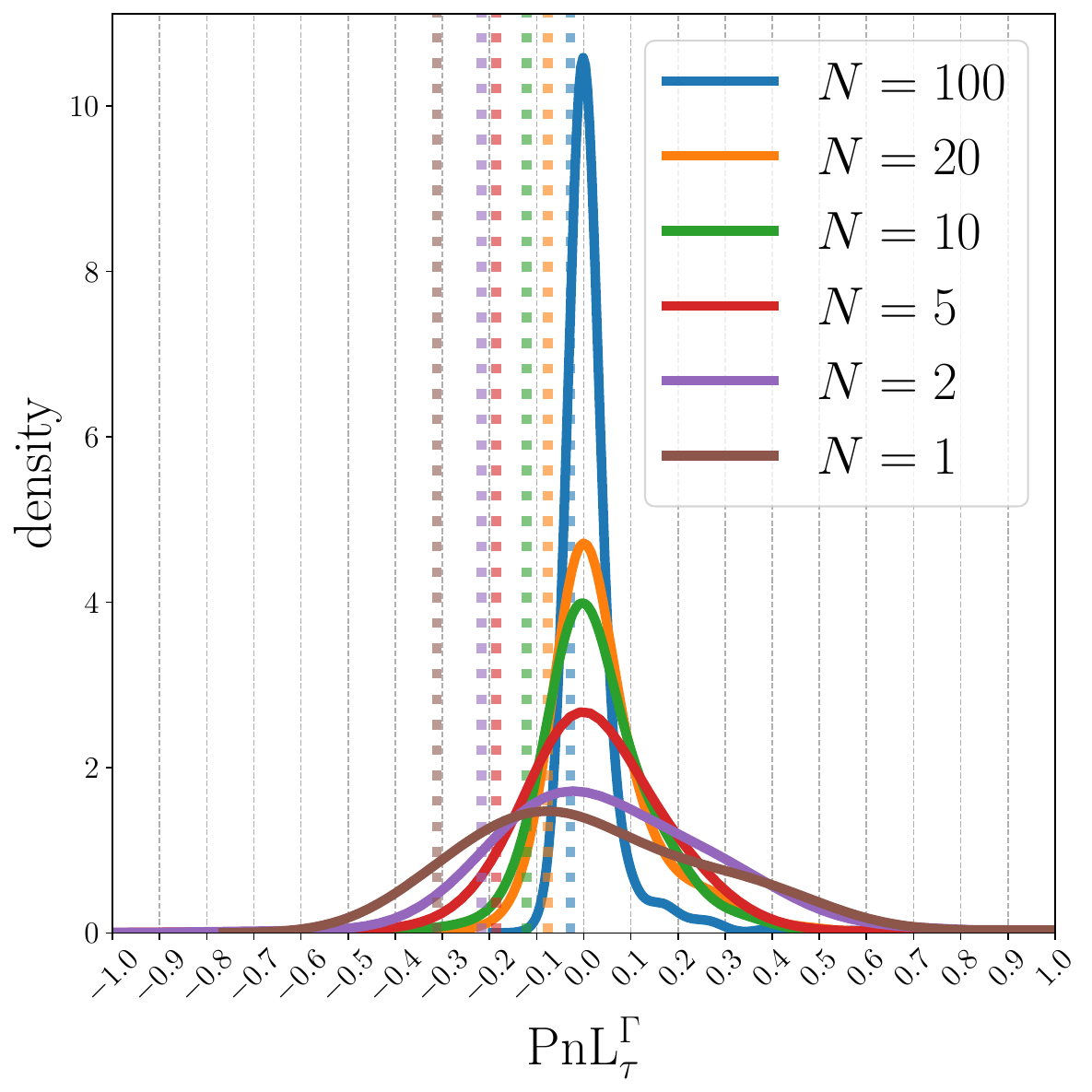}
        \caption{monthly early exercise ($R=5$)}
        \label{fig:bs:reflection:density:R=5}
        \end{subfigure}
        
        \begin{subfigure}[t]{\textwidth}
        \centering
        \includegraphics[width=\sizeonebyone\textwidth]{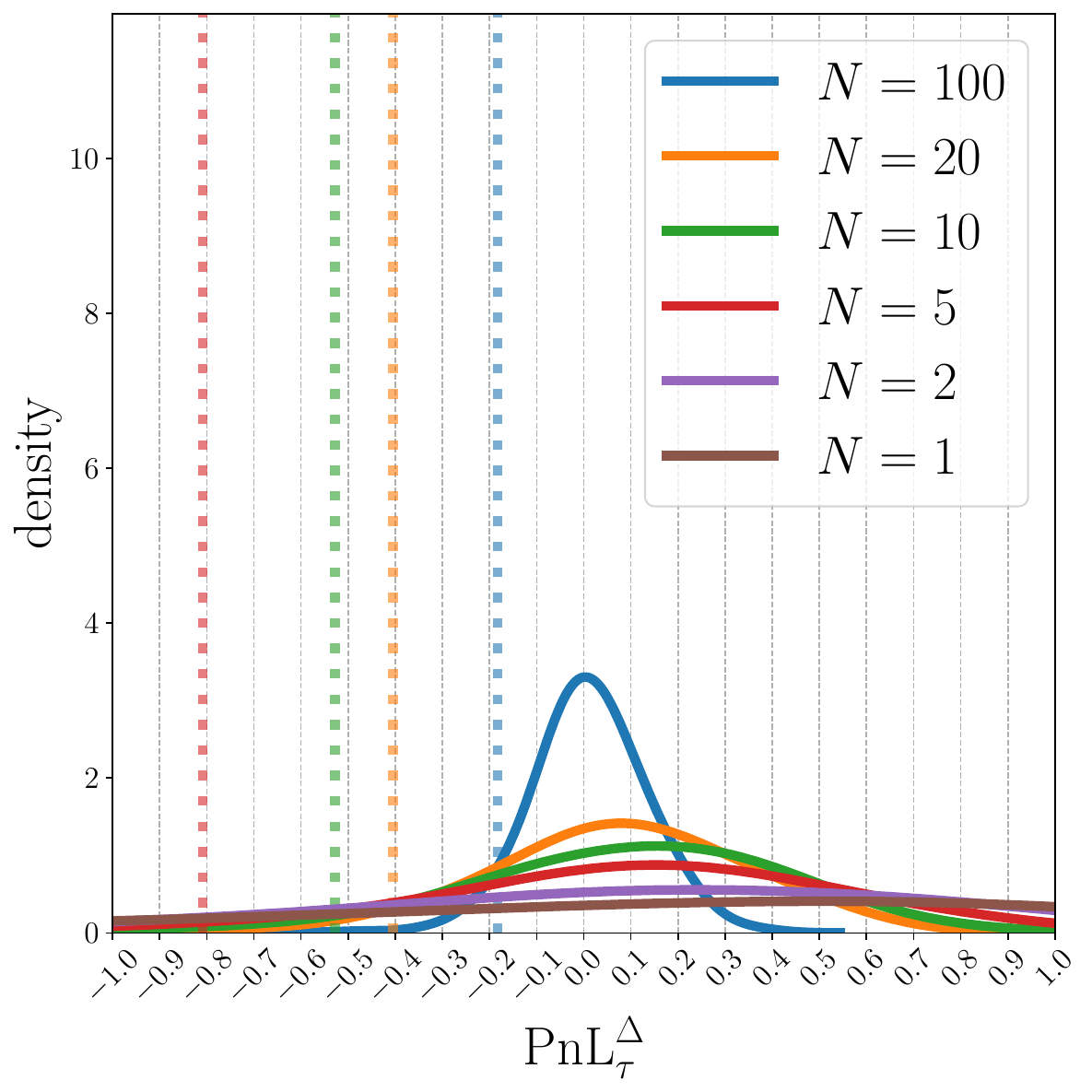}
        \includegraphics[width=\sizeonebyone\textwidth]{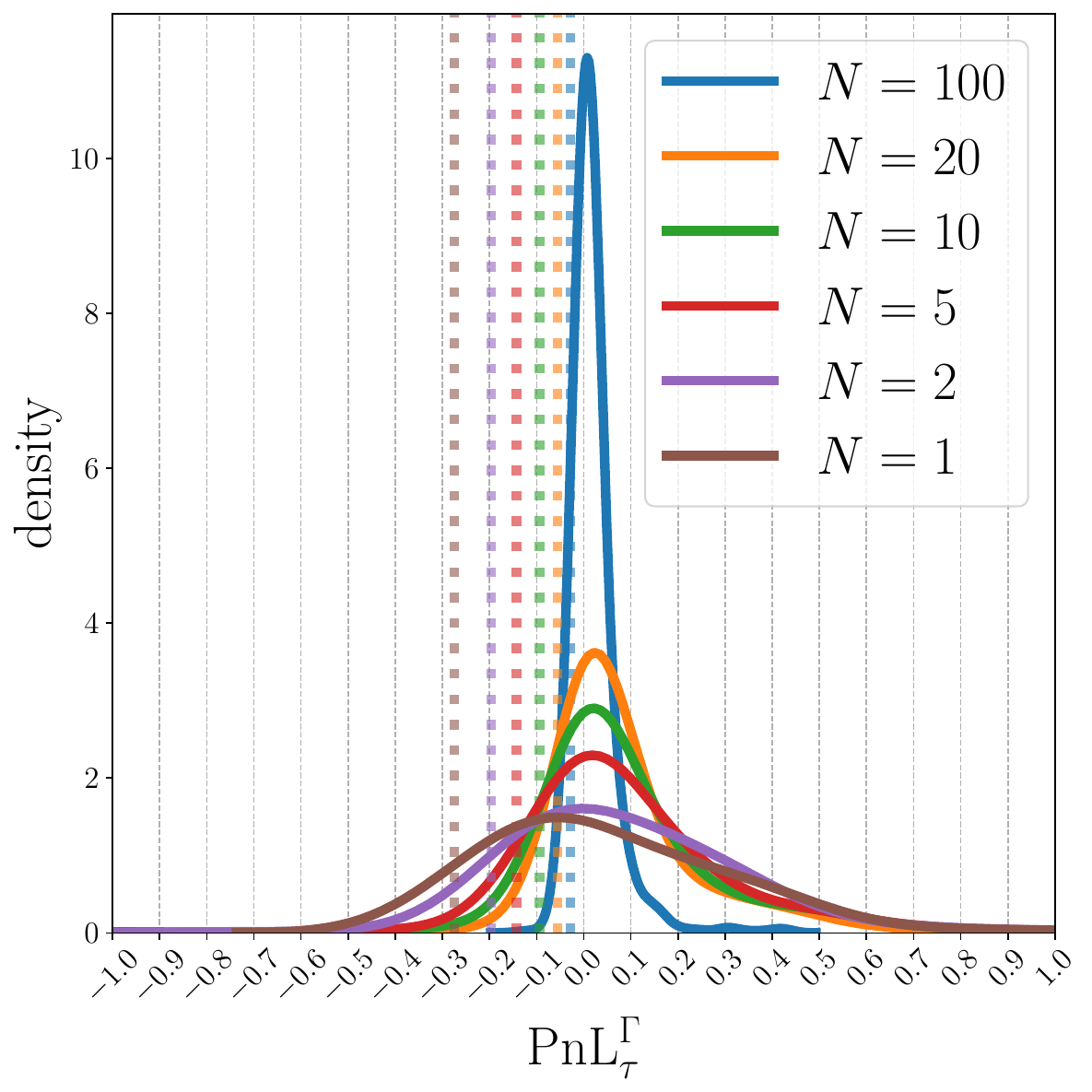}
        \caption{weekly early exercise ($R=20$)}
        \label{fig:bs:reflection:density:R=20}
        \end{subfigure}
        
        \begin{subfigure}[t]{\textwidth}
        \centering
        \includegraphics[width=\sizeonebyone\textwidth]{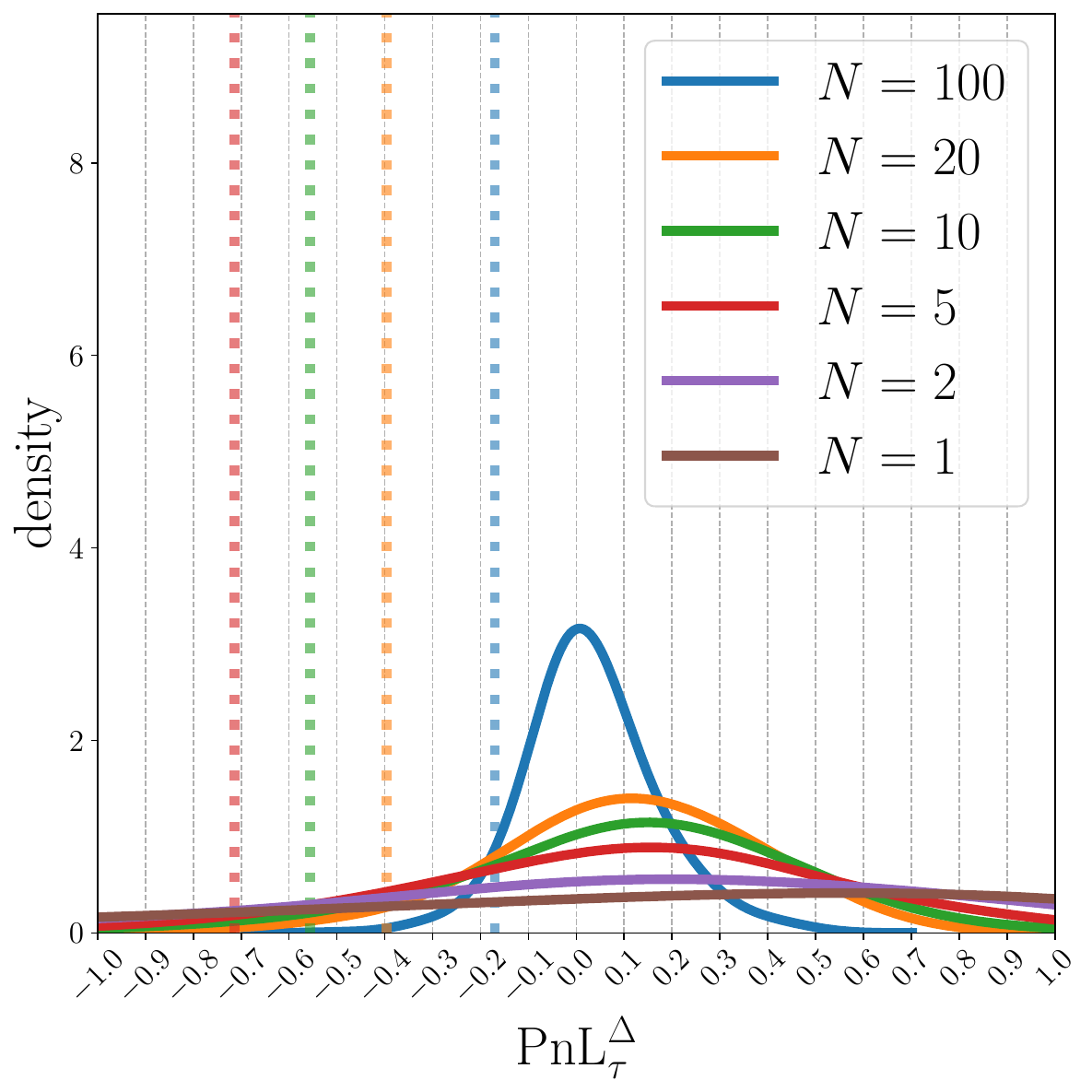}
        \includegraphics[width=\sizeonebyone\textwidth]{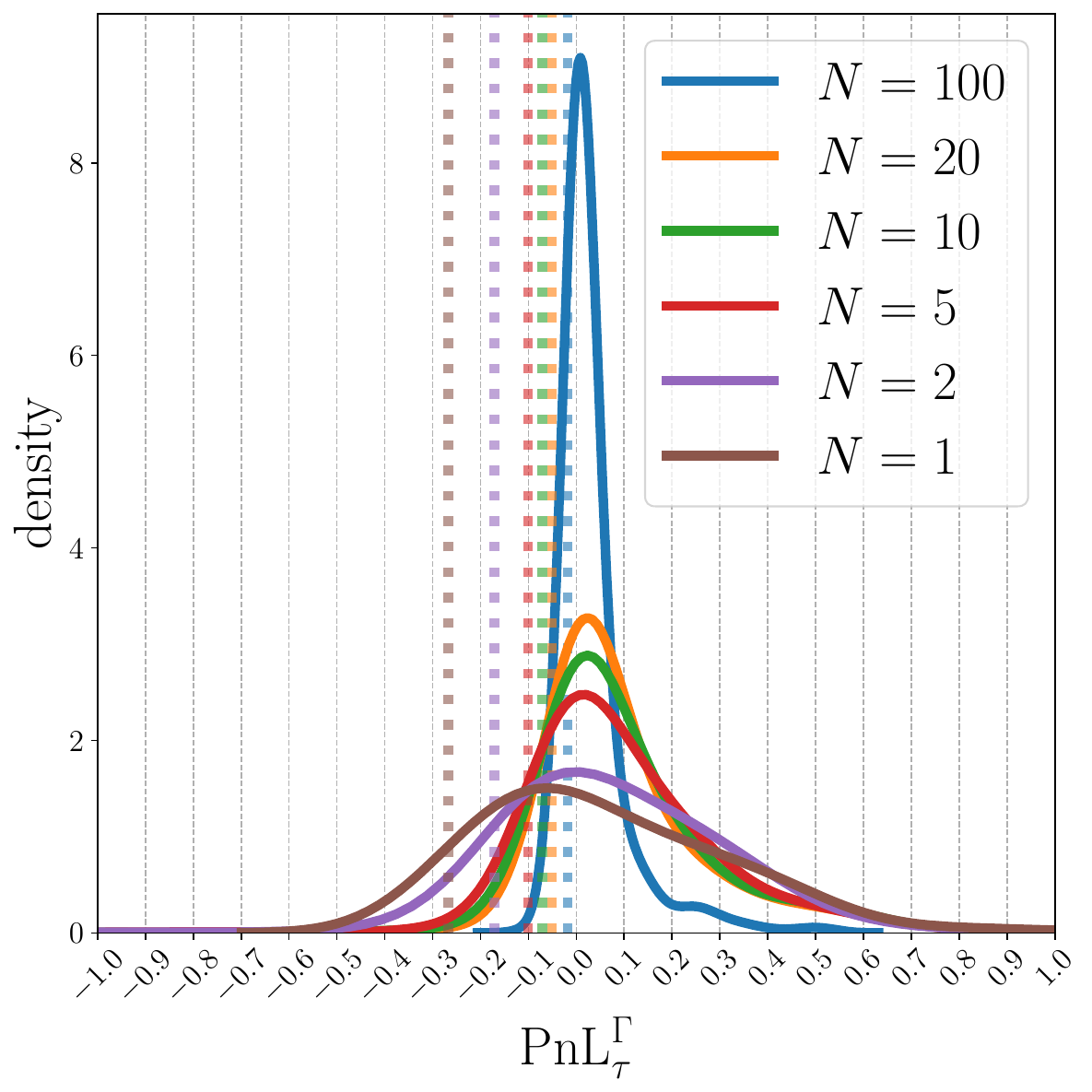}
        \caption{daily early exercise ($R=100$)}
        \label{fig:bs:reflection:density:R=100}
        \end{subfigure}
        \caption{Example 2 in \eqref{eq:example1:black_scholes:physical}. Increasing early exercises rights, from Bermudan to American. PnL densities for several rebalancing frequencies. Left: delta hedging \ref{eq:delta_hedging:foc}, right: delta-gamma hedging \eqref{eq:gamma_hedging:foc-soc}. Dotted vertical lines corresponding to $\text{VaR}_{95}$.}
        \label{fig:bs:reflection:density}
    \end{figure}

    \paragraph{Early exercise rights.} Let us investigate the impact of early exercising on the corresponding replication accuracies. We consider ATM $(K=100)$, Bermudan call options as in \eqref{eq:example1:black_scholes:physical}, issued on $d=m=50$ underlyings, each admitting a volatility of $0.25$. In order to assess the early exercise right's impact on the replication accuracy we vary the equidistant early exercise dates taking values $R=5, 20, 100$, which can be associated with monthly, weekly and daily exercise rights, respectively. We remark that the case $R=100$ can be thought of as a numerical approximation of the American option limit and can thus be directly compared to the results in \cite{chen_deep_2021}. The numerical experiments are summarized in figure \ref{fig:bs:reflection:density} and table \ref{tab:bs:reflection}. As we can see, the OSM hedging strategies are robust and accurate with respect to the number of early exercise dates of the corresponding Bermudan option for both delta- and delta-gamma hedging. In fact, comparing the densities in fig. \ref{fig:bs:reflection:density}, we find that, even though the approximation of the corresponding discretely reflected BSDE becomes more challenging, the resulting delta-gamma hedged PnLs are still sharply distributed around $0$. The scales of the vertical axes show that the replication accuracy goes down as $R$ increases (ceteris paribus), due to the larger number of early exercised paths. Comparing the left and right columns we see that the additional second-order constraints yield a substantial improvement to delta hedging. The delta-gamma hedged portfolio achieves a similar profit and loss distribution with only fortnightly rebalancing ($N=10$) as that of the delta hedging with daily readjusted hedging weights ($N=100$). Risk measures for all hedging strategies are collected in table \ref{tab:bs:reflection} for portfolios rebalanced every fortnight ($N=10$). The delta-gamma hedging strategies enabled by the OSM approximations in algorithm \ref{algorithm:osm} induce close to an order of magnitude improvement in the variance of the associated profit-and-loss distributions compared to mere delta-hedging, irrespective of early-exercise features, in the high-dimensional option setting with $d=m=50$ underlyings.

    \begin{table}[t]
        \centering
        \begin{tabular}{l|ccc|ccc}
             & \multicolumn{3}{c}{Delta} & \multicolumn{3}{c}{Delta-Gamma}\\
             & $R=5$ & $R=20$ & $R=100$ & $R=5$ & $R=20$ & $R=100$\\
             \hline
            mean & \num{2.4E-02} & \num{7.5E-02} & \num{7.9E-02} & \num{2.9E-02} & \num{9.1E-02} & \num{9.7E-02}\\
            variance & \num{1.3E-01} & \num{1.3E-01} & \num{1.2E-01} & \num{1.5E-02} & \num{3.5E-02} & \num{3.2E-02}\\
            $\text{VaR}_{95}$ & \num{-6.2E-01} & \num{-5.3E-01} & \num{-5.7E-01} & \num{-1.2E-01} & \num{-9.3E-02} & \num{-7.2E-02}\\
            $\text{ES}_{95}$ & \num{-9.0E-01} & \num{-8.6E-01} & \num{-8.3E-01} & \num{-2.1E-01} & \num{-1.9E-01} & \num{-1.8E-01}\\
            semivariance & \num{7.7E-02} & \num{8.2E-02} & \num{7.4E-02} & \num{4.7E-03} & \num{5.7E-03} & \num{5.8E-03}
        \end{tabular}
        \caption{Example 2 in \eqref{eq:example1:black_scholes:physical}. Comparison of risk measures across increasing early exercise rights, from monthly to daily exercising. Fortnightly rebalancing ($N=10$).}
        \label{tab:bs:reflection}
    \end{table}

\begin{figure}[t]
        \centering
        
        \begin{subfigure}[t]{\textwidth}
        \centering
        \includegraphics[width=\sizeonebyone\textwidth]{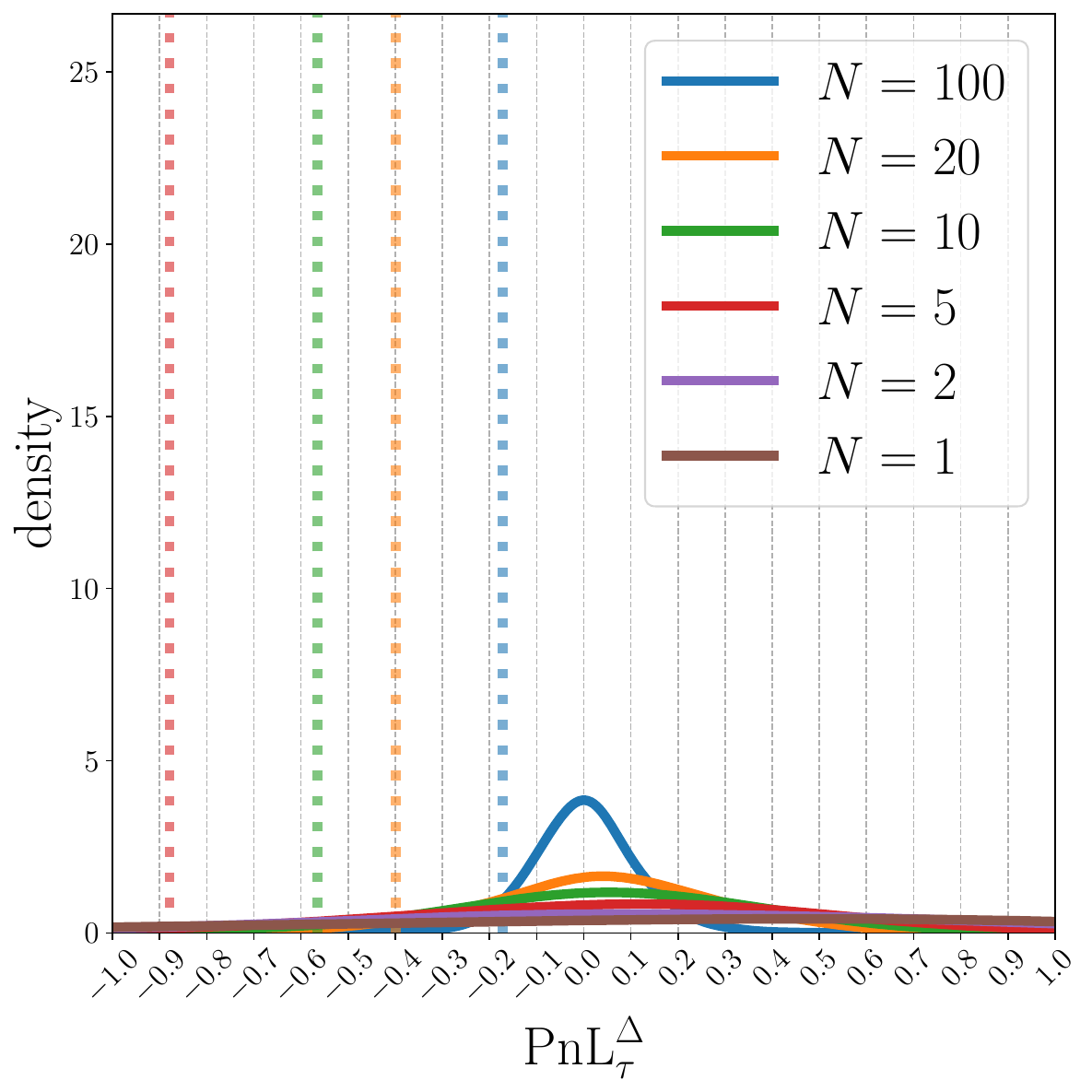}
        \includegraphics[width=\sizeonebyone\textwidth]{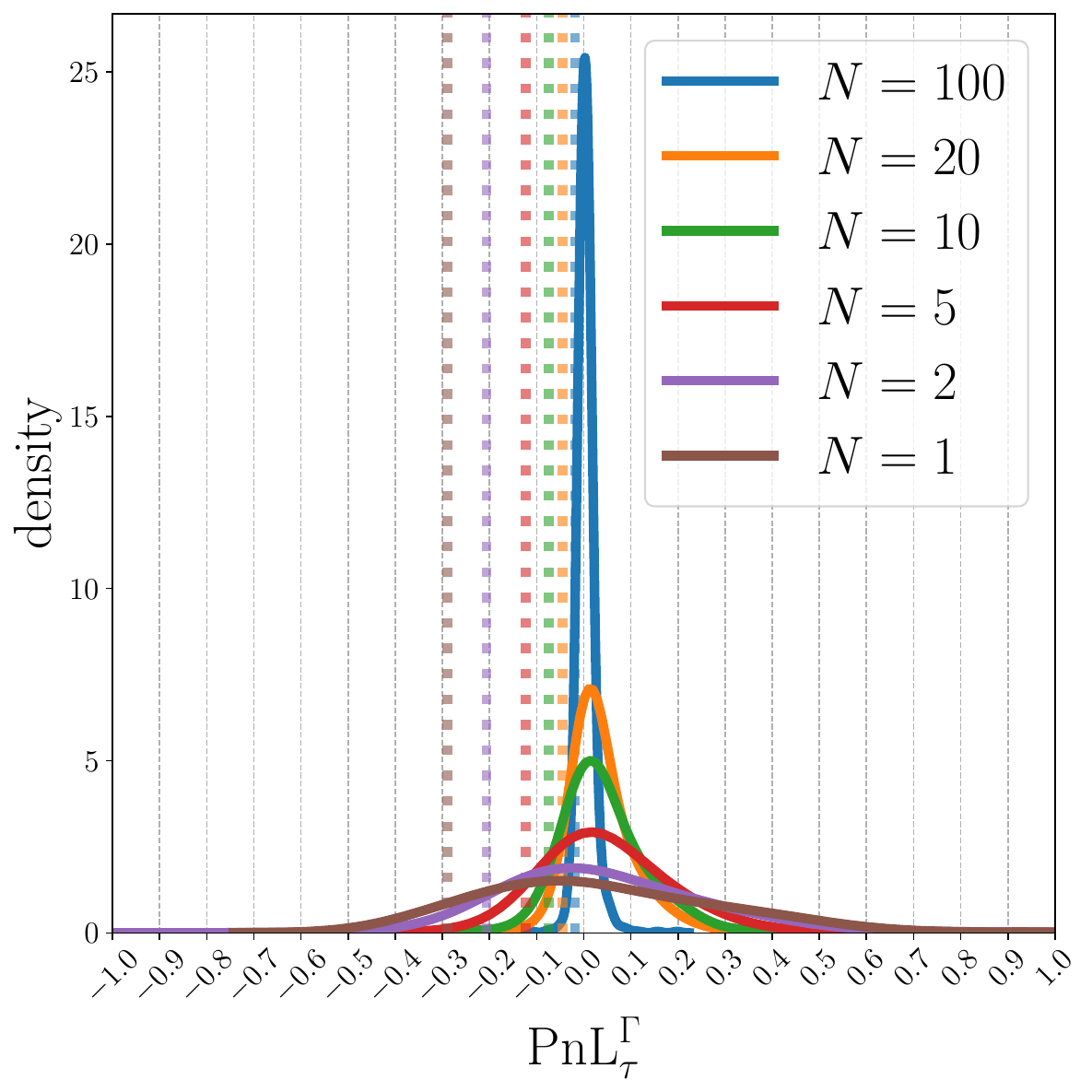}
        \caption{$d=5$}
        \end{subfigure}
        
        \begin{subfigure}[t]{\textwidth}
        \centering
        \includegraphics[width=\sizeonebyone\textwidth]{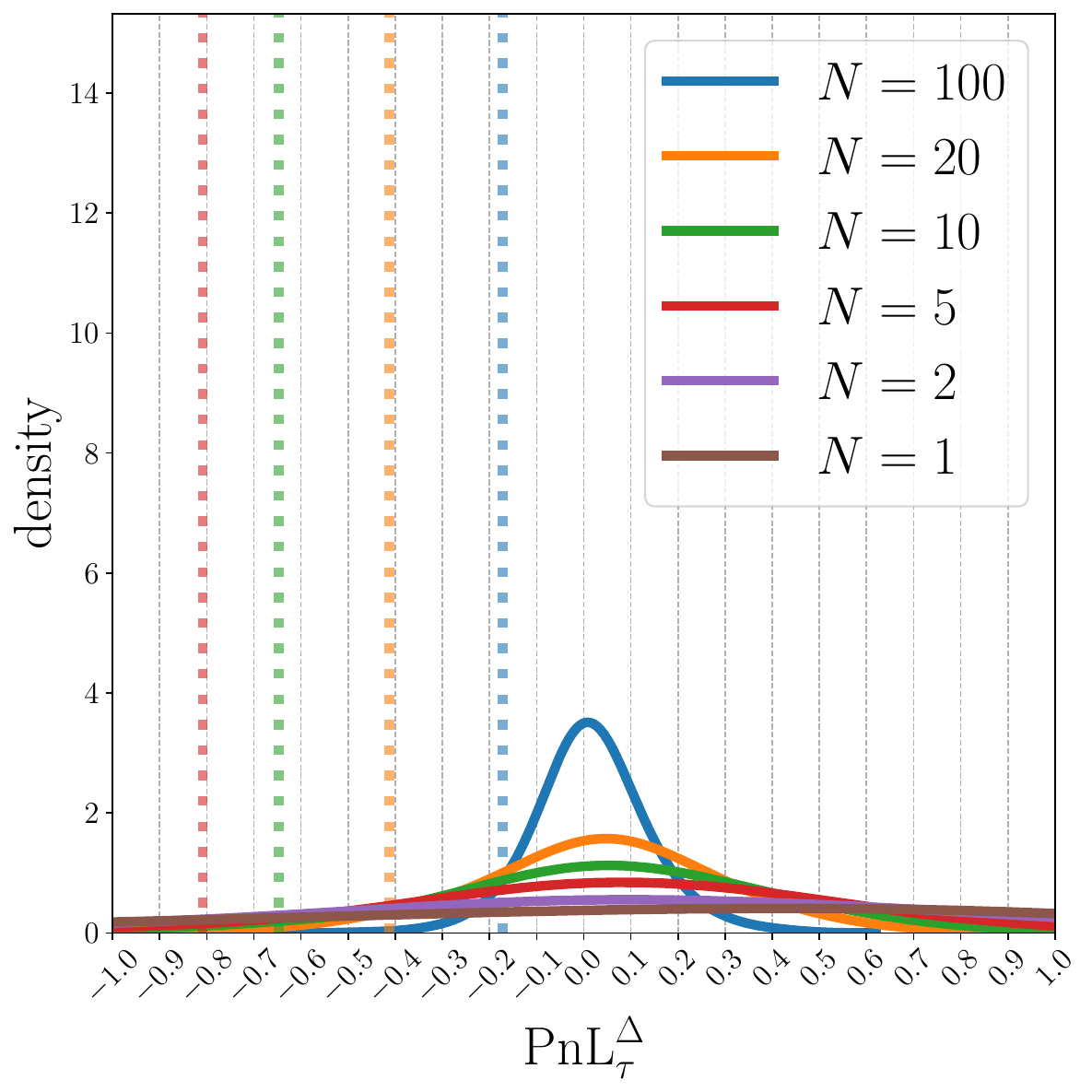}
        \includegraphics[width=\sizeonebyone\textwidth]{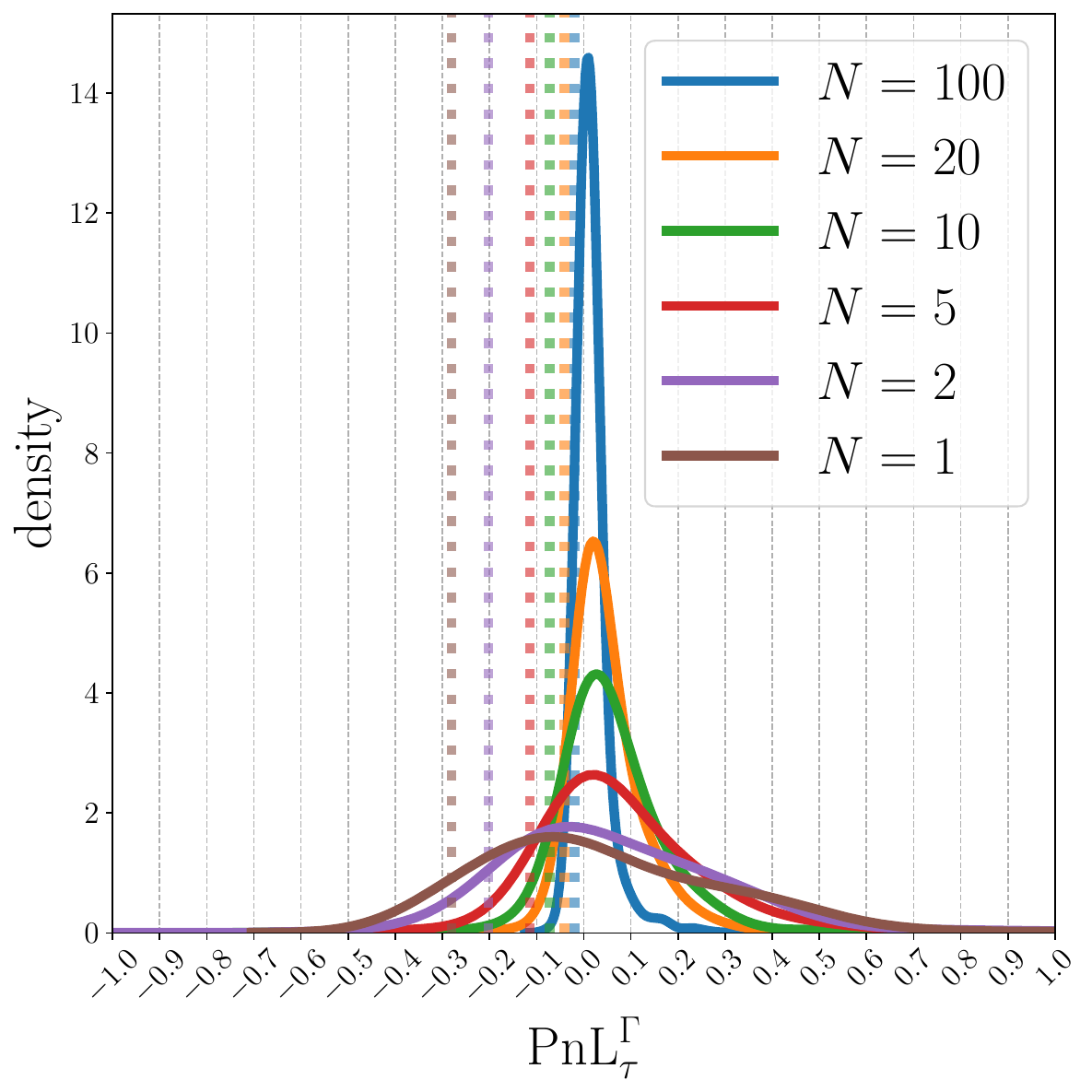}
        \caption{$d=20$}
        \end{subfigure}

        \begin{subfigure}[t]{\textwidth}
        \centering
        \includegraphics[width=\sizeonebyone\textwidth]{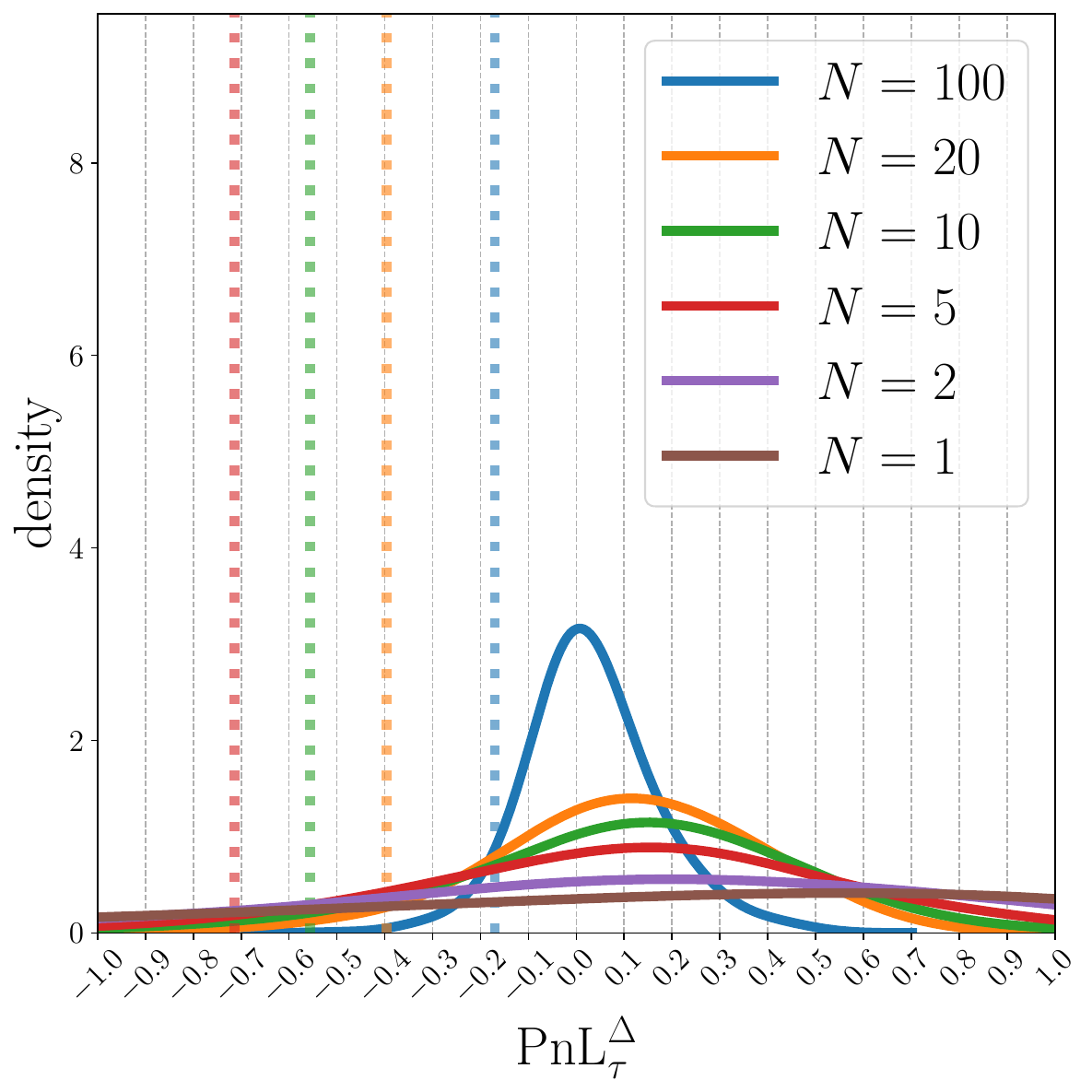}
        \includegraphics[width=\sizeonebyone\textwidth]{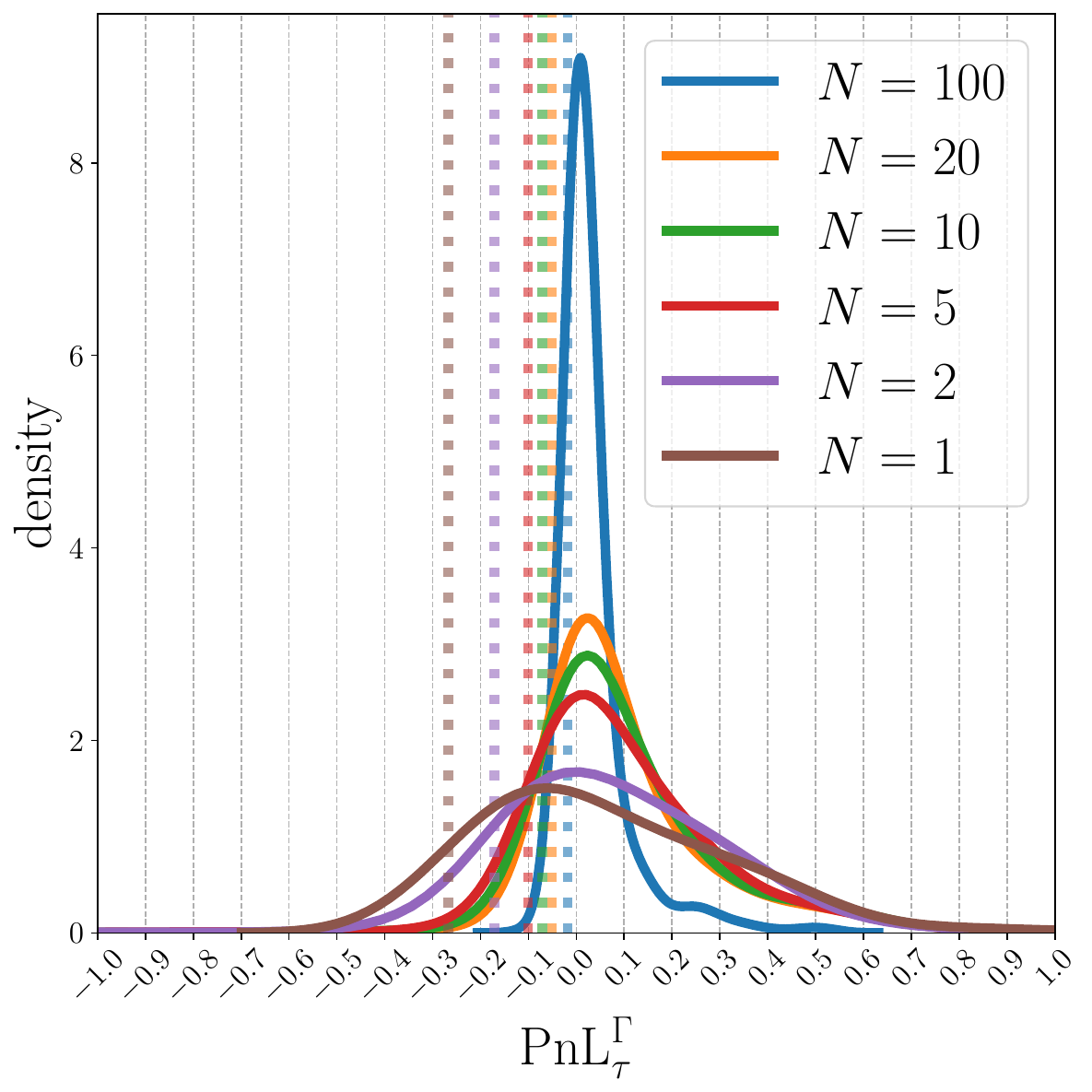}
        \caption{$d=100$}
        \end{subfigure}
        
        \caption{Example 2 in \eqref{eq:example1:black_scholes:physical}. Comparison on dimensionality, increasing number of underlying assets. PnL densities for several rebalancing frequencies. Left: delta hedging \ref{eq:delta_hedging:foc}, right: delta-gamma hedging \eqref{eq:gamma_hedging:foc-soc}. Dotted vertical lines corresponding to $\text{VaR}_{95}$.}
        \label{fig:bs:dim:density}
    \end{figure}

\begin{figure}
        \centering
        \begin{subfigure}[t]{\sizeonebyone\textwidth}
        \centering
        \includegraphics[width=\textwidth]{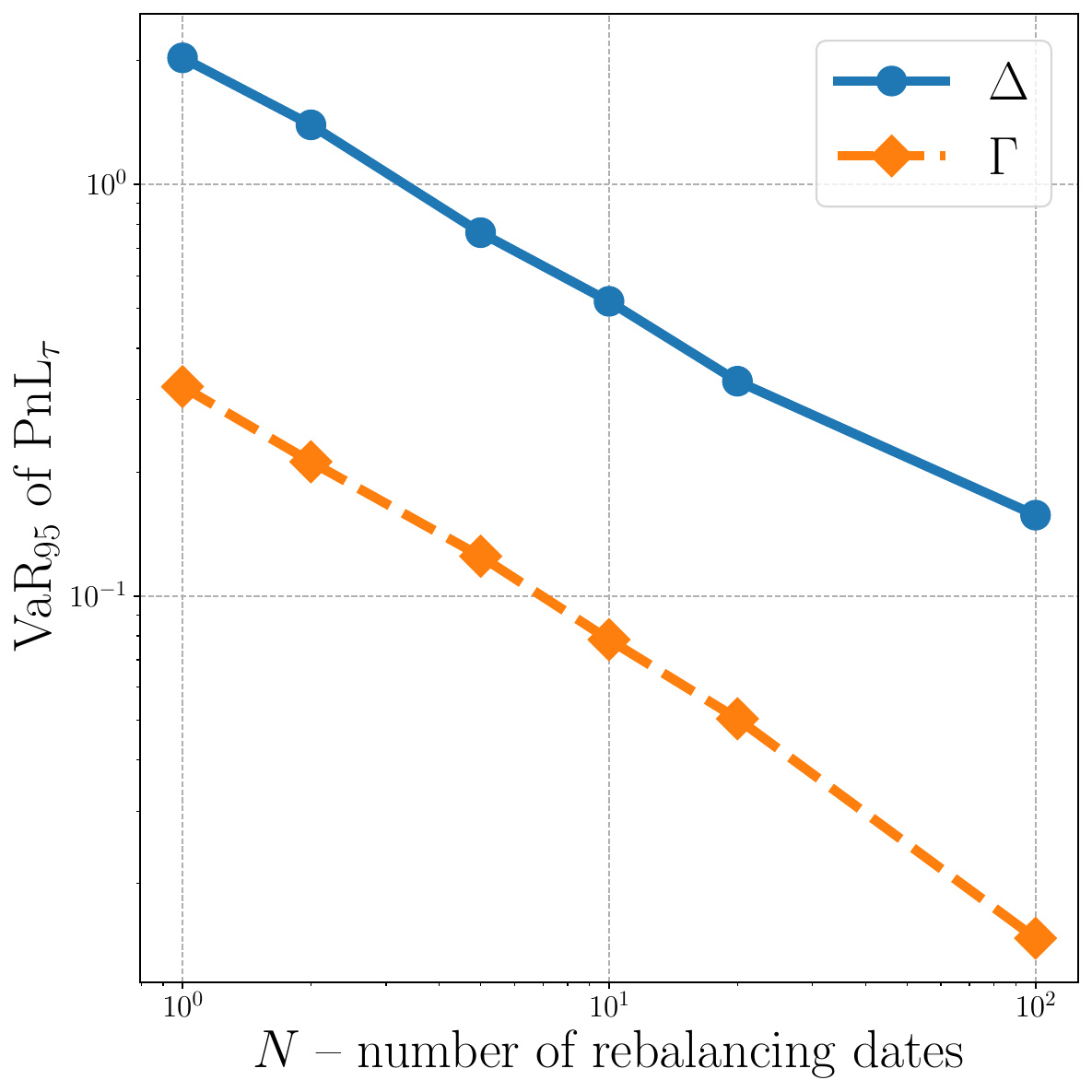}
        \caption{$d=1$}
        \end{subfigure}
        \begin{subfigure}[t]{\sizeonebyone\textwidth}
        \centering
        \includegraphics[width=\textwidth]{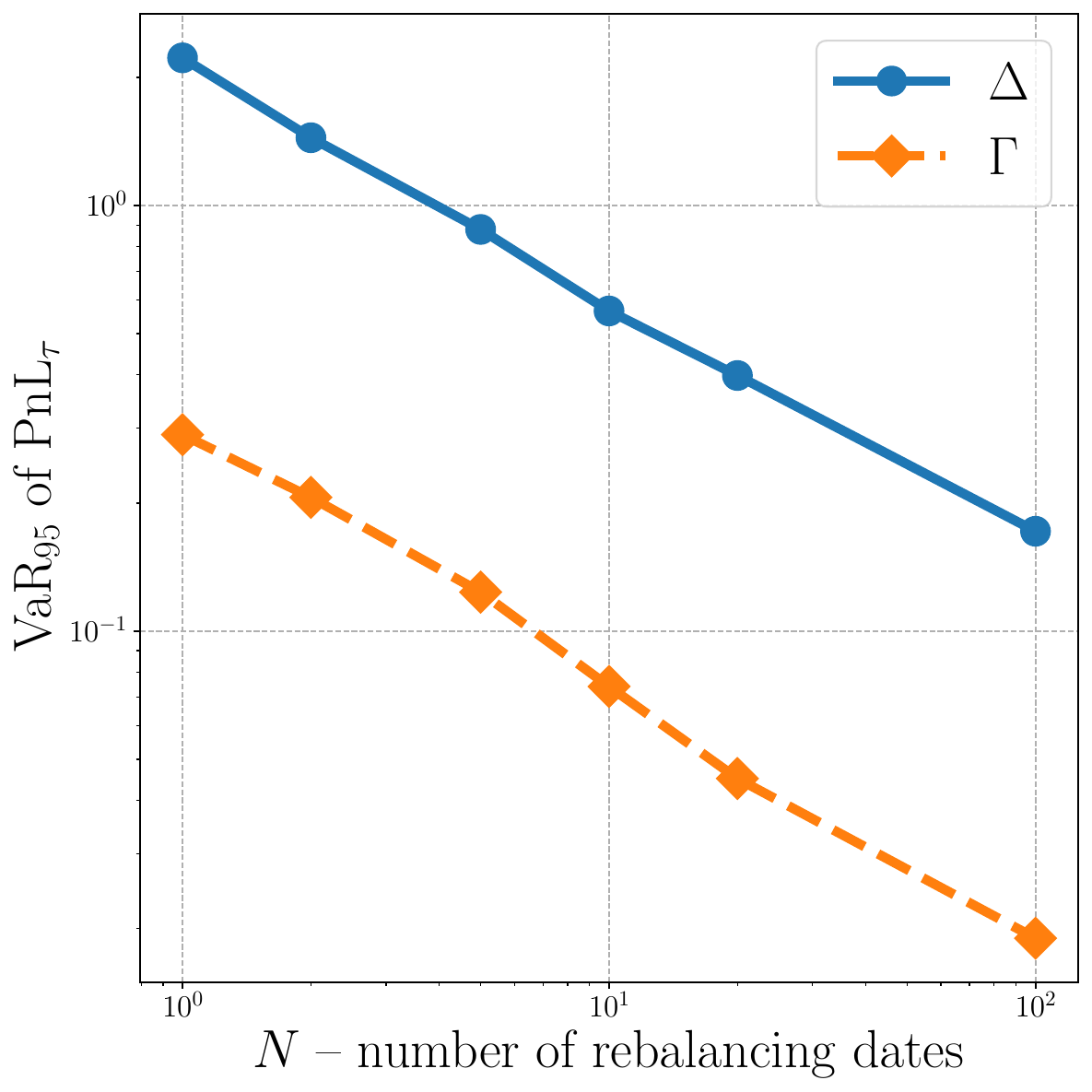}
        \caption{$d=5$}
        \end{subfigure}
        \begin{subfigure}[t]{\sizeonebyone\textwidth}
        \centering
        \includegraphics[width=\textwidth]{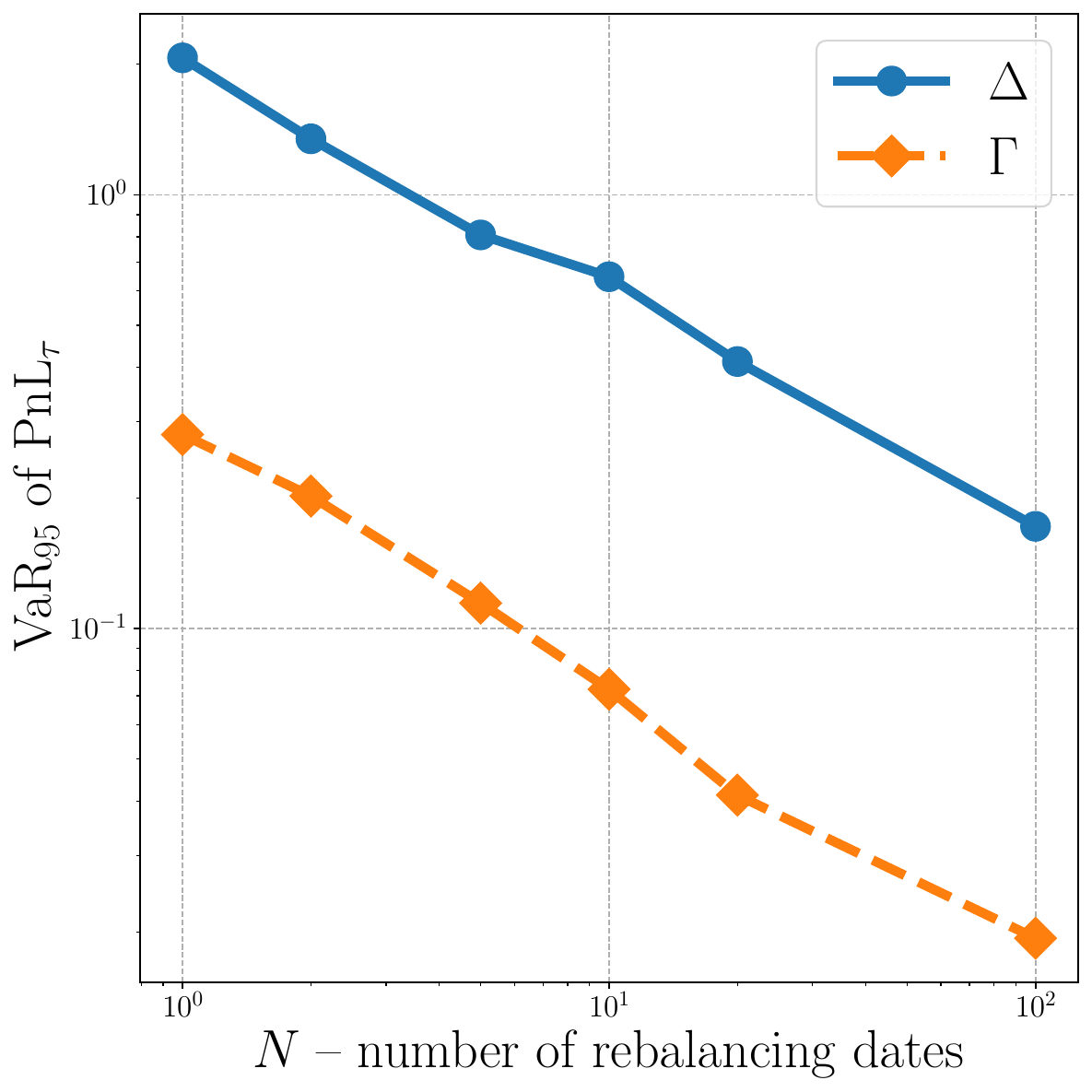}
        \caption{$d=20$}
        \end{subfigure}
        \begin{subfigure}[t]{\sizeonebyone\textwidth}
        \centering
        \includegraphics[width=\textwidth]{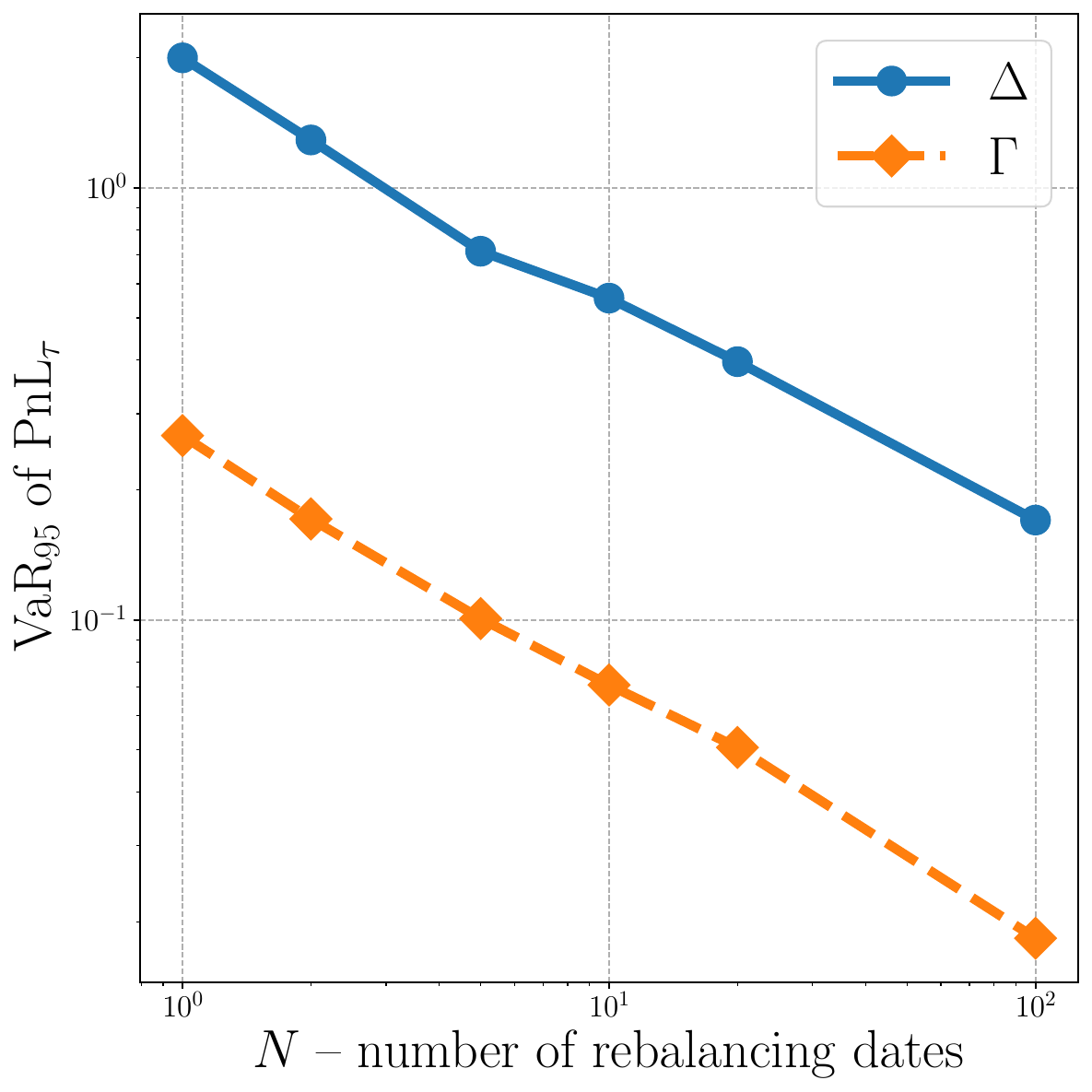}
        \caption{$d=100$}
        \end{subfigure}
        \caption{Example 2 in \eqref{eq:example1:black_scholes:physical}. Comparison on dimensionality, increasing number of underlying assets. Convergence of $\text{VaR}_{95}$ against the discrete number of rebalancing dates in \eqref{eq:delta_hedging:portfolio_value} and \eqref{eq:gamma_hedging:portfolio_value}.}
        \label{fig:bs:convergence}
    \end{figure}

    \paragraph{Dimensionality.} The main motivation behind the deep BSDE approximations for the One Step Malliavin scheme in algorithm \ref{algorithm:osm} is to enable the treatment of high-dimensional basket options, issued on many (correlated) underlyings in \eqref{eq:example1:black_scholes:physical}. We fix $R=100$ corresponding to the American option limit, and vary the number of risk factors $d=m$ in \eqref{eq:example1:black_scholes:physical} between $1, 5, 20, 100$. Recall that these results can directly be compared to all aforementioned results given on $50$ assets. The numerical results are collected in figure \ref{fig:bs:dim:density} and table \ref{tab:bs:dim}. In figure \ref{fig:bs:dim:density}, we see that the OSM enabled hedging strategies are robust with respect to the number of underlying assets, and both in case of delta and delta-gamma hedging the shape of the corresponding profit-and-loss distributions are similar across $d=5, 20, 100$. Observing the vertical scales from top to bottom, we see that both in the case of first- and second-order hedging, the peak of the distribution decreases, as the associated replication problem becomes more difficult due to dimensionality. Comparing the left and right columns indicates that including the additional second-order constraints in \eqref{eq:gamma_hedging:foc-soc} brings a substantial improvement to the discrete replication accuracy. The delta-gamma hedged strategy reaches the same replication error with an order of magnitude less frequent rebalancing, as the mere delta hedging portfolio. The results above illustrate that delta-gamma replication enabled by OSM achieves a given risk tolerance with significantly less number of rebalancing dates compared to delta hedging, irrespective of the number of underlying assets. This is further demonstrated by figure \ref{fig:bs:convergence}, where the convergence of $\text{VaR}_{95}$ is plotted against the number of rebalancing dates. As can be seen, for all considered number of underlying assets, the delta-gamma hedging strategy achieves a risk tolerance level of $\text{VaR}_{95}\leq 10\%$ for all rebalancing frequencies higher than $N=10$. On the contrary, the delta hedged portfolio does not reach this accuracy, not even in case of daily rebalancing $(N=100)$. This implies that the approximation errors in alg. \ref{algorithm:osm} are negligible compared to the time discretization. Moreover, in practical applications, where rebalancing is undesirable due to potential transaction costs, the delta-gamma hedging enabled by the Gamma approximations of OSM may achieve a given risk tolerance more efficiently. We collect all risk measures table \ref{tab:bs:dim} for fortnightly ($N=10$) rebalanced replicating portfolios. For each strategy the risk measures grow as $d$ increases. Nonetheless, both in case of delta and delta-gamma replication, the corresponding risk measures are robust with respect to the number of underlying assets. Across all values of $d$, the Gamma hedging strategies bring an order of magnitude improvement both in the variance and the tail risk measures of the PnL. These numbers demonstrate that the deep BSDE approximation for the OSM scheme efficiently deal with Bermudan basket options in \eqref{eq:example1:black_scholes:physical} issued on a large number of underlyings. We emphasize that the second-order sensitivities in the OSM scheme are given by the matrix-valued $\Gamma$ process in \eqref{eq:scheme:osm:time} which takes values in $\mathbb{R}^{d\times d}$, meaning that in case of $d=m=100$ the OSM scheme accurately approximates $10^4$ gammas simultaneously.

    \begin{table}[t]
        \centering
        \resizebox{\textwidth}{!}{
        \begin{tabular}{l|cccc|cccc}
             & \multicolumn{4}{c}{Delta} & \multicolumn{4}{c}{Delta-Gamma}\\
           $d$  & $1$ & $5$ & $20$ & $100$ & $1$ & $5$ & $20$ & $100$\\
             \hline
                mean & \num{4.1E-03} & \num{1.7E-03} & \num{1.5E-02} & \num{3.0E-02} & \num{1.4E-03} & \num{3.9E-03} & \num{1.7E-02} & \num{3.2E-02}\\
                variance & \num{1.0E-02} & \num{1.2E-02} & \num{1.4E-02} & \num{1.8E-02} & \num{2.6E-04} & \num{3.7E-04} & \num{1.6E-03} & \num{5.3E-03}\\
                $\text{VaR}_{95}$ & \num{-1.6E-01} & \num{-1.7E-01} & \num{-1.7E-01} & \num{-1.7E-01} & \num{-1.5E-02} & \num{-1.9E-02} & \num{-1.9E-02} & \num{-1.8E-02}\\
                $\text{ES}_{95}$ & \num{-2.3E-01} & \num{-2.3E-01} & \num{-2.5E-01} & \num{-2.5E-01} & \num{-2.5E-02} & \num{-3.3E-02} & \num{-2.8E-02} & \num{-3.3E-02}\\
                semivariance & \num{4.7E-03} & \num{5.0E-03} & \num{6.4E-03} & \num{6.2E-03} & \num{9.9E-05} & \num{1.4E-04} & \num{1.7E-04} & \num{2.3E-04}
        \end{tabular}
        }
        \caption{Example 2 in \eqref{eq:example1:black_scholes:physical}. Comparison of risk measures across increasing early exercise rights, from monthly to daily exercising. Fortnightly rebalancing ($N=10$).}
        \label{tab:bs:dim}
    \end{table}

\subsection{Example 3: portfolio of several derivatives with different early exercise rights}
In order to demonstrate the accuracy and robustness of the proposed hedging strategies in the context of a portfolio of multiple options, in our last example we investigate a high-dimensional portfolio of derivatives. We take $d=m=20$ Black-Scholes type underlyings under the physical measure as in \eqref{eq:example1:black_scholes:physical} , with parameters $\bar{\mu}=(0.2, 0.19, \dots, 0.01)$, $\bar{\sigma}=(0.4, 0.25, 0.2, 0.15, 0.1, \dots, 0.4, 0.25, 0.2, 0.15, 0.1)$, $r=0.04$, $q=(0, \dots, 0)$, $X_0=(100, \dots, 100)$ and pairwise correlation $c_{ij}=0.25, i\neq j$. We consider a time horizon of $T=1$ year, over which $J=25$ Bermudan derivatives are held. The types of contracts are collected in table \ref{tab:ex:portfolio}.
\begin{table}[t]
    \centering
    \begin{tabular}{l|c|c}
        & contract & payoff\\
        \hline
      $j=1$   &  geometric put & $\max[K_1-(\prod_{i=1}^m x_i)^{1/m}, 0]$\\
        $j=2$ & arithmetic put on $(S_1, \dots, S_{m/2})$& $\max[K_2 - 2/m\sum_{i=1}^{m/2} x_i, 0]$\\
        $j=3$ & call on max $(S_{m/2+1}, \dots, S_m)$ & $[\max_{i=m/2+1, \dots, m}{x_i} - K_3]^+$\\
        $j=4$ & cash or nothing & $\prod_{i=1}^m \mathds{1}_{50\leq x_i\leq 150}$\\
        $j=5$ & put on min & $[K_5 - \min_{i=m/2+1, \dots, m} x_i]^+$\\
        $j=6, \dots, 25$ & call & $[x_i-K_j]^+$
    \end{tabular}
    \caption{Example 3. Derivatives in the portfolio.}
    \label{tab:ex:portfolio}
\end{table}
In terms of moneyness and early exercise rights, we distinguish three different versions of the portfolio corresponding to
\begin{enumerate}
    \item Case 1: all ATM $(K_j=100, j=1, \dots, J)$, European $(R_j=1, j=1, \dots, J)$ contracts;
    \item Case 2: \begin{itemize}
        \item varying moneyness: $K_1=100, K_2=120, K_3=80, K_5=50, K_j=150, j=6, \dots, 25$,
        \item Bermudan options with uniform, monthly early exercise rights: $R_j=5, j=1, \dots, J$;
    \end{itemize}
    \item Case 3: \begin{itemize}
        \item varying moneyness: $K_1=100, K_2=120, K_3=80, K_5=50, K_j=150, j=6, \dots, 25$,
        \item varying early exercise rights: $R_1=20, R_2=5, R_3=2, R_4=1, R_5=10, R_6=R_7=5, R_8=R_9=10, R_{10}=\dots=R_{15}=100, R_{16}=2, R_{17}=\dots=R_{19}=20, R_{20}=\dots=R_{25}=100$.
    \end{itemize}
\end{enumerate}
For each choice above the solutions of the collection of discretely reflected BSDEs in \eqref{eq:bsde:discretely_reflected:collection} take values in $\mathbb{R}^{J}$, $\mathbb{R}^{J\times d}$ and $\mathbb{R}^{J\times d\times d}$ for the prices, Deltas and Gammas, respectively. Indeed, the Gamma process approximated by the One Step Malliavin scheme in alg. \ref{algorithm:delta_gamma_hedging} has $10^4$ elements. As this portfolio consists of multiple contracts, which may be exercised at different points in time, we provide the profit and loss distributions at maturity $T$, instead of $\tau$ in \eqref{def:tau}. This means that in case one of the options in \eqref{eq:delta_hedging:portfolio_value} or \eqref{eq:gamma_hedging:portfolio_value} is exercised before $T$, the corresponding payoff is collected and the weights are computed with the remaining derivatives only.

\begin{figure}[t]
        \centering
        
        \begin{subfigure}[t]{\textwidth}
        \centering
        \includegraphics[width=\sizeonebyone\textwidth]{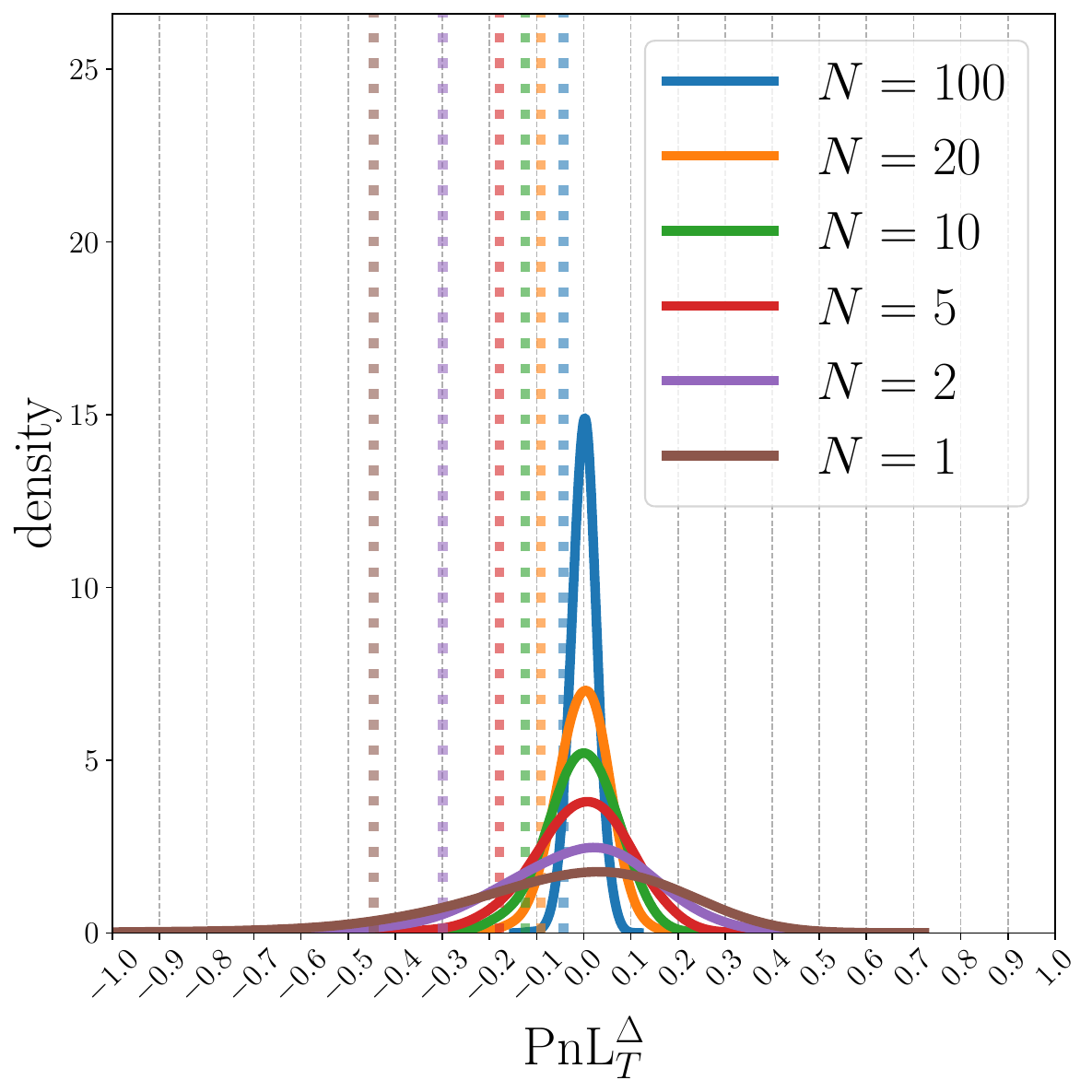}
        \includegraphics[width=\sizeonebyone\textwidth]{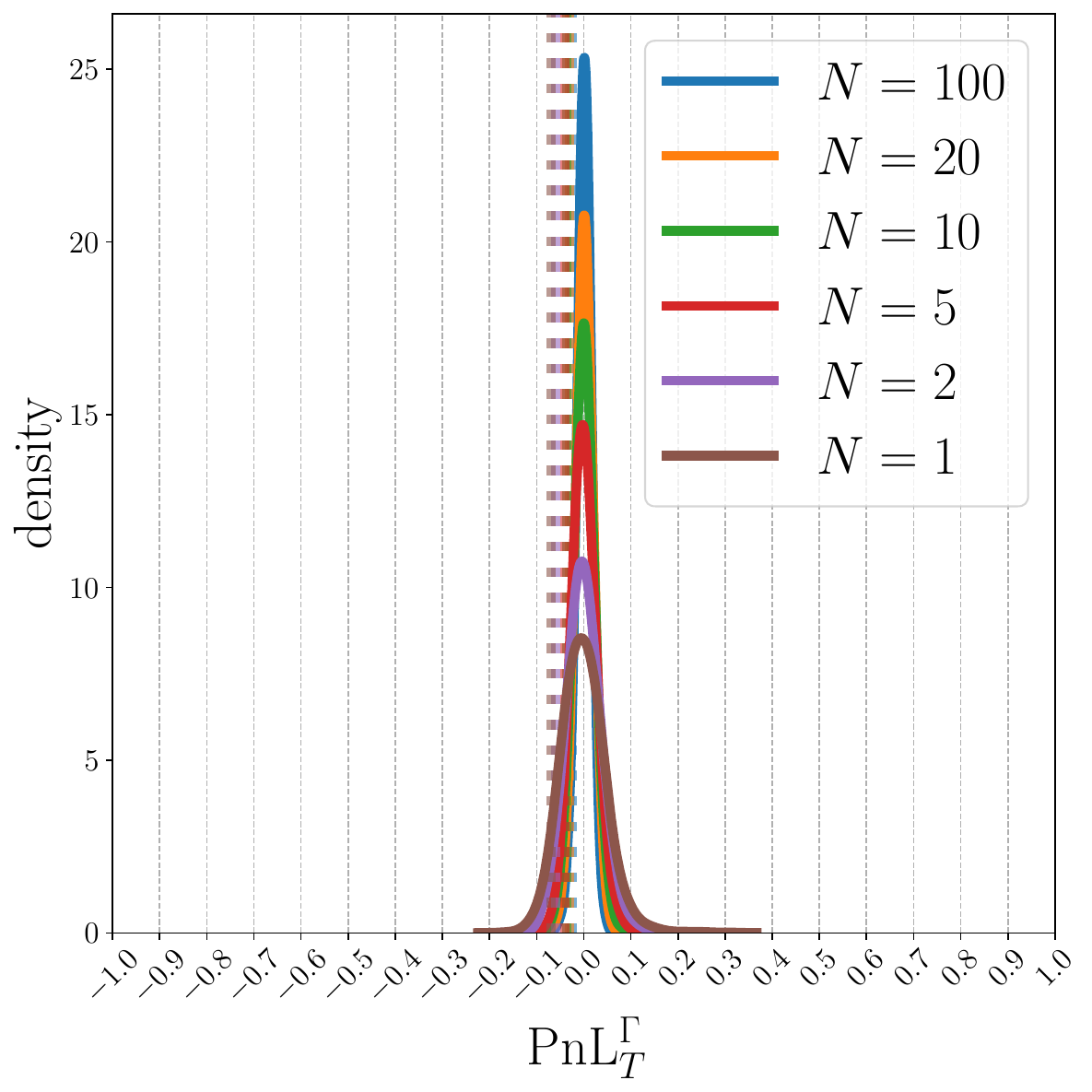}
        \caption{Case 1: uniformly European, ATM}
        \label{fig:portfolio:density:atm}
        \end{subfigure}
        
        \begin{subfigure}[t]{\textwidth}
        \centering
        \includegraphics[width=\sizeonebyone\textwidth]{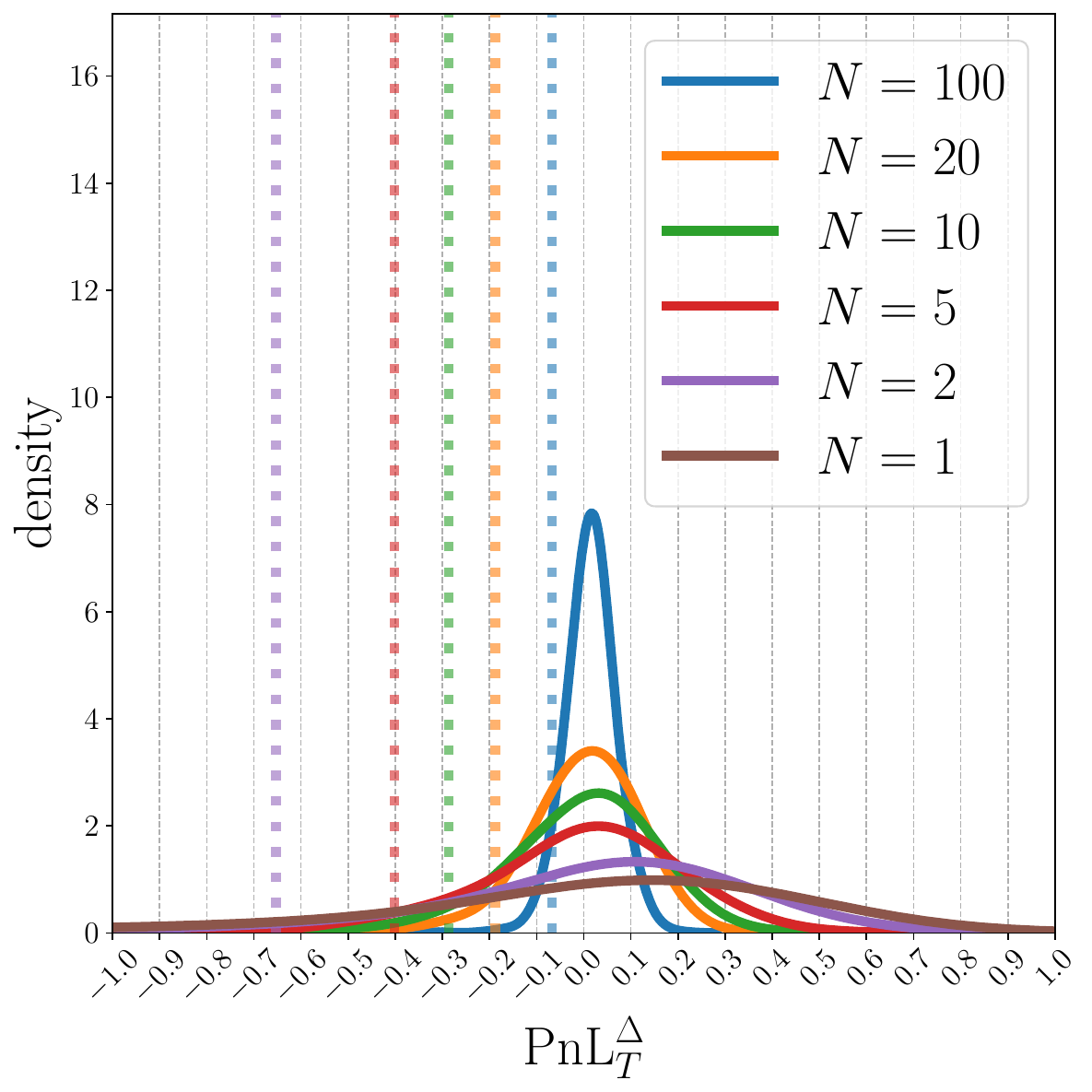}
        \includegraphics[width=\sizeonebyone\textwidth]{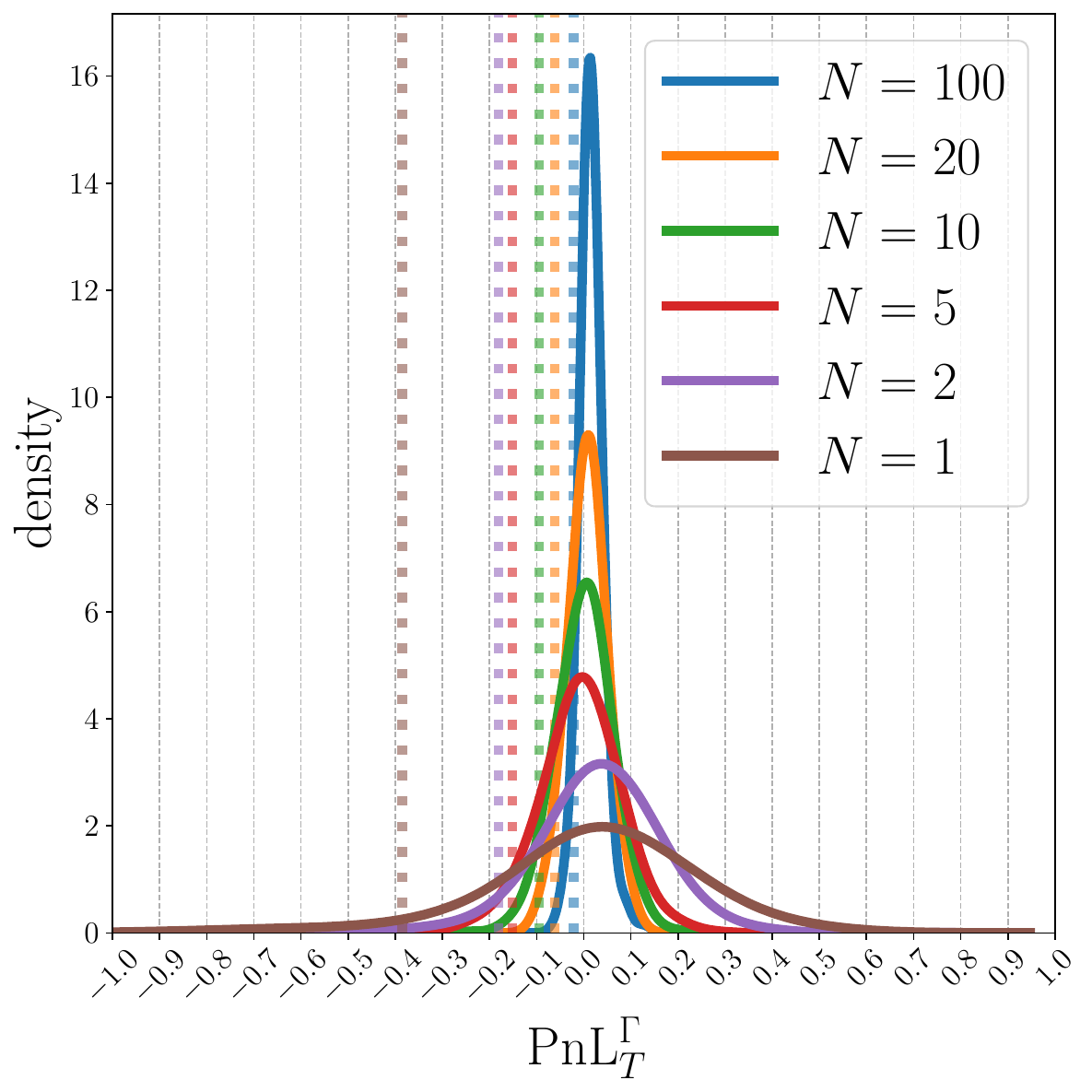}
        \caption{Case 2: uniform monthly $(R=5)$ early exercise rights, mixed moneyness}
        \label{fig:portfolio:density:bermudan}
        \end{subfigure}

        \begin{subfigure}[t]{\textwidth}
        \centering
        \includegraphics[width=\sizeonebyone\textwidth]{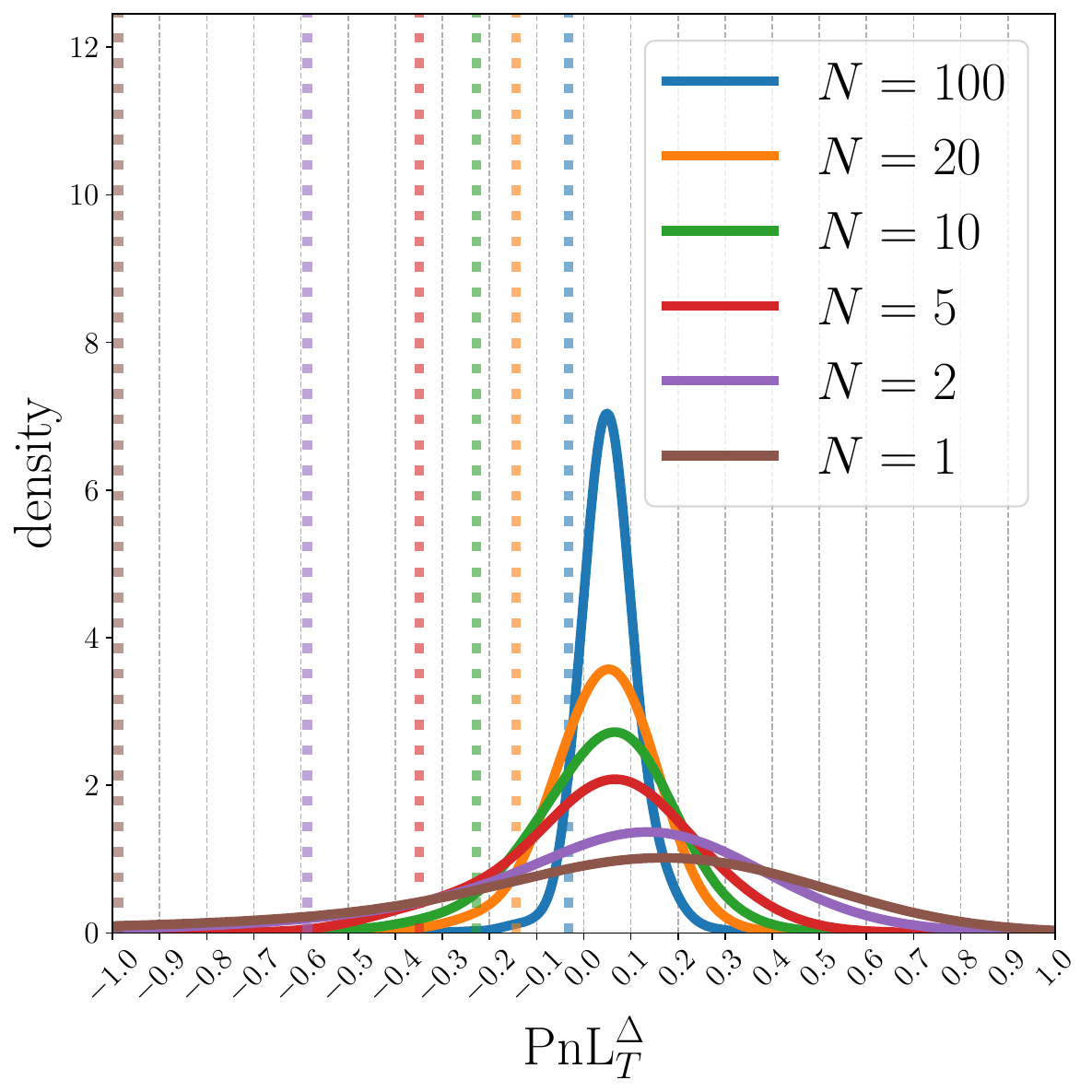}
        \includegraphics[width=\sizeonebyone\textwidth]{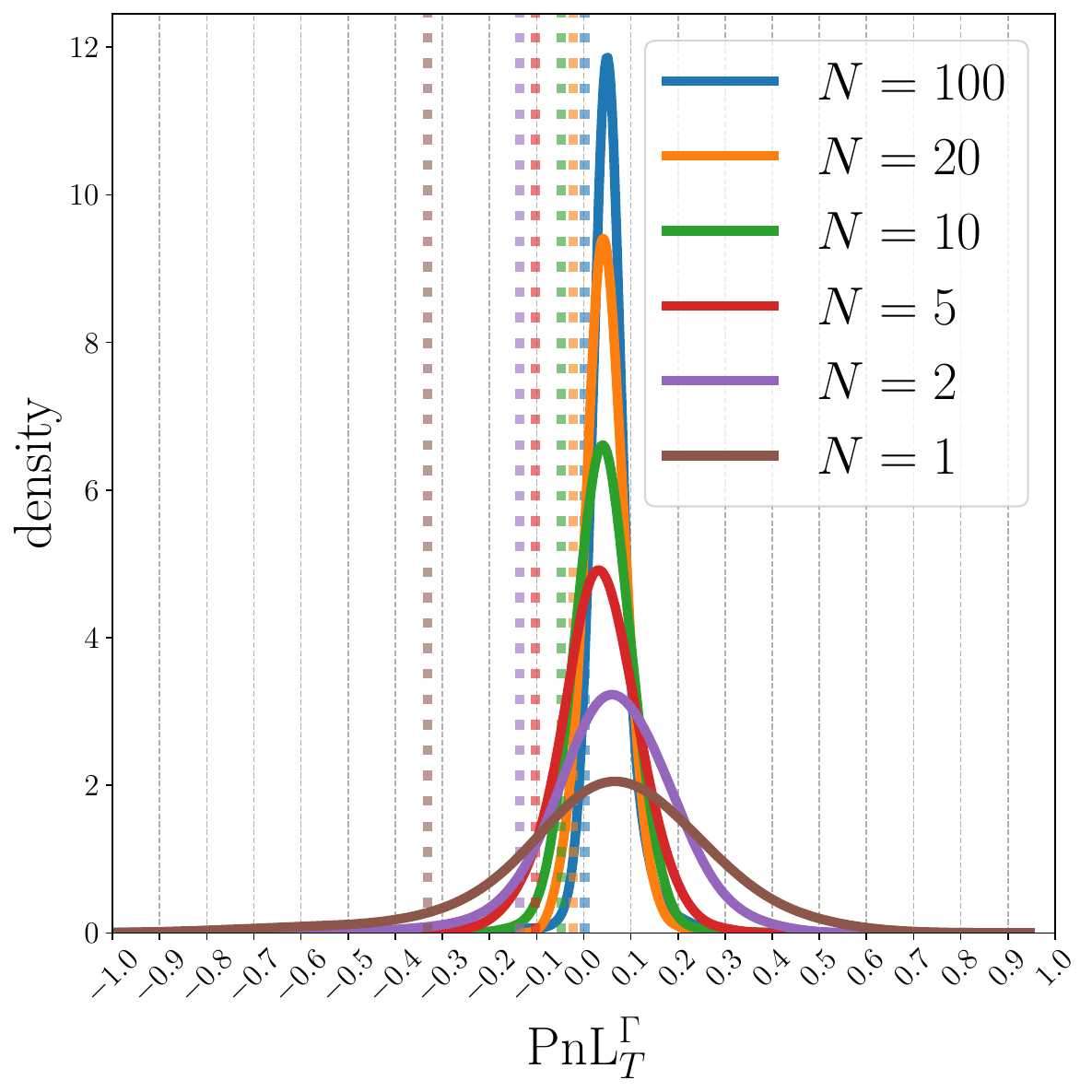}
        \caption{Case 3: varying early exercise rights, mixed moneyness}
        \label{fig:portfolio:density:varying}
        \end{subfigure}
        
        \caption{Example 3. PnL densities for several rebalancing frequencies. Left: delta hedging \eqref{eq:delta_hedging:foc}, right: delta-gamma hedging \eqref{eq:gamma_hedging:foc-soc}. Dotted vertical lines corresponding to $\text{VaR}_{95}$.}
        \label{fig:portfolio:density}
    \end{figure}

\begin{figure}
        \centering
        \begin{subfigure}[t]{\sizeonebyone\textwidth}
        \centering
        \includegraphics[width=\textwidth]{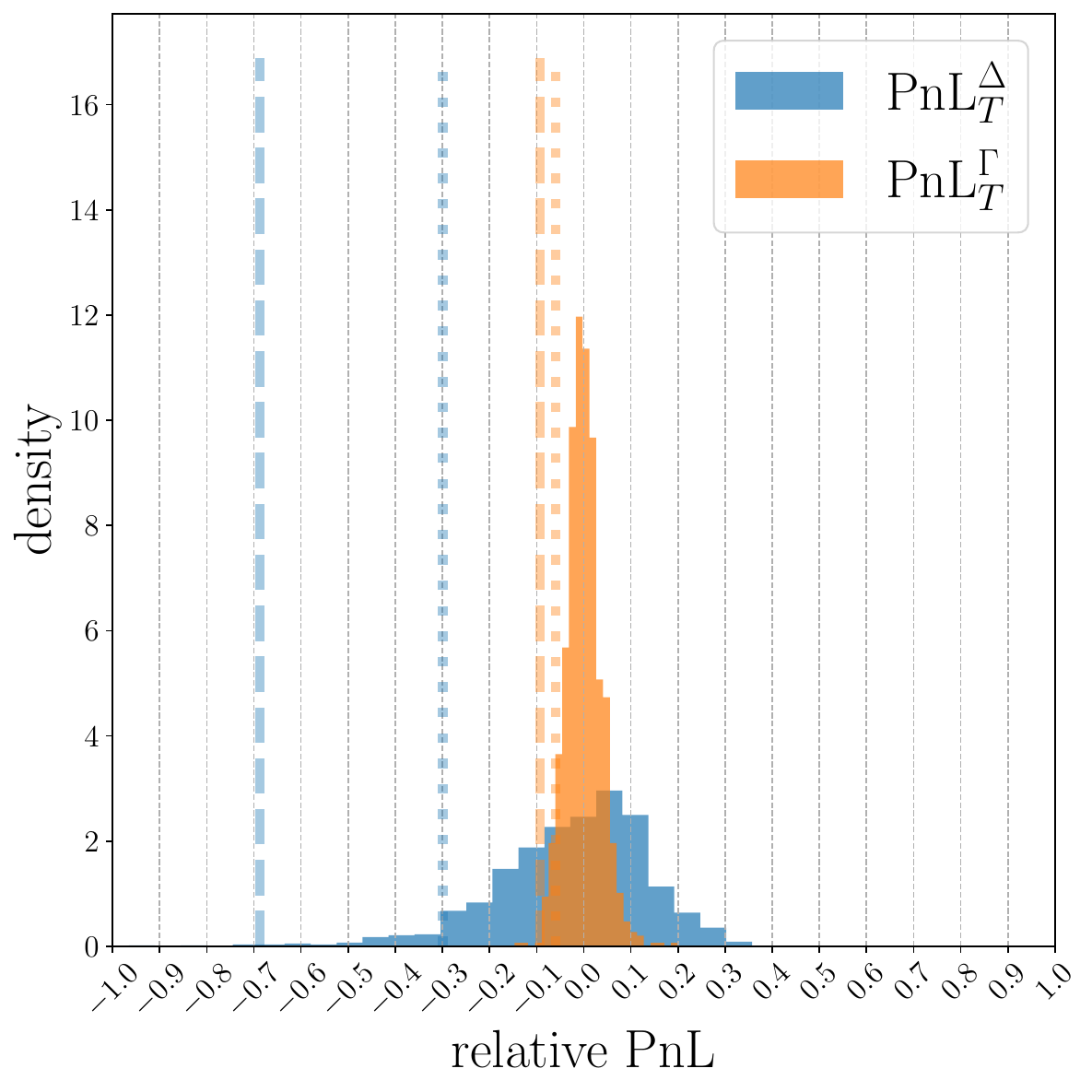}
        \includegraphics[width=\textwidth]{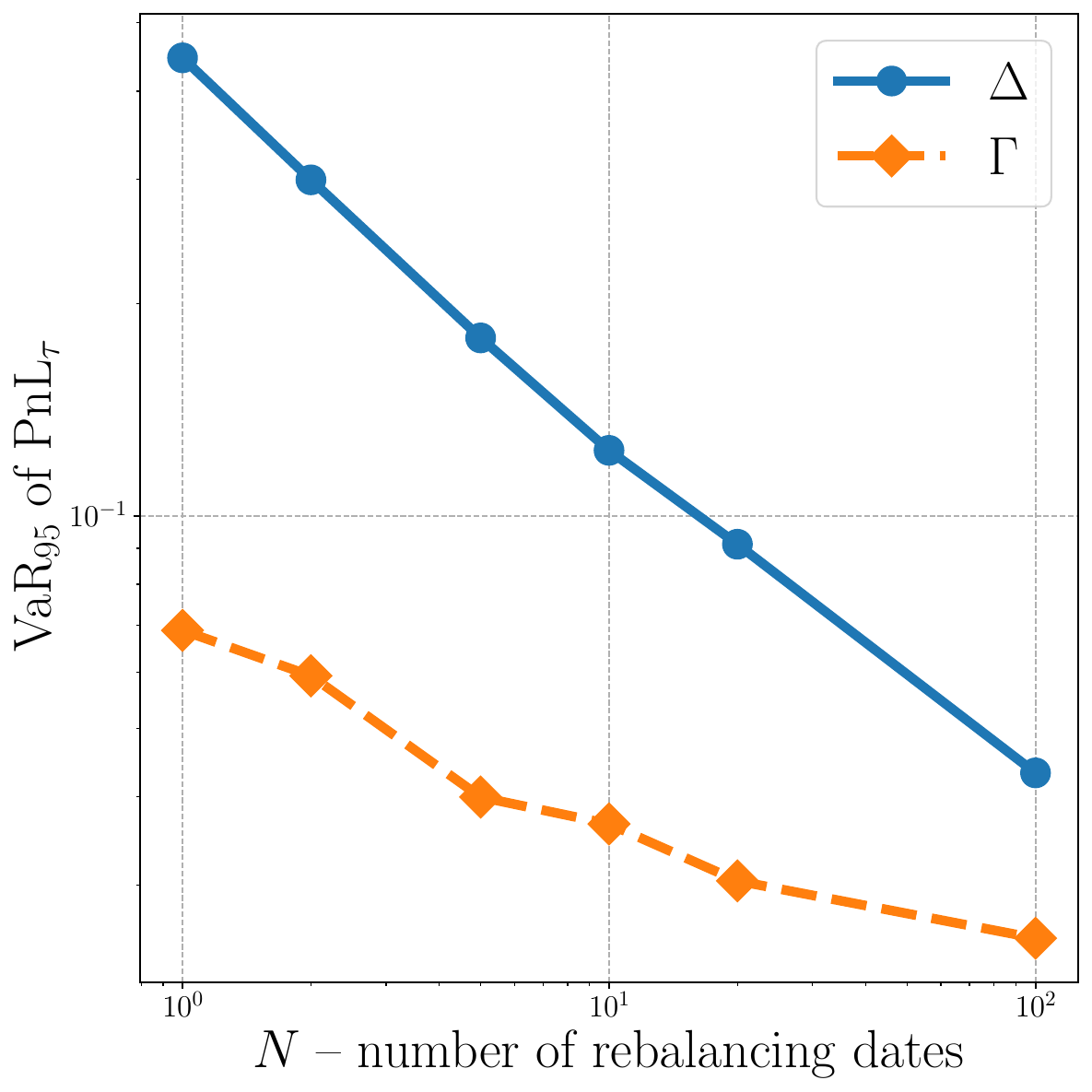}
        \caption{Case 1}
        \end{subfigure}
        \begin{subfigure}[t]{\sizeonebyone\textwidth}
        \centering
        \includegraphics[width=\textwidth]{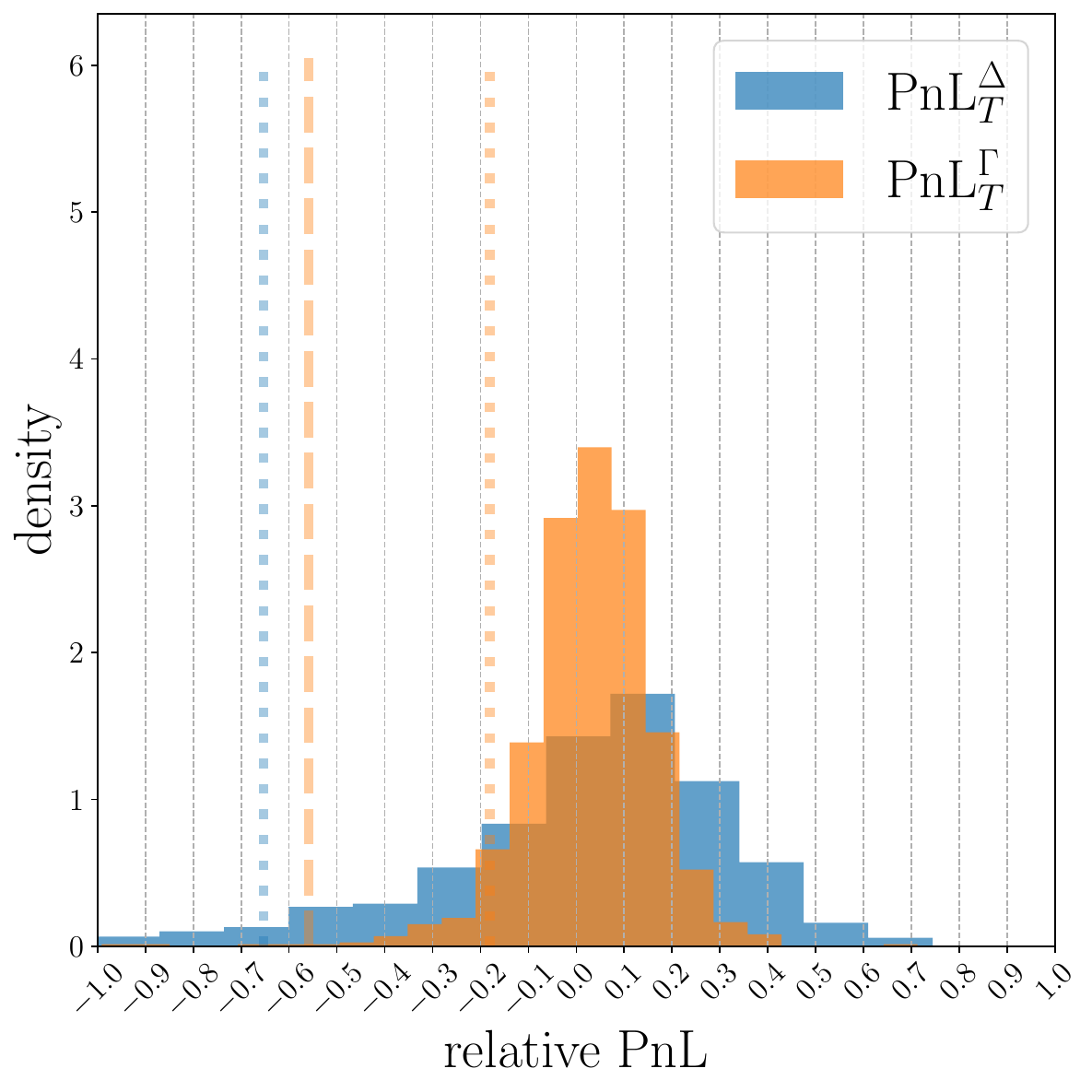}
        \includegraphics[width=\textwidth]{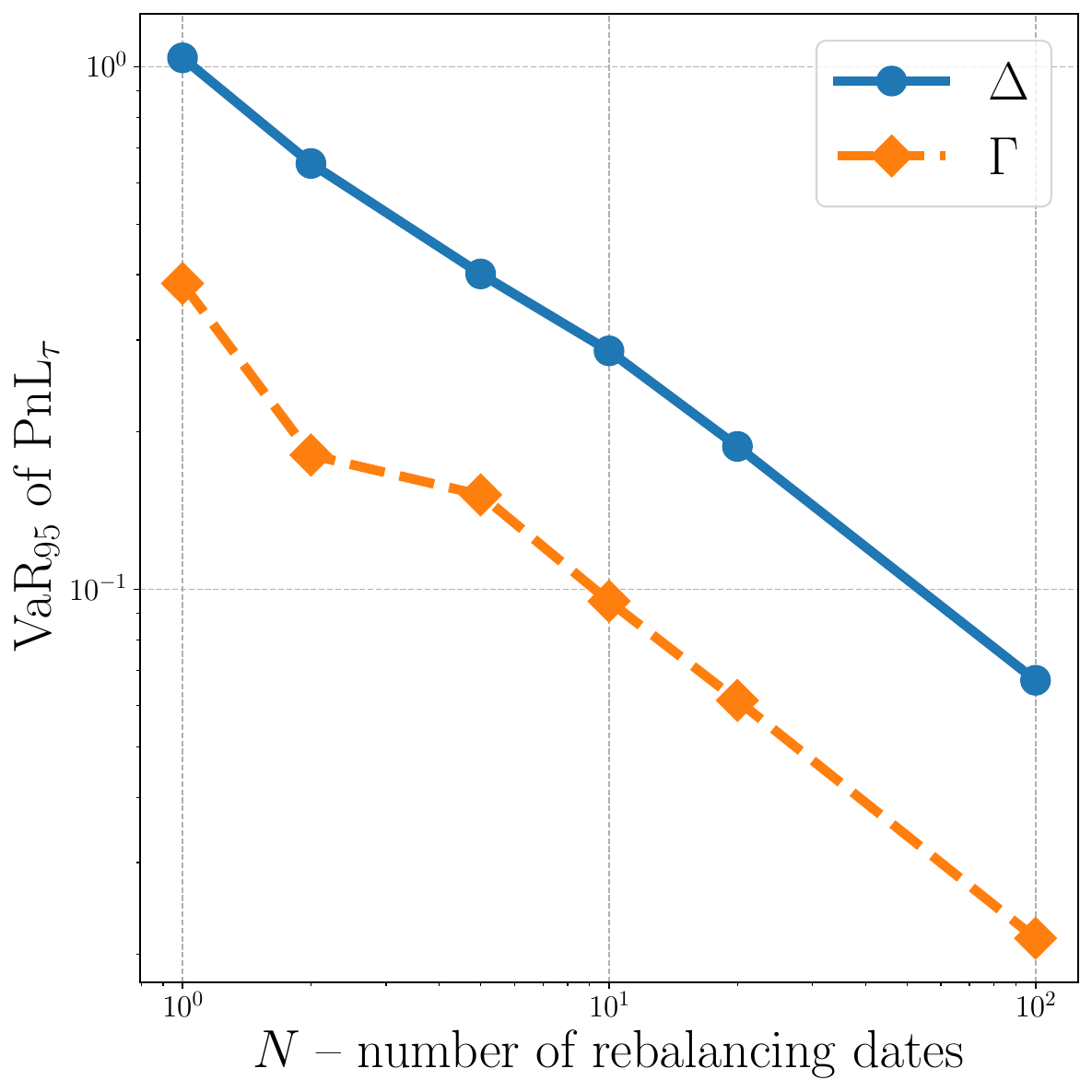}
        \caption{Case 2}
        \end{subfigure}
        \begin{subfigure}[t]{\sizeonebyone\textwidth}
        \centering
        \includegraphics[width=\textwidth]{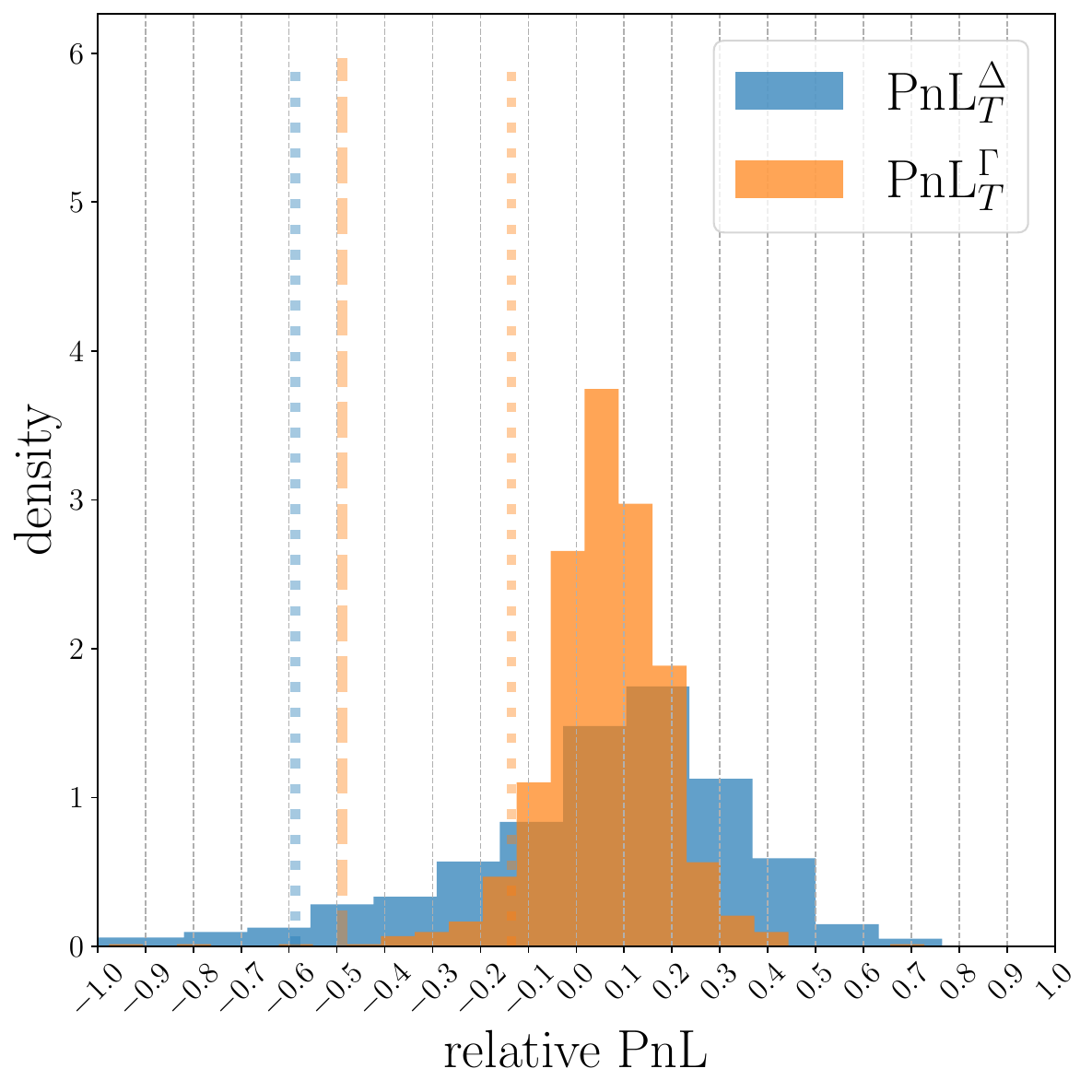}
        \includegraphics[width=\textwidth]{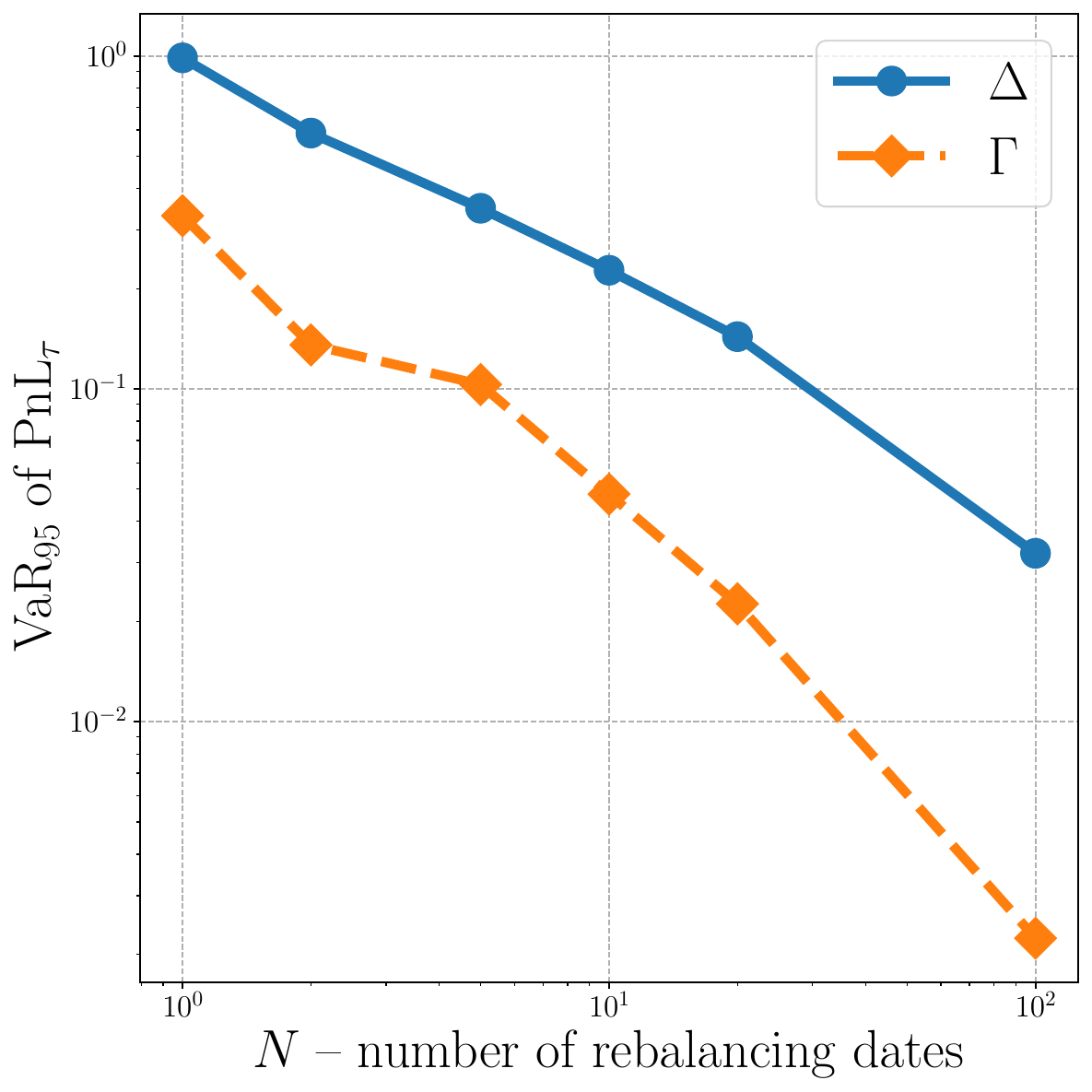}
        \caption{Case 3}
        \end{subfigure}
        
        \caption{Example 3. Top: histograms for delta \eqref{eq:delta_hedging:foc} and delta-gamma replication \eqref{eq:gamma_hedging:foc-soc} with quarterly ($N=2$) rebalancing. Bottom: convergence of $\text{VaR}_{95}$ with respect to the number of rebalancing dates. Dotted and dashed vertical lines corresponding to $\text{VaR}_{95}$ and $\text{ES}_{99}$, respectively.}
        \label{fig:portfolio:histogram}
    \end{figure}

The numerical results are collected in figures \ref{fig:portfolio:density}, \ref{fig:portfolio:histogram} and table \ref{tab:portfolio}. Looking at the approximated PnL densities depicted in fig. \ref{fig:portfolio:density}, we find that the OSM scheme accurately approximates all Deltas and Gammas for all derivatives in the portfolio simultaneously. The replication is most accurate in the case of European contracts without early exercise features, nevertheless, the approximations remain accurate with varying early exercise rights for each contract separately, and also for different levels of moneyness. Comparing the left and right columns, we find that for all three cases outlined above, the gamma hedging strategy yields a substantial improvement in terms of replication accuracy compared to the standard delta hedging. For both first- and second-order hedging, the PnL distributions are distributed around the origin with a decreasing variance as the number of rebalance dates increases. In particular, consistently throughout all examples the gamma hedged strategies reach the same shape of the PnL distribution with fortnightly rebalancing as the delta hedged positions with daily rebalancing. In terms of tail risk, $\text{VaR}_{95}$ is approximately $50$ percentage points better across the delta-gamma hedged portfolios. 

Similar conclusions can be drawn from the histograms collected in the top row of figure \ref{fig:portfolio:histogram}. In line with the earlier results, the delta-gamma hedged OSM portfolios significantly outperform standard delta-hedging even for this portfolio of multiple derivatives. This is accentuated with less frequent rebalancing, where the delta-gamma hedging strategies achieve a much sharper PnL in the discrete time framework. These observations hold across all three cases outlined above, irrespective of the derivatives moneyness and early exercise features. This phenomenon is further demonstrated by the bottom of figure \ref{fig:portfolio:histogram}. Herein, the convergence of the $95$ percentile value-at-risk against the number of rebalancing dates is collected. We find that $\text{VaR}_{95}$ corresponding to the delta-gamma hedged PnL is consistently an order of magnitude smaller than in case of delta hedging, across all different number of rebalancing dates. In other words, any prespecified risk tolerance level -- measured e.g. by value-at-risk -- is achieved by the delta-gamma hedging strategy in an order of magnitude less frequent rebalancing. In particular, we find that the multi-dimensional portfolio extension of the OSM scheme yields practically identical conclusions as in case of single options, implying that the corresponding deep BSDE approximations truly excel in case of these high-dimensional equations.

Finally, risk measures corresponding to each hedging strategy are collected in table \ref{tab:portfolio} for $N=2$, i.e. portfolios rebalanced once every quarter. All risk measures confirm our findings above. In particular, delta-gamma hedging the portfolio of Bermudan derivatives brings an order of magnitude improvement compared to delta hedging in terms of the variance of the profit-and-loss distribution at maturity for all three cases of early exercise features and moneyness. Additionally, we find that offsetting second-order sensitivities in \eqref{eq:gamma_hedging:foc-soc} enabled by the Gamma approximations of the One Step Malliavin scheme, brings a substantial improvement also in terms of tail risk measured both by value-at-risk and expected shortfall. In particular, both for the $95$ and $99$ percentile tails, the delta-gamma hedged distributions provide a factor of $3$ improvement compared to the standard delta strategies. In light of the dimensionality considered in this last example, with a portfolio consisting of $J=25$ derivative contracts issued on $d=m=20$ underlyings with varying drift and volatility parameters, we can conclude that the deep BSDE approximations of the OSM scheme provide a robust and accurate approximations for all option Greeks in the portfolio up to second order.

    \begin{table}[t]
        \centering
        \begin{tabular}{l|ccc|ccc}
             & \multicolumn{3}{c}{Delta} & \multicolumn{3}{c}{Delta-Gamma}\\
             & ATM & Bermudan & Mixture & ATM & Bermudan & Mixture\\
             \hline
            mean & \num{-2.6E-02} & \num{-2.2E-03} & \num{3.0E-02} & \num{-2.9E-04} & \num{2.9E-02} & \num{6.0E-02}\\
            variance & \num{2.7E-02} & \num{1.2E-01} & \num{1.1E-01} & \num{1.3E-03} & \num{1.8E-02} & \num{1.7E-02}\\
            $\text{VaR}_{95}$ & \num{-3.0E-01} & \num{-6.5E-01} & \num{-5.9E-01} & \num{-5.9E-02} & \num{-1.8E-01} & \num{-1.4E-01}\\
            $\text{ES}_{95}$ & \num{-4.5E-01} & \num{-1.0E+00} & \num{-9.3E-01} & \num{-7.2E-02} & \num{-3.1E-01} & \num{-2.6E-01}\\
            semivariance & \num{1.7E-02} & \num{1.1E-01} & \num{1.0E-01} & \num{4.5E-04} & \num{1.2E-02} & \num{1.0E-02}
        \end{tabular}
        \caption{Example 3. Convergence of $\text{VaR}_{95}$ against the discrete number of rebalancing dates in \eqref{eq:delta_hedging:portfolio_value} and \eqref{eq:gamma_hedging:portfolio_value}.}
        \label{tab:portfolio}
    \end{table}

\section{Conclusion}
In this paper we proposed a novel deep BSDE based approach for the simultaneous pricing, delta and delta-gamma hedging of large, high-dimensional portfolios of Bermudan options. First, we gave a vector-valued extension to the One Step Malliavin scheme in \cite{negyesi_one_2024, negyesi_reflected_2024}. This way, we casted the pricing, delta- and delta-gamma hedging of a portfolio of Bermudan options into the framework of a \emph{system of discretely reflected BSDEs}. Subsequently, we proposed a deep BSDE approach for the accurate numerical solution of this collection of equations, which is robust and efficient when the number of underlying risk factors and/or options is large. In fact, our approach includes not only prices and Deltas but also second-order Greeks, Gammas of all options in the portfolio, simultaneously approximated, throghout the entire spacetime. We demonstrated the performance of our algorithm on several examples, highlighting key features of our technique. Our findings suggest that, the hereby proposed OSM approximations outperform reference methods \cite{hure_deep_2020, chen_deep_2021} even in the context of delta hedging, when the risk factors are highly volatile. Most importantly, by performing delta-gamma hedging enabled by the Gamma approximations of the OSM scheme, we managed to improve on the discrete replication accuracy compared to standard delta hedging. Our results demonstrate that the algorithm is robust and accurate, for different levels of moneyness, high volatility, and early exercise rights up to the American option limit even in case high dimensional basket options issued on $d=100$ assets.
	
\paragraph{Acknowledgments} B.N. acknowledges financial support from the Peter Paul Peterich Foundation via the TU Delft University Fund.
	
\begin{appendix}
\section{Beyond delta-gamma hedging}\label{sec:beyond_delta_gamma}

In principle, our framework is more general than mere delta-gamma hedging, as it allows for the treatment of hedging uncertain volatility. In that case, one needs to complete the market with an extra set of securities that have non-vanishing \emph{Vegas},  such that volatility risk can be hedged, in a similar fashion to \eqref{eq:gamma_hedging:portfolio_value} with first-order constraints only. Hence, an appropriate delta-vega hedging strategy on the portfolio \eqref{eq:gamma_hedging:portfolio_value} would therefore yield the following first-order conditions
\begin{subequations}\label{eq:delta_vega_hedging:foc}\noeqref{eq:delta_vega_hedging:foc:vega, eq:delta_vega_hedging:foc:delta}
    \begin{align}
    \sum_{k=1}^K\beta_t^k \partial_l u^k(t, X_t) &= \sum_{j=1}^J \partial_l v^j(t, X_t),&& l=d-m+1, \dots, d,\label{eq:delta_vega_hedging:foc:vega}\\
    \alpha_t^{i} &= \sum_{j=1}^J \partial_i v^j(t, X_t) - \sum_{k=1}^K \beta_t^k \partial_i u^k(t, X_t), &&i=1, \dots, m.\label{eq:delta_vega_hedging:foc:delta}
\end{align}
\end{subequations}

Similar to the previous discussion, \eqref{eq:delta_vega_hedging:foc} only offsets first-order sensitivities of the portfolio with respect to asset prices and uncertain volatility. In order to mitigate replication errors stemming from the discrete rebalancing, one has the conditions determining the optimal hedging weights by appropriate second-order constraints including second-order sensitivities corresponding to not just asset prices (Gammas) but also to uncertain volatility such as \emph{Vommas} and \emph{Vannas}. Therefore, in case such second-order sensitivities are also taken into account in the presence of non-hedgeable risk factors such as stochastic volatility, one can impose the following extra second-order conditions
\begin{align}\label{eq:delta_vega_hedging:soc}
    \sum_{k=1}^K \beta^k_t \partial_{il}^2 u^k(t, X_t) = \sum_{j=1}^J \partial_{il}^2 v^j(t, X_t),\quad il\in\mathcal{I},
\end{align}
with some arbitrarily chosen index set $\mathcal{I}$. Note that in this case \eqref{eq:delta_vega_hedging:soc} and \eqref{eq:delta_vega_hedging:foc:vega} together form a linear system of size $(d-m+|\mathcal{I}|)\times K$. Natural choices for the index set $\mathcal{I}$ include
\begin{itemize}
    \item $\mathcal{I}=\{il: 1\leq i=l\leq m\}$ -- diagonal elements of the $\Gamma$ matrix are hedged, corresponding to second order sensitivities of the tradeable assets;
    \item $\mathcal{I}=\{il: 1\leq i, l\leq m\}$ -- the whole $\Gamma$ matrix is hedged, with cross-gammas included;
    \item $\mathcal{I}=\{il: m+1\leq i, l\leq d\}$ -- second-order sensitivities with respect to the uncertain volatility, i.e. \emph{Vomma}s, including cross-vommas,
    \item $\mathcal{I}=\{il: 1\leq i, l\leq d\}$ -- all second-order Greeks with respect to each underlying risk factor, e.g. including mixed partial derivatives in asset prices and volatilities \emph{Vanna}s.
\end{itemize}

We emphasize that the focus above on volatility risk is mere illustration. In fact, the abstract FBSDE framework allows for the treatment of second-order Greeks of all types. For instance, one could consider stochastic interest rates and compute the Vera/rhova as the mixed partial derivative with respect to volatility and interest rate; as long as the corresponding risk factor forms part of the It\^{o} diffusion in \eqref{eq:introduction:sde}. As is made clear by the discrete backward recursion in \eqref{eq:bsde:euler}, the only type of second-order Greeks where one needs to take additional measures is derivatives including time, e.g. \emph{Charm}. Nonetheless, provided that the time partition used for the discrete time resolution of the discretely reflected FBSDE system associated with the portfolio is fine enough, one can approximate such Greeks by finite difference type approximations.\footnote{e.g. $\text{Charm}_n^\pi = -\frac{\alpha_n^\pi - \alpha_{n-1}^\pi}{t_n-t_{n-1}}$} Therefore, the framework built on the One Step Malliavin scheme allows for the hedging of all Greeks up to second order, as long as the corresponding risk factors are incorporated in associated discretely reflected FBSDE.
\section{Margrabe formula with dividends}\label{sec:appendix:margrabe}
	We have the standard Black-Scholes model \eqref{eq:example1:black_scholes:physical} in which all assets follow a geometric Brownian motion
	\begin{align}
		\mathrm{d}S_t^i/S_t^i = (r - q^i)\mathrm{d}t + \sigma^i \mathrm{d}W_t^i,\quad 1\leq i\leq d=\prod_{j=1}^J d_j.
	\end{align}
	The Brownian motions are pairwise correlated with a correlation parameter $\varrho^{ij}$.
	The Margrabe formula \cite{margrabe_value_1978} then gives an explicit, closed-form analytical expression for the price of a European type contract whose payoff is as follows
	\begin{align}
		g(t, S) = [S_T^k - K^{kj}S_T^j]^+,
	\end{align}
	at some terminal time $T$, where the parameter $K^{kj}$ is an exchange strike.
	The price of such a contract then satisfies
	\begin{align}
		C^{kj}(t, S) = e^{-q^k(T-t)}S_t^k\Phi(d_1^{kj}(t, S)) - e^{-q^j(T-t)}K^{kj}S_t^j\Phi(d_2^{kj}(t, S)),
	\end{align}
	where $\Phi$ denotes the standard normal cumulative distribution function and
	\begin{align}
		d_1^{kj}(t, S) &= \frac{1}{\sigma^{kj}\sqrt{T-t}}\left(\ln(\frac{S_t^k}{K^{kj}S_t^j}) + (q_j - q_k + (\sigma^{kj})^2/2)(T-t)\right),\\
		d_2^{kj}(t, S) &= d_1^{kj}(t, S) - \sigma^{kj}\sqrt{T-t},\\
		\sigma^{kj} &= \sqrt{(\sigma^k)^2 + (\sigma^j)^2 -2\varrho^{kj}\sigma^k\sigma^j}.
	\end{align}
	Straightforward computations lead to the first-order derivatives satisfying
	\begin{align}\label{eq:exchange:delta}
		\partial_k C^{kj}(t, S) &= e^{-q^k(T-t)}\Phi(d_1^{kj}(t, S)),\quad 
		\partial_j C^{kj}(t, S) = -e^{-q^j(T-t)}K^{kj}\Phi(d_2^{kj}(t, S)).
	\end{align}
	Further differentiation yields the second-order derivatives
	\begin{align}\label{eq:exchange:gamma}
		\partial_{kk}^2 C^{kj}(t, S) &= \frac{e^{-q^k(T-t)}}{\sigma^{kj}\sqrt{T-t}}\frac{\phi(d_1^{kj}(t, S))}{S_t^k},\quad &&\partial_{jk}^2 C^{kj}(t, S) = -\frac{e^{-q_k(T-t)}}{\sigma^{kj}\sqrt{T-t}}\frac{\phi(d_1^{kj}(t, S))}{S_t^j},\\
		\partial_{kj}^2 C^{kj}(t, S) &= -\frac{e^{-q_j(T-)}}{\sigma^{kj}\sqrt{T-t}}\frac{K^{kj}\phi(d_2^{kj}(t, S))}{S_k},\quad
		&&\partial_{jj}^2 C^{kj}(t, S) = \frac{e^{-q_j(T-t)}}{\sigma^{kj}\sqrt{T-t}}\frac{K^{kj}\phi(d_2^{kj}(t, S))}{S_t^j},
	\end{align}
    where $\phi$ is the standard normal density.
	Note that $\partial_{jk}^2 C^{kj}(t, S)\equiv \partial_{kj}^2 C^{kj}(t, S)$.

 \end{appendix}
	
\printbibliography[heading=bibintoc, title={References}]

\end{document}